\documentclass[11pt,a4paper,oneside]{article}
\pdfoutput=1

\usepackage[affil-it]{authblk}

\usepackage[a4paper,left=2.5cm,right=2.5cm,top=2.8cm,bottom=3.7cm]{geometry}
	
\usepackage[]{amsmath}
	\numberwithin{equation}{section}
\usepackage[]{amssymb}
\usepackage{amsfonts}
\usepackage{mathrsfs}
\usepackage{amsthm}
	\newtheorem{stm}{Statement}[section]
	\newtheorem*{ansatz}{Ansatz}

\theoremstyle{definition}
	\newtheorem{defin}[stm]{Definition}
	\newtheorem{lemma}[stm]{Lemma}

\newenvironment{digr}{\begin{proof}[Digression]}{\end{proof}}

\DeclareMathOperator{\R}{\mathbb{R}}
\DeclareMathOperator{\C}{\mathbb{C}}
\DeclareMathOperator{\Z}{\mathbb{Z}}
\DeclareMathOperator{\N}{\mathbb{N}}
\DeclareMathOperator{\cs}{\mathbb{S}}
\newcommand{\mz}{\mathcal{Z}}

\newcommand{\mN}{\mathcal{N}}
\newcommand{\mec}{\mathfrak{M}}

\newcommand{\dd}{\mathrm{d}}
\newcommand{\ii}{\mathtt{i}}
\newcommand{\ee}{\mathrm{e}}
\newcommand{\De}{\mathsf{D}}
\newcommand{\kk}{\mathsf{k}}

\def\Xint#1{\mathchoice
   {\XXint\displaystyle\textstyle{#1}}%
   {\XXint\textstyle\scriptstyle{#1}}%
   {\XXint\scriptstyle\scriptscriptstyle{#1}}%
   {\XXint\scriptscriptstyle\scriptscriptstyle{#1}}%
   \!\int}
\def\XXint#1#2#3{{\setbox0=\hbox{$#1{#2#3}{\int}$}
     \vcenter{\hbox{$#2#3$}}\kern-.5\wd0}}
\def\dashint{\Xint-}

\usepackage[utf8]{inputenc}
\usepackage[english]{babel}

\usepackage[pdftex]{graphicx}
\usepackage[dvipsnames]{xcolor}

\usepackage[linktocpage=true,colorlinks=true,citecolor=BlueViolet,linkcolor=blue,urlcolor=BlueViolet]{hyperref}

\usepackage[toc,page]{appendix}
\usepackage[nottoc]{tocbibind}
\usepackage{caption}
\usepackage{enumerate}
\usepackage{ytableau}

\usepackage{tikz}
		\usetikzlibrary{er,positioning,arrows.meta,calc}

\usepackage[numbers,compress,square,comma]{natbib}

\begin{document}
\bibliographystyle{myJHEP}
\captionsetup[figure]{labelfont={bf,small},labelformat={default},labelsep=period,font=small}
\captionsetup[table]{labelfont={bf,small},labelformat={default},labelsep=period,font=small}

{\pagenumbering{roman} 

		\renewcommand*{\thefootnote}{\fnsymbol{footnote}}
	\title{\textbf{\huge Linear Quivers at Large-\texorpdfstring{$N$}{N}}}

\author[$\spadesuit$]{Carlos Nunez\footnote{c.nunez@swansea.ac.uk}}
\affil[$\spadesuit$]{\small Department of Physics, Swansea University,\protect\\ \small Swansea SA2 8PP, United Kingdom \vspace{0.2cm}}

\author[$\heartsuit$]{Leonardo Santilli\footnote{santilli@tsinghua.edu.cn}}
\affil[$\heartsuit$]{\small Yau Mathematical Sciences Center,\protect\\ \small Tsinghua University, Beijing, 100084, China \vspace{0.2cm}}

\author[$\diamondsuit,\clubsuit$]{Konstantin Zarembo\footnote{zarembo@kth.se}}
\affil[$\diamondsuit$]{\small Nordita, KTH Royal Institute of Technology and Stockholm University,\protect\\ \small Hannes Alfv\'ens V\"ag 12, 106 91 Stockholm, Sweden \vspace{0.2cm}}
\affil[$\clubsuit$]{\small Niels Bohr Institute, Copenhagen University,\protect\\ \small Blegdamsvej 17, 2100 Copenhagen, Denmark \vspace{0.2cm}}

	\date{\hspace{8pt}}

	\maketitle
	\thispagestyle{empty}

	\begin{abstract}
		Quiver theories constitute an important class of supersymmetric gauge theories with well-defined holographic duals. Motivated by holographic duality, we use localisation on $\mathbb{S}^d$ to study long linear quivers at large-$N$. 
		The large-$N$ solution shows a remarkable degree of universality across dimensions, including $d=4$ where quivers are genuinely superconformal. In that case we upgrade the solution of long quivers to quivers of any length.
	\end{abstract}

	\clearpage
	\tableofcontents
	\thispagestyle{empty}
}

	\clearpage
	\pagenumbering{arabic}
	\setcounter{page}{1}
		\renewcommand*{\thefootnote}{\arabic{footnote}}
		\setcounter{footnote}{0}

\section{Introduction}
Four-dimensional supersymmetric gauge theories with eight supercharges ($\mN=2$) are an invaluable source of lessons for physicists and mathematicians alike. The amount of supersymmetry allows to extract a wealth of exact results, without forcing a rigid structure and thus including a wide variety of phenomena. Renowned examples of these powerful features are Seiberg--Witten theory \cite{Seiberg:1994rs,Seiberg:1994aj} and the wall-crossing phenomenon for BPS states \cite{Kontsevich:2008fj,Gaiotto:2010okc,Gaiotto:2009hg}. We refer for instance to \cite{Labastida:2005zz,Tachikawa:2013kta,Martone:2020hvy,Akhond:2021xio} for reviews of the many facets of four-dimensional $\mN=2$ theories.\par 
Embedding these quantum field theories in M-theory or Type II string theory yields a top-down perspective on them \cite{Witten:1997sc}. If, on the one hand, the higher-dimensional approach leads to the geometric engineering method \cite{Katz:1996fh,Katz:1997eq} to explore strongly coupled dynamics in four dimensions, on the other hand it entails the holographic principle at large-$N$ \cite{Maldacena:1997re}, which is especially well-suited for explicit computations.\par
Building on \cite{Maldacena:2000mw,Lin:2004nb}, a class of asymptotically-AdS$_5$ supergravity solution was constructed by Gaiotto and Maldacena \cite{Gaiotto:2009gz}, and further explored and generalised in \cite{Reid-Edwards:2010vpm,Aharony:2012tz,Lozano:2016kum,Nunez:2018qcj,Nunez:2019gbg}. The holographic dual superconformal field theories (SCFTs) are four-dimensional $\mN=2$ quiver theories. As will be manifest in our discussion, a necessary condition and a trademark for a four-dimensional quiver to admit a weakly coupled supergravity dual is that the number of gauge nodes of the quiver goes to infinity. The resulting SCFTs are dubbed \emph{long quiver theories}.\par
Supersymmetric localisation \cite{Pestun:2007rz} guarantees that, placing the $\mN=2$ theory on a four-sphere, the computation of several observables reduces to evaluate ordinary, finite-dimensional integrals. The large-$N$ results thus obtained are then instrumental to test the AdS/CFT correspondence, or to make predictions on the holographic dual (see \cite{Zarembo:2016bbk} for a review).\par
\medskip
Long quiver theories are not exclusive to four dimensions: SCFTs with eight supercharges that correspond to conformal fixed points of linear quiver gauge theories exist in all possible dimensions ($d=1,\dots, 6$). It is therefore natural to pursue a field theoretical treatment that works uniformly in $d$ spacetime dimensions. This is one of the main achievements of this paper, whose focus is on $d \ge 3$.\par
A formulation of AdS$_{d+1}$/SCFT$_d$ in the Gaiotto--Maldacena electrostatic formalism was provided in \cite{Akhond:2021ffz} for $d=3$ (see also \cite{Assel:2011xz} for related earlier work); in \cite{Legramandi:2021uds} for $d=5$ (see also \cite{DHoker:2016ujz,DHoker:2016ysh,DHoker:2017mds,DHoker:2017zwj} for related work using different methods) and in \cite{Filippas:2019puw} for $d=6$ (see also \cite{Apruzzi:2013yva,Cremonesi:2015bld,Bobev:2016phc,Bergman:2020bvi} for related work). On the quantum field theory side, the large-$N$ limit of long linear quivers in odd dimensions $d \in \{ 3,5 \}$ has been initiated by Uhlemann \cite{Uhlemann:2019ypp} and further extensively analysed in the subsequent works \cite{Uhlemann:2019lge,Coccia:2020cku,Coccia:2020wtk,Akhond:2022awd,Akhond:2022oaf}.\par 
The intermediate case $d=4$ has not been addressed systematically so far, and we amend the situation in this work.\footnote{Instances of four-dimensional $\mN=2$ conformal theories at large-$N$ have been considered in \cite{Rey:2010ry,Passerini:2011fe,Russo:2012ay,Fiol:2015mrp,Pini:2017ouj}. The  $N_f=2N_c$ superconformal QCD, solved at large-$N$ in \cite{Passerini:2011fe}, can be regarded as the simplest example of quiver CFT where the quiver has but one node.} We tackle long linear quivers comprehensively and extend the existing bibliography in two major ways:
\begin{itemize}
	\item We solve the large-$N$ limit of arbitrary four-dimensional $\mN=2$ linear quiver gauge theories;
	\item We provide a uniform derivation in $d \ge 3$ of the large-$N$ and long quiver limit.
\end{itemize}
In this way, we shed light on the mechanism governing the holographic correspondence, its functioning across dimensions, and the relationship between length of the quiver and supergravity dual.

\subsection{Main results and organisation}
This is a technically dense paper. Here we provide a roadmap of the main results. The rest of this paper is organised in two main blocks:
\begin{itemize}
	\item Section \ref{sec:d} is devoted to long linear quivers in arbitrary dimension $d \ge 3$ (not necessarily integer). The main outcome of this first part is to show that long quivers behave uniformly in $d \ge 3$.
		\begin{itemize}
			\item[$\triangleright$] After preliminaries to set up the analytic continuation in $d$ (Subsections \ref{sec:prelimd}-\ref{sec:rkfn}), we consider the matrix models for the partition functions of linear quiver SCFTs on the $d$-sphere. 
			\item[$\triangleright$] The long quiver limit is solved for arbitrary $d$ in Subsections \ref{sec:LQlimit}-\ref{sec:LQsolved}. We find that the density of eigenvalues, which is a central tool to characterize the matrix model, satisfies a Poisson equation \eqref{eq:Poisson} uniformly for $d \notin 2\N$.
			\item[$\triangleright$] We then compute a quantity $\tilde{F}_d$ that interpolates between the central charges ($d$ even) and free energies ($d$ odd) in Subsection \ref{sec:aF}. A universal expression valid for real $d \ge 3$ is given in \eqref{eq:Ftilded}.
			\item[$\triangleright$] The interpolation across dimensions is extended to defect one-point functions in the long quiver limit.
			\item[$\triangleright$] It is shown in Subsection \ref{sec:mirror} that a certain transformation of the gauge ranks produces identities between the physical observables for the corresponding long quiver gauge theories.
		\end{itemize}
	\item Section \ref{sec:4dLQ} is devoted to linear quivers in $d=4$. The main outcome of this second part is to show that two different large-$N$ regimes exist. In the infinite coupling regime, suitable for comparison with supergravity, we recover the long quiver description.
		\begin{itemize}
			\item[$\triangleright$] $d=4$ is special in the sense that the theory is conformal at arbitrary gauge coupling. We thus first solve the long quiver limit at arbitrary finite 't Hooft coupling (Subsections \ref{sec:4dZlong}-\ref{sec:a4d}), finding that the procedure differs from $d \ne 4$. 
			\item[$\triangleright$] We compute the anomaly coefficient $a$ in $d=4$ obtained from this procedure in Subsection \ref{sec:a4d}, showing agreement both with the result from counting supermultiplets as well as with the limit $d \to 4$ of the interpolating quantity $\tilde{F}_d$ found previously in Subsection \ref{sec:aF}.
			\item[$\triangleright$] Then, in Subsection \ref{sec:4dStrong} we take the strong 't Hooft coupling limit at large-$N$. The ensuing procedure is distinct from the finite-coupling solution, and it is much closer to the long quiver limit in $d \ne 4$. We show that the $d=4$ eigenvalue density at infinite 't Hooft coupling satisfies the same Poisson equation found in Subsection \ref{sec:LQlimit}.
			\item[$\triangleright$] In this strong coupling regime we are able to go beyond the computation of $a$, and we obtain the free energy in Subsection \ref{sec:4dlogZstrong}. Despite the latter quantity being scheme-dependent, we are able to identify and isolate a universal dependence on the length of the quiver and on the gauge ranks.
			\item[$\triangleright$] In Subsection \ref{sec:toysugra} we propose a comparison of the free energy with a supergravity-inspired calculation, finding agreement. Instead of computing the on-shell supergravity action, we take a shortcut in this subsection, which makes the argument only heuristic.
			\item[$\triangleright$] Finally we provide a large-$N$ solution beyond the long quiver limit in $d=4$ in Subsection \ref{sec:4dshortQ}.
		\end{itemize}
\end{itemize} 
The results from both main parts are showcased in concrete examples in Section \ref{sec:ex}. We conclude in Section \ref{sec:outlook} with a summary of our main findings and an outlook on avenues for future research. The main text is complemented with two appendices.

\section{Long quiver SCFTs in arbitrary \texorpdfstring{$d$}{d}}
\label{sec:d}

In this section we study supersymmetric gauge theories in spacetime dimension $d \ge 3$ and compute quantities that interpolate across dimensions.\par
Subsections \ref{sec:prelimd}-\ref{sec:setup} serve to set the stage and to introduce generalities on balanced linear quivers, and in Subsection \ref{sec:rkfn} we introduce the rank function, which is a central quantity in the ensuing computations. The core of the derivation is in Subsections \ref{sec:LQlimit} to \ref{sec:aF}, where the long quiver limit is addressed in arbitrary $d \ge 3$. In the subsequent subsections, further physical observables, such as defect one-point functions, are evaluated.

\subsection{Setup: Supersymmetric gauge theory in \texorpdfstring{$d$}{d} dimensions}
\label{sec:prelimd}

\begin{table}[th]
\centering
	\begin{tabular}{|c| c| c| c| c|}
	\hline
	$d=$ & 3 & 4 & 5 & 6 \\
	\hline
	$\mN=$ & 4 & 2 & 1 & (1,0) \\
	\hline
	\end{tabular}
\caption{Notation for theories with eight supercharges in diverse dimensions. It is customary to indicate the amount of supersymmetry by the number $\mN$ of Killing spinors.}
\label{tab:Nsusyd}
\end{table}
We consider a supersymmetric gauge theory in $d$ Euclidean dimensions with eight supercharges. The notation for this amount of supersymmetry in diverse dimensions is recalled in Table \ref{tab:Nsusyd}. The theory is coupled to the round $d$-sphere $\cs^d$ preserving supersymmetry \cite{Festuccia:2011ws}. The discussion that follows is physically meaningful in the range $3\le d\le 7$ but, as we will show, all the results for physical quantities can be continued to $d \in \R_{\ge 3}$.\par
The gauge group is assumed to be
\begin{equation}
\label{eq:GaugelinearQ}
	G= \mathrm{U}(N_1) \times \mathrm{U} (N_2) \times \cdots \times \mathrm{U} (N_{P-1})
\end{equation}
and the matter content consists in hypermultiplets in the (bi)fundamental representation of the simple factors $\mathrm{U}(N_j)$ of the gauge group. The $\mathrm{U}(N_j)$ gauge factors can be replaced with special unitary gauge groups $\mathrm{SU}(N_j)$, 
\begin{equation*}
	G= \mathrm{SU}(N_1) \times \mathrm{SU} (N_2) \times \cdots \times \mathrm{SU} (N_{P-1})
\end{equation*}
without changing our results. We assume unitary groups for notational simplicity throughout Section \ref{sec:d}, and use special unitary groups later on in Section \ref{sec:4dLQ}, dedicated to $d=4$.\par
The gauge theory is conveniently encoded in a quiver. More precisely, the gauge and matter content of a theory with eight supercharges is encoded in the representation of a framed doubled quiver, in which the nodes indicate gauge groups, the framing (drawn as square nodes) indicate the flavour groups, and unoriented edges (which correspond to pairs of edges with opposite orientation) indicate hypermultiplet in the bifundamental representation.\par
Physically, a framing node of rank $K_j$ gives rise to $K_j$ hypermultiplets in the fundamental representation. From the choice of gauge group \eqref{eq:GaugelinearQ} we can label the nodes of the quiver with an index $j \in \left\{ 1, \dots, P-1 \right\}$, and introduce the arrays 
\begin{equation*}
	\vec{N}=(N_1, \dots, N_{P-1}), \qquad \vec{K}=(K_1, \dots, K_{P-1}) . 
\end{equation*}
We denote by $\mathsf{M}$ the Cartan matrix associated to the quiver, namely twice the identity matrix minus the adjacency matrix. 
\begin{defin}
A quiver gauge theory with eight supercharges is said to be \emph{balanced} if 
\begin{equation*}
	\vec{K} = \mathsf{M} \vec{N} . 
\end{equation*}
We will refer to this condition as the \emph{balancing condition}, and interpret it as a constraint on the flavour ranks for a given choice of gauge group \eqref{eq:GaugelinearQ}.
\end{defin}
Balanced theories yield the unique choice of flavour ranks for which the gauge theory is either conformal, or connected to a CFT via an RG flow controlled by the $d$-dimensional gauge couplings $g_{d,j}$, in the whole window $3 \le d \le 6$. We now quickly recall how this comes about.\footnote{There exist many more SCFTs which do not have a Lagrangian description, and in particular they are not quiver gauge theories. In this work we only consider Lagrangian theories.}
\begin{itemize}
	\item[---] In $d=3$, a UV gauge theory flows at low energies to a strongly coupled fixed point if it is ugly or good in the Gaiotto--Witten classification \cite{Gaiotto:2008ak}, meaning $K_j \ge \sum_{k} \mathsf{M}_{jk} N_k -1$. The IR fixed point of an ugly theory is the direct sum of the interacting SCFT of a good quiver and decoupled free fields. Without loss of generality, we can thus consider good theories only, for which balanced theories provide a lower bound.
	\item[---] In $d=4$, all the Lagrangian SCFTs must be balanced. Indeed, the beta function, which is one-loop exact thanks to the $\mathcal{N}=2$ supersymmetry, reads:
		\begin{equation}
 			\beta (g_{4,j})=\frac{g_{4,j}^3}{16\pi ^2}\left(K_j- \sum_{k} \mathsf{M}_{jk} N_k\right) 
		\end{equation}
		and vanishes as soon as the balancing condition is imposed.
	\item[---] In $d=5$, a UV complete gauge theory is obtained from a relevant deformation of a strongly coupled SCFT. A sufficient condition for the gauge theory to sit in the IR of one such SCFT is that $K_j \le  \sum_{k} \mathsf{M}_{jk} N_k  $ \cite{Intriligator:1997pq}. This is equivalent to demanding the convergence of the perturbative partition function. Slightly less stringent conditions can be imposed in some cases \cite{Jefferson:2018irk} (see \cite{Bhardwaj:2019ngx} for lists of examples). However, due to the lack of analytic control over instantons, we will have to work with the perturbative partition function throughout, which forces the previous inequality.
	\item[---] In $d=6$, cancellation of gauge anomalies in any gauge theory obtained by a relevant deformation of a SCFT requires the balancing condition.
\end{itemize}\par
\bigskip
Supersymmetric localisation reduces the partition function on $\cs^d$ to a matrix model \cite{Pestun:2007rz}, schematically of the form
\begin{equation}
\label{eq:Zd}
	\mz_{\cs^d} = \int_{\mathfrak{t}} \dd \vec{\phi} ~ \ee^{- S_{\mathrm{cl}} (\vec{\phi}) } ~ Z_{\mathrm{vec}} Z_{\mathrm{hyp}} Z_{\text{\rm non-pert}}  .
\end{equation}
In this formula, 
\begin{itemize}
	\item $\mathfrak{t} \cong \R^{\mathrm{rk} (G)}$ denotes a Cartan subalgebra of the Lie algebra of the gauge group $G$. The integration variable $\vec{\phi} \in \mathfrak{t}$ is the zero-mode of a real scalar in the vector multiplet, and $\dd \vec{\phi}$ is the Lebesgue measure on $\mathfrak{t}$.
	\item $S_{\mathrm{cl}}$ is the classical action evaluated at the localisation locus. 
	\item $Z_{\mathrm{vec}} $ and $ Z_{\mathrm{hyp}} $ are the one-loop determinants from the vector and hypermultiplet.
	\item In addition, there is a non-perturbative part $Z_{\text{\rm non-pert}}$ from instantons. We set this last term to 1, because the instanton corrections are non-perturbatively suppressed in the planar large-$N$ limit considered below.
\end{itemize}
We use the index $j=1, \dots, P-1 $ to label the gauge node $\mathrm{U}(N_j)$ and the index $a=1, \dots, N_j$ to label the basis of the Cartan of $\mathfrak{u}(N_j)$, so that the integration variables are 
\begin{equation*}
	\vec{\phi} = \left\{ \phi ^{(j)}_a \right\}^{j=1, \dots, P-1}_{a=1, \dots, N_j} .
\end{equation*}\par
\begin{digr}
We are presenting the partition function in the so-called Coulomb branch localization. Expression \eqref{eq:Zd} has been derived for integer $1\le d \le 5$, while more supercharges are required for the Coulomb branch localization in $d \ge 6$. However, we will work with arbitrary $d$, where the integration variables need not have an interpretation as Coulomb branch parameters. As explained in Subsection \ref{sec:aF}, provided a valid analytic continuation of $\mz_{\cs^d}$, the limit $d \to 6^{-}$ will bear physical significance, even though the integral presentation \eqref{eq:Zd} may not be valid. The conclusion is independent of how the analytic continuation has been obtained, and remains valid for CFTs without supersymmetry, for which no localization is available and one the partition function must be computed by other methods.\par
\end{digr}

\subsection{Setup: Linear quivers in \texorpdfstring{$d$}{d} dimensions}
\label{sec:setup}
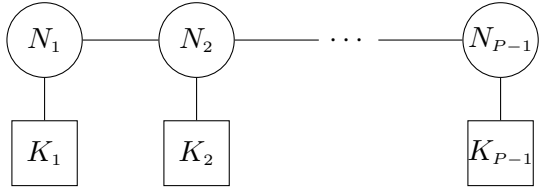
\begin{figure}[t]
\centering
\begin{tikzpicture}[auto,square/.style={regular polygon,regular polygon sides=4}]
	\node[circle,draw] (gauge1) at (3,0) { \hspace{20pt} };
	\node (a1) at (3,0) {$N_{\scriptscriptstyle P-1}$};
	\node[draw=none] (gaugemid) at (1,0) {$\cdots$};
	\node[circle,draw] (gauge3) at (-1,0) { \hspace{20pt} };
	\node[circle,draw] (gauge4) at (-3,0) { \hspace{20pt} };
	\node (a2) at (-3,0) {$N_{\scriptscriptstyle 1}$};
	\node (a3) at (-1,0) {$N_{\scriptscriptstyle 2}$};
	\node[square,draw] (fl1) at (3,-1.5) { \hspace{10pt} };
	\node[square,draw] (fl2) at (-3,-1.5) { \hspace{10pt} };
	\node[square,draw] (fl3) at (-1,-1.5) { \hspace{10pt} };
	\node[draw=none] (aux1) at (3,-1.5) {$K_{\scriptscriptstyle P-1}$};
	\node[draw=none] (aux2) at (-3,-1.5) {$K_{\scriptscriptstyle 1}$};
	\node[draw=none] (aux3) at (-1,-1.5) {$K_{\scriptscriptstyle 2}$};
	\draw[-](gauge1)--(gaugemid);
	\draw[-](gaugemid)--(gauge3);
	\draw[-](gauge4)--(gauge3);
	\draw[-](gauge1)--(fl1);
	\draw[-](gauge4)--(fl2);
	\draw[-](gauge3)--(fl3);
\end{tikzpicture}
\caption{Linear quiver of length $P-1$. Circular nodes indicate gauge groups, square nodes indicate flavour symmetries.}
\label{fig:quiver1}
\end{figure}\par

From now on, we restrict our attention to \emph{linear} quivers. These correspond to quiver gauge theories with gauge group \eqref{eq:GaugelinearQ}, or its version with special unitary instead of unitary groups, and hypermultiplets in the bifundamental representation of $\mathrm{U}(N_j) \times \mathrm{U} (N_{j+1})$, with additional $K_j$ hypermultiplets in the fundamental representation of $\mathrm{U}(N_j)$. These data are encoded in a linear quiver of length $P-1$, shown in Figure \ref{fig:quiver1}.\par 
We moreover restrict our attention to \emph{balanced} quivers. Applying the definition, a linear quiver is balanced if the numbers of flavours $K_j$ satisfy 
\begin{equation}
\label{eq:balance}
	K_j = 2N_j - N_{j+1} - N_{j-1} 
\end{equation}
$\forall~j=1, \dots, P-1$.\par
We will consider Yang--Mills theories with eight supercharges and no Chern--Simons term,\footnote{In $d=5$ a non-vanishing Chern--Simons term generically requires the theory to be unbalanced. For unbalanced $d=5$ theories whose Chern--Simons term is dynamically generated by integrating out hypermultiplets from a balanced linear quiver, our procedure still applies with few modifications. In $d=3$, only a suitably restricted class of Chern--Simons terms preserves eight supercharges. Equalities of partition functions among a subset of these Chern--Simons-matter theories and linear quivers have been derived in \cite{Santilli:2020snh,Li:2023ffx}, thus our results will apply to that particular subset.} and study their sphere partition functions \eqref{eq:Zd}. The classical action evaluated on the localisation locus reads 
\begin{equation*}
	S_{\mathrm{cl}} (\vec{\phi}) = \frac{\pi^{(d-3)/2}}{\Gamma \left( \frac{d-3}{2}\right)} \sum_{j=1} ^{P-1} \frac{8 \pi^2}{g_{d, j}^2} \sum_{a=1} ^{N_j} \left( \phi^{(j)} _a \right)^2 ,
\end{equation*}
where $g_{d,j}$ is the gauge coupling for the $j^{\text{th}}$ gauge node in dimension $d$, and we understand $\frac{1}{g_{d <4}} =0$. This is because, while the localised partition function is calculated in the gauge theory, it is a protected quantity and the theory flows to strong coupling in the IR if $d<4$. In integer dimension, $\frac{1}{g_{d <4}} =0$ is a consequence of the $Q$-exactness of the Yang--Mills Lagrangian, and we extend this condition to real $3 \le d <4$.\par
In view of the large-$N$ planar limit, we define the 't Hooft couplings $\left\{ t_{d,j} \right\}$ in $d \ge 4$ according to 
\begin{equation}
\label{eq:deftHoofttj}
	t_{d,j} := \frac{Ng_{d,j}^2}{16 \pi^2} , \qquad \forall~j=1, \dots, P-1 .
\end{equation}
To lighten the formulae, we also adopt the shorthand notation
\begin{equation*}
	u_d := \frac{\pi^{(d-3)/2}}{\Gamma \left( \frac{d-3}{2}\right)} .
\end{equation*}\par
Altogether we write 
\begin{equation}
\label{eq:ZSeff}
	\mz_{\cs^d} = \int_{\mathfrak{t}} \dd \vec{\phi} ~ \exp \left\{ -S_{\mathrm{eff}} (\vec{\phi}) \right\}~ Z_{\mathrm{non-pert}} 
\end{equation}
where
\begin{equation*}
\begin{aligned}
	S_{\mathrm{eff}} (\vec{\phi}) = \sum_{j=1} ^{P-1} & \left\{ \frac{u_d}{2 t_j} N\sum_{a=1} ^{N_j} (\phi ^{(j)}_a)^2  + \sum_{1 \le a \ne b \le N_j} f_{\mathrm{v}} \left( \phi^{(j)} _a  -  \phi^{(j)} _b \right)  \right. \\
	& \qquad \left. + K_j \sum_{a=1} ^{N_j} f_{\mathrm{h}} \left( \phi^{(j)} _a \right) + (1- \delta_{j,P-1}) \sum_{a=1} ^{N_j} \sum_{b=1} ^{N_{j+1}}  f_{\mathrm{h}} \left( \phi^{(j)}_a - \phi^{(j+1)}_b \right)\right\} .
\end{aligned}
\end{equation*}
In this expression, $f_{\mathrm{h}}$ captures the one-loop contribution from each hypermultiplet zero-mode and $f_{\mathrm{v}}$ includes the one-loop contribution of the W-bosons and their fermionic superpartners, as well as a Vandermonde term from the measure. These functions have been explicitly derived by localisation for $d \in [3,7] \cap \N$, and expressions valid for $d \in \R_{\ge 3}$ that interpolates between the known functions on the integers were provided by Minahan \cite{Minahan:2015any}. We collect them in Appendix \ref{app:Minahan}.\par
For later convenience, let us introduce the indicator function  
\begin{equation}
\label{eq:defthetaS}
	\theta_{\mathscr{S}} := \begin{cases} 1 & \text{if $\mathscr{S}$ is true}\\ 0 & \text{otherwise} \end{cases}
\end{equation}
for any condition $\mathscr{S}$. We define the function
\begin{equation}
\label{eq:deff0}
	f_0 (\phi) := f_{\mathrm{h}} (\phi) + f_{\mathrm{v}} (\phi) \theta_{\phi \ne 0} . 
\end{equation}
It is also convenient to isolate the term proportional to $\delta_{j,P-1}$. It is a contribution from the tail of the quiver, which avoids overcounting of the bifundamental hypermultiplets:
\begin{equation*}
	\mathfrak{B} := - \sum_{j=1}^{P-1} \delta_{j,P-1} \sum_{a=1} ^{N_j} \sum_{b=1} ^{N_{j+1}}  f_{\mathrm{h}} \left( \phi^{(j)}_a - \phi^{(j+1)}_b \right) = -  \sum_{a=1} ^{N_{P-1}}  f_{\mathrm{h}} \left( \phi^{(P-1)}_a \right) .
\end{equation*}\par

\subsection{Setup: Rank function}
\label{sec:rkfn}
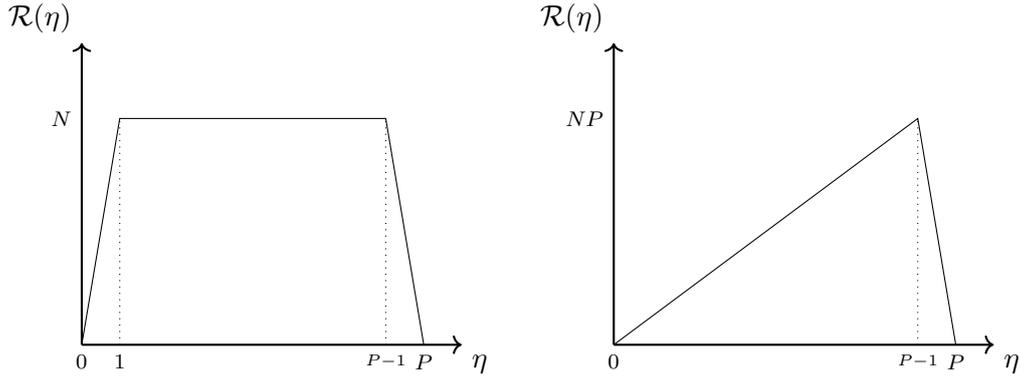
\begin{figure}[t]
\centering
\begin{tikzpicture}
	\draw[->,thick] (-6,0) -- (-1,0);
	\draw[->,thick] (1,0) -- (6,0);
	\draw[->,thick] (-6,0) -- (-6,4);
	\draw[->,thick] (1,0) -- (1,4);
	
	
	\draw[-] (1,0) -- (5,3) -- (5.5,0);
	\draw[-] (-6,0) -- (-5.5,3) -- (-2,3) -- (-1.5,0);
	
	\node[anchor=north west] at (-1,0) {$\eta$};
	\node[anchor=north west] at (6,0) {$\eta$};
	\node[anchor=south east] at (-6,4) {$\mathcal{R} (\eta)$};
	\node[anchor=south east] at (1,4) {$\mathcal{R} (\eta)$};
	\node[anchor=east] at (-6,3) {$\scriptstyle N$};
	\node[anchor=east] at (1,3) {$\scriptstyle NP$};
	\node[anchor=north] at (-6,0) {$\scriptstyle 0$};
	\node[anchor=north] at (1,0) {$\scriptstyle 0$};
	\node[anchor=north] at (-1.5,0) {$\scriptstyle P$};
	\node[anchor=north] at (-2,0) {$\scriptscriptstyle P-1$};
	\draw[thin,dotted] (-2,0) -- (-2,3);
	\node[anchor=north] at (-5.5,0) {$\scriptstyle 1$};
	\draw[thin,dotted] (-5.5,0) -- (-5.5,3);
	\node[anchor=north] at (5.5,0) {$\scriptstyle P$};
	\node[anchor=north] at (5,0) {$\scriptscriptstyle P-1$};
	\draw[thin,dotted] (5,0) -- (5,3);

\end{tikzpicture}
\caption{Two sample plots of the rank function $\mathcal{R} (\eta)$. Left: first example, also known as $+_{N,P}$ theory. Right: second example, also known as $T_P$ theory if $N=1$.}
\label{fig:rankplusTN}
\end{figure}\par

We introduce a continuous variable $0 \le \eta \le P$ and let $\mathcal{R} (\eta)$ be the rank function. 
\begin{defin}
The rank function $\mathcal{R} (\eta)$ is the piece-wise linear function on $[0,P]$ that satisfies 
\begin{equation*}
	\mathcal{R} (0)=0=\mathcal{R} (P), \qquad \mathcal{R} (j)=N_j .
\end{equation*}
\end{defin}
In terms of the rank function, the balancing condition \eqref{eq:balance} reads 
\begin{equation}
\label{eq:d2rdeta2}
	\partial_\eta ^2 \mathcal{R} (\eta) = - \sum_{j=1}^{P-1} K_j \delta\left( \eta - j \right) .
\end{equation}
Two examples of rank function for balanced linear quivers are shown in Figure \ref{fig:rankplusTN}.\par
With the large-$N$ limit in mind, it is convenient to fix $N \in \mathbb{N}$ and write $N_j = \nu_j N$ for some rational number $\nu_j$. We then introduce the scaled variable 
\begin{equation*}
	z:= \frac{\eta}{P} , \qquad 0 \le z \le 1 , 
\end{equation*}
and the scaled rank function 
\begin{equation}
	\nu (z) := \frac{1}{N} \mathcal{R} (P z) .
\end{equation}
To work in the planar limit, we also introduce the Veneziano parameters 
\begin{equation*}
	\zeta_j := \frac{K_j}{N} , 
\end{equation*}
and gather them in the scaled flavour rank function 
\begin{equation*}
	\zeta (z)= \frac{1}{P} \sum_{j=1}^{P-1} \zeta_j \delta\left( z- \frac{j}{P}\right).
\end{equation*}
Imposing the balancing condition \eqref{eq:d2rdeta2}, these redefinitions are subject to 
\begin{center}\noindent\fbox{\parbox{0.98\linewidth}{%
\begin{equation}
\label{eq:dnudziszeta}
	\partial_z ^2 \nu (z) = - P^2 \zeta (z) .
\end{equation}}}\end{center}
The normalisation of $\nu(z)$ and $\zeta (z)$ is chosen so that all the quantities that follow have a well-defined large-$N$ and large-$P$ limit. In the same vein, we replace the 't Hooft couplings $\left\{ t_{d,j}\right\}$ with a continuous function $t_d (z)$ on $[0,1]$.\par
\bigskip
According to the definition, the Fourier expansion of a generic $\mathcal{R} (\eta)$ is given by
\begin{equation}\label{Fourier-R}
	\mathcal{R} (\eta) = \sum_{k=1}^{\infty} R_k \sin \left( \frac{k \pi \eta}{P} \right).
\end{equation}
From here, it follows that 
\begin{equation}
\label{eq:Fouriernu}
	\nu (z) = \frac{1}{N} \sum_{k=1}^{\infty} R_k \sin (k \pi z)
\end{equation}
and 
\begin{equation}
\label{eq:Fourierzeta}
	P^2\zeta (z) = \frac{\pi^2}{N} \sum_{k=1}^{\infty} k^2 R_k \sin (k \pi z) .
\end{equation}
We also define the Fourier expansion of the inverse gauge coupling:
\begin{equation}
\label{eq:Fourierg}
	\frac{1}{g_d (z)} = \sum_{k=1} \left[ \gamma_k \sin (k \pi z)  + \tilde{\gamma}_k \cos (k \pi z) \right] .
\end{equation}

\subsection{Long quiver limit}
\label{sec:LQlimit}

Our goal is to address the planar limit of \eqref{eq:ZSeff}, extending \cite{Uhlemann:2019ypp}. For the linear quivers of Figure \ref{fig:quiver1}, we take a planar limit in which moreover $P \to \infty$. It goes under the name of \emph{long quiver} limit. 
\begin{defin}
The long quiver limit of a linear quiver of $P-1$ nodes is the planar large-$N$ and large-$P$ limit with 't Hooft scaling of the gauge couplings. We use the symbol $\simeq$ to indicate asymptotic equality in the long quiver limit.
\end{defin}
The goal of this subsection and the next one is to prove that the structure observed in \cite{Legramandi:2021uds,Akhond:2021ffz,Akhond:2022oaf} for the long quiver limit in $d\in \left\{3,5\right\}$ extends to arbitrary $d$.
\begin{stm}
	Let $d \in \R_{\ge 3} \setminus 2\Z$. The long quiver limit of a balanced linear quiver is governed by a function 
	\begin{equation*}
		\varrho \ : \ [0,1] \times \R \ \to \ \R_{\ge 0} 
	\end{equation*}
	that solves the Poisson equation 
	\begin{equation*}
		\frac{1}{4} \partial_x ^2 \varrho (z,x) + \partial_z^2 \varrho (z,x) - \partial_z^2 \nu (z)\delta (x) = 0 ,
	\end{equation*}
	for $0<z<1$ and $x \in \R$. 
\end{stm}\par
\bigskip
To prepare the stage, we introduce the eigenvalue densities 
\begin{equation*}
	\rho_j (\phi) := \frac{1}{N} \sum_{a=1} ^{N_j} \delta \left( \phi - \phi ^{(j)}_a \right) , \qquad \phi \in \R.
\end{equation*}
Note that they carry an overall factor $1/N$, as opposed to the customary $1/N_j$. Hence, they are not canonically normalised to 1. We can collect all the eigenvalue densities for each gauge factor in the function of two variables $\rho (z,\phi)$ with $\rho \left(\frac{j}{P},\phi \right) = \rho_j (\phi)$. This new eigenvalue density is normalised as 
\begin{equation*}
	\int_{\R} \dd \phi \rho (z,\phi) = \nu (z) \qquad \forall ~0 < z < 1 .
\end{equation*}
In the latter expressions $\phi \in \R$ is a one-dimensional integration variable, related to but different from the $\mathrm{rk}(G)$-dimensional vector $\vec{\phi}$.\par
With these definitions, the matrix model effective action in \eqref{eq:ZSeff} is written as a functional of $\rho$, in the form 
\begin{equation}
	\begin{aligned}
	S_{\mathrm{eff}} [\rho] = N^2 P \int_0^{1} \dd z \int \dd \phi \rho (z,\phi)  & \left\{  \frac{u_d}{2 t_d (z)} \phi^2  +  \dashint \dd \sigma   \rho (z,\sigma) f_{\mathrm{v}} (\phi - \sigma) \right. \\
	& \left. + \zeta (z) f_{\mathrm{h}} (\phi) +  \int \dd \sigma \rho (z+\delta z, \sigma ) f_{\mathrm{h}} (\phi - \sigma ) \right\}  + \mathfrak{B} .
\end{aligned}
\label{eq:Seffdlong1}
\end{equation}
Replacing each sum over $a=1, \dots, N_j$ with an integral over $\phi$ yields an overall factor of $N$, and the factor of $P$ comes from replacing the sum over $j=1, \dots, P-1$ with an integral over $z$. The symbol $\dashint$ stands for the principal value integral. Besides, we are using the notation $\delta z =\frac{1}{P}$ in the contribution from bifundamental hypermultiplets.\par
To put \eqref{eq:Seffdlong1} in a more convenient form, we symmetrise this last term:
\begin{equation}
\label{eq:symmbifund}
	 \rho (z+\delta z, \sigma ) \ \mapsto  \ \frac{1}{2} \left[  \rho (z+\delta z, \sigma ) + \rho (z-\delta z, \sigma ) \right] , \qquad \forall z \in (0,1) .
\end{equation}
One may check directly the invariance of $S_{\mathrm{eff}} [\rho]$ under this replacement, noting that $f_{\mathrm{h}}$ is an even function for every $d$. A more straightforward way to see it, is that we have labelled the bifundamental hypermultiplets with the $j$ of the gauge node at its left, leading to discard the term at $j=P-1$. We could have simply labelled them with the $j$ from right, which then would lead to not account for the term $j=1$. The more symmetric option is to divide the contribution of each bifundamental by $2$ and label half from the left and half from the right, with the boundary term written as 
\begin{equation*}
	\mathfrak{B} = - \frac{1}{2}\sum_{j=1}^{P-1} \left[   \delta_{j,P-1}   \sum_{a=1} ^{N_j} \sum_{b=1} ^{N_{j+1}} f_{\mathrm{h}} \left( \phi^{(j)}_a \right) + \delta_{j,1}  \sum_{a=1} ^{N_j} \sum_{b=1} ^{N_{j-1}} f_{\mathrm{h}} \left( \phi^{(j)}_a \right)\right]  .
\end{equation*}
We use the trivial identities $\delta_{j,P-1}=\delta_{j+1,P}$, $\delta_{j,1}=\delta_{j-1,0}$, and the replacements $ \delta_{j+1,P} \mapsto \frac{1}{P} \delta (z + \delta z -1)$, $ \delta_{j-1,0} \mapsto \frac{1}{P} \delta (z - \delta z)$ inside $\mathfrak{B}$. These manipulations produce a manifest factor $1/P$ compared to the contributions form the rest of the quiver, suited for the planar limit.
Inserting the eigenvalue density, this expression becomes
\begin{equation*}
		\mathfrak{B} = - N \left[ \lim_{z \to 0^{+}} \int \dd \phi \rho (z,\phi)  f_{\mathrm{h}} (\phi  ) + \lim_{z \to 1^{-}} \int \dd \phi \rho (z,\phi)  f_{\mathrm{h}} (\phi ) \right] .
\end{equation*}
Here we have written the limit $\delta z \to 0^{+}$, evaluated at $z-\delta z =0$, as $z \to 0^{+}$, and likewise for $z \to 1^{-}$ in the second term. This expression for $\mathfrak{B}$ is equivalent to the contribution of the bifundamental hypermultiplets at the first and last node, if we define $\rho (z,\sigma):= \delta (\sigma)$ if $z \notin (0,1)$.\par
Back to the leading contribution \eqref{eq:Seffdlong1}, we analyse the piece \eqref{eq:symmbifund}. At leading order in the large-$P$ limit, we have \cite{Uhlemann:2019ypp}
\begin{equation*}
	 \frac{1}{2} \left[ \rho (z+\delta z, \sigma ) + \rho (z-\delta z, \sigma ) \right] - \rho (z, \phi) \ \to \ \frac{1}{2P^2} \partial_z ^2 \rho (z, \phi) ,
\end{equation*}
and we arrive at 
\begin{equation}
\begin{aligned}
	S_{\mathrm{eff}} [\rho] \simeq N^2 P \int_{0} ^{1} \dd z  & \int \dd \phi \rho (z, \phi) \left\{ \frac{u_d }{2 t_d (z)}  \phi^2 + \zeta (z) f_{\mathrm{h}} (\phi) \right. \\
	& \left. + \int \dd \sigma  \left[  \rho (z, \sigma) f_{0} (\phi - \sigma)  + \frac{1}{2 P^2} \partial^2_z \rho (z,\sigma) f_{\mathrm{h}} (\phi - \sigma)  \right] \right\}  \ + \mathfrak{B},
\end{aligned}
\label{eq:Seffdlong2}
\end{equation}
where $f_0$ was defined in \eqref{eq:deff0}, and $f_0 =f_{\mathrm{v}}+f_{\mathrm{h}}$ except when the argument is zero. The piece $\mathfrak{B}$ from the boundaries of the quiver is negligible compared to the $\mathcal{O}(N^2P)$ leading contributions from the rest of nodes of the quiver. Stated more precisely, the pieces proportional to $\delta_{j,P-1}$ and $\delta_{j,1}$ have measure 0 in the long quiver limit, and therefore do not appear in the effective action as long as we can neglect the more fine-grained discretised structure of the $z$-interval.\par
\bigskip
The planar limit of the long linear quivers above has been studied in $d=5$ and $d=3$. Here, we leverage the knowledge of the analytic continuation in the dimension $d$ of the main ingredients $f_{\mathrm{v}}, f_{\mathrm{h}}$ \cite{Minahan:2015any}.\par
Instead of writing the formulae of \cite{Minahan:2015any} for the one-loop determinants (cf. Appendix \ref{app:Minahan}), we note that the functions $f_{\mathrm{h}}$ and $f_{0}$ behave at large argument as 
\begin{equation}
\label{eq:flargephi}
	f_{\mathrm{h}} (\phi) \simeq s_d \lvert \phi \rvert^{d-2} , \qquad f_0 (\phi) \simeq \tilde{s}_d \partial_{\phi} ^2 f_{\mathrm{h}} (\phi) ,
\end{equation}
where the $d$-dependent coefficients are 
\begin{equation}
\label{eq:sdcoefs}
	s_d = 2 \cos \left( \frac{\pi d}{2}\right) \Gamma (2-d) , \qquad  \tilde{s}_d = \frac{1}{8} (d-2)^2  .
\end{equation}
The coefficient $s_d$ diverges at even integer $d$, which signals that the expression homogeneous in $d$ is not suitable there. Indeed, in $d=4$ the naive $\phi^2$ scaling of $f_{\mathrm{h}}$ should be replaced by $\phi^2 \ln \lvert \phi \rvert$ \cite{Pestun:2007rz}, see for instance \cite[Eq.(2.8)]{Russo:2012ay}.\par
In the planar limit, we are led to study the saddle points of the effective action. The saddle point equation is 
\begin{equation}
\label{eq:spe1}
	\int \dd \sigma ~ \partial_{\bar{\phi}}  \left[ f_{0} (\bar{\phi}) + \frac{1}{2 P^2}  f_{\mathrm{h}} (\bar{\phi}) \partial^2_z  ~ \right]_{\bar{\phi} =\phi - \sigma} \rho (z, \sigma)  = - \frac{u_d}{2 t_d (z)} \phi  - \frac{\zeta (z)}{2} \partial_{\phi} f_{\mathrm{h}} (\phi) .
\end{equation}
The derivation of this equation is standard, thus we omit the details. See Appendix \ref{app:3SPE} for additional comments on \eqref{eq:spe1}.\par
\medskip
To proceed further, we make an ansatz about the long quiver behaviour of the eigenvalues.
\begin{ansatz} In the long quiver limit, the leading contribution comes from the eigenvalues that scale as 
\begin{equation}
\label{eq:LQansatz}
	\phi = (d-2)P^{\chi} x , \qquad \chi >0 
\end{equation}
with $x$ fixed and $\mathcal{O}(1)$ in the limit. The overall factor $(d-2)$ is just for convenience, whereas the crucial aspect of the ansatz is $\chi >0$, with strict inequality.
\end{ansatz}
The ansatz is an extension of the know behaviour at $d=5$, and we allow $\chi = \chi (d)$ to possibly depend on $d$. The value of $\chi$ and the self-consistency of \eqref{eq:LQansatz} will be established in the following.\par
We utilise the ansatz \eqref{eq:LQansatz}, and also define a new density of eigenvalues $\varrho (z,x)$ to make the substitution of measures 
\begin{equation*}
	\varrho (z,x) \dd x = \rho (z,\phi) \dd \phi .
\end{equation*}
Inserting the ansatz \eqref{eq:LQansatz} into the asymptotic expressions \eqref{eq:flargephi}, we get that the quantities in \eqref{eq:spe1} scale as 
\begin{align*}
	\left. \partial_{\bar{\phi}} f_{\mathrm{h}} (\bar{\phi}) \right\rvert_{\bar{\phi} = \phi - \sigma} & \approx \left. s_d P^{\chi (d-3)} (d-2)^{d-3}\partial_{\xi} \lvert \xi \rvert^{d-2} \right\rvert_{\xi=x-y} \\
	\left. \partial_{\bar{\phi}} f_0 (\bar{\phi})  \right\rvert_{\bar{\phi} = \phi - \sigma}	 	& \approx \left. \tilde{s}_d  s_d P^{\chi (d-5)} (d-2)^{d-5} \partial_{\xi} ^3 \lvert \xi \rvert^{d-2}  \right\rvert_{\xi=x-y} = \frac{1}{8}  \left. s_d P^{\chi (d-5)} (d-2)^{d-3} \partial_{\xi} ^3 \lvert \xi \rvert^{d-2}  \right\rvert_{\xi=x-y} .
\end{align*}
For the 't Hooft coupling, we impose the long quiver scaling 
\begin{equation*}
	t_d =P^{\tau} \tilde{t}_d , \qquad \tau >0 
\end{equation*}
and get the contribution 
\begin{equation*}
	\frac{\phi}{2 t_d (z)} = P^{\chi - \tau } (d-2) \frac{x}{2 \tilde{t}_d (z)} .
\end{equation*}
The powers $\chi, \tau$ will be fixed demanding self-consistency of the equilibrium equation.\par
Integrating by parts twice the $f_0$ term in \eqref{eq:spe1} after the substitution, we arrive at 
\begin{equation}
\label{eq:sped-1}
\begin{aligned}
	\int \dd y ~ & (d-2)^{d-3} s_d \partial_x \lvert x-y \rvert^{d-2} ~ \left[  P^{(d-3)\chi -2 }  \cdot \frac{1}{2} \partial_z^2  \varrho (z,y) +  P^{(d-5)\chi } \cdot \frac{1}{8} \partial_y ^2 \varrho (z,y)\right] \\
	& = - P^{\chi - \tau } (d-2) u_d \frac{ x }{2 \tilde{t}_d (z)} - s_d (d-2)^{d-3} P^{(d-3)\chi -2 } \frac{ [P^2 \zeta (z)]}{2}  \partial_x \lvert x\rvert^{d-2} .
\end{aligned}
\end{equation}
We are using the fact that $P^2 \zeta(z) $ is finite in the large-$P$ limit. We have also implicitly used the assumption that $\varrho (z,x)$ falls of faster than any polynomial as $\vert x \vert \to \infty$, to show that the boundary terms from integration by part vanish. This assumption is a necessary condition for the Borel measure $\varrho (z,x)\dd x$ to have finite moments, meaning that gauge-invariant combinations of the Coulomb branch scalar have normalizable expectation value on $\mathbb{S}^d$. This assumption is satisfied in all known examples (not only for long quivers), and will indeed be satisfied by the solution we find below.\par
Note that we work in $d \ge 3$, thus we always have 
\begin{equation*}
	\left. \partial_{\xi} \lvert \xi \rvert^{d-2}  \right\rvert_{\xi=x-y} = (d-2) (x-y)^{d-3} \mathrm{sign} (x-y) .
\end{equation*}
We will loosely write the above as $\lvert x-y \rvert^{d-3}$, with the understanding that in the limit $d \to 3^{+}$ we have the sign function. Furthermore, working in $d \ge 3$, we can always divide both sides by $\frac{1}{2} P^{(d-3)\chi -2} (d-2)^{d-2}$ and define 
\begin{equation}
\label{eq:wdcoefs}
	w_d := - (d-2)^{3-d} u_d = - (d-2)^{3-d} \frac{\pi^{(d-3)/2}}{\Gamma \left( \frac{d-3}{2}\right)} .
\end{equation}
With these elementary manipulations, the saddle point equation \eqref{eq:spe1} is cast in the form 
\begin{equation}
\label{eq:sped-2}
	s_d \int \dd y \lvert x-y \rvert^{d-3} ~ \left[  \partial_z^2  \varrho (z,y) +  \frac{1}{4} P^{2-2\chi } \partial_y ^2 \varrho (z,y) \right] = P^{(4-d)\chi +2 - \tau } w_d \frac{x}{\tilde{t}_d (z)} - s_d P^2 \zeta (z) \lvert x\rvert^{d-3} .
\end{equation}\par
We need to inspect this equation and determine $\chi, \tau$ such that it admits a non-trivial solution. First, note that the first and last term are $\mathcal{O}(1)$. Therefore, to cancel the powers of $P$, we demand 
\begin{equation*}
	2 \chi -2=0, 
\end{equation*}
that is, $\chi=1$. It satisfies the self-consistency condition $\chi>0$. To keep the 't Hooft couplings as well, we impose $\tau= (4-d)\chi+2= 6-d$, consistent with $\tau>0$ in the region $d<6$. To sum up, 
\begin{itemize}
	\item[---] If $3 \le d <4$ we set $t_{d<4} (z)= \infty$;
	\item[---] If $4 \le d <6$ the long quiver admits a finite 't Hooft coupling with scaling $t_{4 \le d<6} (z) =P^{6-d} \tilde{t}_d (z)$;
	\item[---] If $d>6$ we set $t_{d>6} (z)= \infty$.
\end{itemize}
From now on, we set $\chi=1$ and $\tau=6-d$. We stress that, while $t_{d<4} (z)= \infty$ is a natural consequence of the $Q$-exactness of the Yang--Mills Lagrangian in lower dimensions, we have to impose $t_{d>6} (z)= \infty$ for consistency with the procedure.\par

\subsection{Long quiver solution in arbitrary \texorpdfstring{$d$}{d}}
\label{sec:LQsolved}
The various terms in the saddle point equation \eqref{eq:sped-2} bear a coefficient $s_d$, defined in \eqref{eq:sdcoefs}, except for the Yang--Mills contribution, which has a coefficient $w_d$, defined in \eqref{eq:wdcoefs}. We thus need to study the zeros and poles of these coefficients as functions of $d$ to understand the structure of the solution to \eqref{eq:sped-2}.\par
The numerical coefficient $w_d$ of the gauge coupling is a smooth real function on $d\ge 3$. It is negative on the whole $3<d<\infty$ and vanishes at $d=3$ and $d \to \infty$; it has a unique minimum near $d \approx 4.2$ and moreover $w_4=- \frac{1}{2}$.\par
Therefore, at $d=3$, we have $w_3=0$, the gauge coupling drops out, and all the remaining terms in \eqref{eq:sped-2} have the same coeficient $s_{3}=\pi$, which can be simplified. On the other hand, for every $3<d<\infty$, we can divide both sides of \eqref{eq:sped-2} by $w_d$ and define $C_d:=\frac{s_d}{w_d}$, explicitly\footnote{Recall that we have absorbed a coefficient $16\pi^2$ in the definition of $t_d$ in \eqref{eq:deftHoofttj}. We might as well have defined $t_d =Ng_d$ and eventually have reabsorbed the $16\pi^2$ in the definition of $w_d$ and $C_d$.}  
\begin{equation}
\label{eq:totcoefCd}
	C_d  = 2 \left( \frac{d-2}{\sqrt{\pi}} \right)^{d-3} \Gamma \left( \frac{d-3}{2} \right) \Gamma \left( 2-d \right) \cos \left( \frac{\pi d}{2} \right) .
\end{equation}
The saddle point equation becomes 
\begin{equation}
\label{eq:sped-f}
	C_d \int \dd y \lvert x-y \rvert^{d-3} ~ \left[  \partial_z^2  \varrho (z,y) +  \frac{1}{4} \partial_y ^2 \varrho (z,y)  + P^2 \zeta (z)  \delta (y)\right] =  \frac{x}{\tilde{t}_d (z)}  .
\end{equation}
As a function of $d$, $C_d$ is singular at $d=d_{\star}$, for any $d_{\star}\in 2\N \cap \R_{\ge 3}$. Therefore, we cannot analytically continue \eqref{eq:sped-f} across the even integer values $d=d_{\star}$, which subdivide the $d$-axis in distinct regions.
\begin{defin} An admissible region is a connected component of $\R_{d \ge 3} \setminus 2 \N$. Two distinct scenarios are identified:
\begin{itemize}
\item [i)] $d_{\star} < d < d_{\star} +2$, in which the long quiver limit of \cite{Uhlemann:2019ypp} can be adapted.
\item[ii)] $3 \le d < 4$, where the adaptation of the long quiver limit \cite{Uhlemann:2019ypp} holds, with moreover $\hat{t}_{3\le d<4}=\infty$. Eq.~\eqref{eq:sped-f} is understood with both sides multiplied by $w_d$ if $d \to 3^{+}$.
\end{itemize}
Let us stress that localisation requires maximal supersymmetry in $d \ge 6$, so it falls outside our framework. We can nevertheless study \eqref{eq:sped-f} for arbitrary $d \ge 3$, although its field theoretical interpretation is unclear for $d \ge 6$.
\end{defin}\par
In the admissible regions, a solution to \eqref{eq:sped-f} is obtained by a direct generalisation of \cite{Uhlemann:2019ypp,Akhond:2022oaf}. Let us define the operator $\De^{n}$ as follows. Acting on a power $x^m$, 
\begin{equation*}
	\De^n x^m = \begin{cases} \frac{\Gamma (m+1)}{\Gamma (m-n+1)} x^{m-n} & n \le m \\ 0 & n>m \end{cases}
\end{equation*}
where $n,m \in \R_{\ge 0}$ are arbitrary non-negative numbers. This operator is a generalisation of the ordinary derivative, $\frac{\dd \ }{\dd x^n} x^m$ when $n,m\in \N$.
\begin{digr}
	One may give a more exhaustive mathematical definition of $\De^n$, but we will not need it. Also, had we set $t_d (z)=\infty$ for all $0<z<1$ and all $d \ge 3$, we may use the textbook definition of fractional derivative for $\De^n$, see for instance \cite{Herrmann:2011zza}.
\end{digr}\par
Now, we let $\De^{d-3}$ act on both sides of \eqref{eq:sped-f} and use $\De^{d-3} x=0$ if $d>4$, while we set $\frac{1}{\tilde{t}_d}=0$ if $d<4$, according to our previous assumption. We obtain for $d \ne 4$:
\begin{center}\noindent\fbox{\parbox{0.98\linewidth}{%
\begin{equation}
\label{eq:Poisson}
	\frac{1}{4} \partial_y ^2 \varrho (z,y) + \partial_z^2 \varrho (z,y) + P^2 \zeta (z)\delta (y) = 0 .
\end{equation}}}\end{center}
This is the Poisson equation already found in $d=3$ and $d=5$. We have proved it for every admissible region (including admissible regions with $d>6$). We use the balancing condition \eqref{eq:dnudziszeta} and the Fourier expansion \eqref{eq:Fourierzeta} of $\zeta (z)$ in terms of the Fourier coefficients $\left\{R_k\right\}_{k \ge 1}$ of the rank function $\mathcal{R} (\eta)$. The solution $\varrho (x,z)$ to \eqref{eq:Poisson} is 
\begin{equation}
\label{eq:solrho}
	\varrho (z,x) = \frac{\pi}{N} \sum_{k=1}^{\infty} k R_k \sin (\pi k z) \ee^{- 2\pi k \lvert x \rvert}.
\end{equation}
In particular, the measure $\varrho (z,x)\dd x$ has falls off exponentially at $\lvert x \rvert \to \infty$ and has finite moments, as anticipated.\par
In the rest of the current section, we will use this solution to compute several physical observables at arbitrary $d$. The results will interpolate and extend known expressions in the literature.

\subsection{Interpolating between \texorpdfstring{$a$}{a} and \texorpdfstring{$F$}{F}}
\label{sec:aF}

We rewrite the effective action \eqref{eq:Seffdlong2} after the substitution \eqref{eq:LQansatz} and the ensuing change of variables:
\begin{equation*}
\begin{aligned}
	S_{\mathrm{eff}} [\varrho] \simeq N^2 P^{d-3} (d-2)^{d-2} s_d \int_{0} ^{1} \dd z & \int \dd x \varrho (z, x) \left\{ \frac{(d-2) }{2 C_d \tilde{t}_d (z)}  x^2 + [P^2\zeta (z)] \lvert x \rvert^{d-2} \right. \\
	+ &  \int \dd y  \left[  \varrho (z, y) \frac{ \lvert x-y \rvert^{d-4}}{8}  + \frac{ \lvert x-y \rvert^{d-2}}{2} \partial^2_z \varrho (z,y) \right] \\
	- & \left. \lim_{\delta z \to 0^{+}}\left[ \delta (z- \delta z) + \delta (z-1 + \delta z) \right] \frac{\lvert x\rvert^{d-2}}{2NP} \right\} .
\end{aligned}
\end{equation*}
Here we have used \eqref{eq:flargephi}, and recall that the numerical coefficient $C_d$ appearing in the first line was defined in \eqref{eq:totcoefCd} and satisfies $\frac{(d-2)^{d-1} s_d}{C_d}= u_d$. Moreover, the last line proceeds from the term $\mathfrak{B}$, which imposes $Pz \in \left\{1,P-1 \right\}$ and is manifestly suppressed compared to the rest. To evaluate $ S_{\text{eff}} $ on-shell in the planar limit we insert the solution \eqref{eq:solrho} and discard all the sub-leading contributions.\par

\begin{stm}
Consider a balanced linear quiver. With the notation from Subsection \ref{sec:rkfn}, in the long quiver limit it holds that 
\begin{center}\noindent\fbox{\parbox{0.98\linewidth}{%
\begin{equation*}
\begin{aligned}
	\ln \mz_{\cs^d} \simeq P^{(d-3)}  \sum_{k=1} ^{\infty} \left[ c_{\mathrm{univ.}}(d)~  k^{4-d} R_k ^2  + c_{\mathrm{gauge}}(d)~ k^{-2} R_k \gamma_k \right] .
\end{aligned}
\end{equation*}}}\end{center}
with $d$-dependent coefficients 
\begin{align*}
	c_{\mathrm{univ.}} (d) & = (d-2)^{d-2} \frac{\pi^{5-d}}{2^d \sin \left( \frac{\pi d}{2}\right)} , \\
	c_{\mathrm{gauge}} (d) & = \begin{cases} - (d-2)^{2} \frac{ \pi^{\frac{d-3}{2}}}{\Gamma \left(\frac{d-3}{2}\right)}  & \ 4 \le d <6 \\ 0 & \ \text{otherwise}.\end{cases}
\end{align*}
\end{stm}
\begin{proof} At leading order in the long quiver limit we have 
\begin{equation*}
	\ln \mz_{\cs^d} \simeq - \left. S_{\text{eff}} \right\rvert_{\text{on-shell}} ,
\end{equation*}
with the on-shell effective action evaluated using the solution \eqref{eq:solrho}. To reduce clutter, let us introduce the shorthand notation
\begin{equation*}
	\tilde{w}_d := - (d-2) w_d = (d-2)^{4-d} u_d .
\end{equation*}
A computation akin to \cite{Santilli:2021qyt,Akhond:2022oaf} (which is the adaptation to long quivers of the standard calculation of the free energy in matrix models, reviewed for instance in \cite{Marino:2004eq}) gives:
\begin{equation*}
	\left. S_{\text{eff}} \right\rvert_{\text{on-shell}}  = \frac{1}{2} N^2 P^{d-3} \cdot (d-2)^{d-2} \int_0 ^{1} \dd z \int_{- \infty} ^{+ \infty} \dd x \varrho (z,x) \left[  P^2 \zeta (z) s_d \lvert x \rvert^{d-2} + \frac{ \tilde{w}_d }{2 \tilde{t}_d (z)} x^2  \right] .
\end{equation*}
We now insert the Fourier expansions of $\varrho (z,x)$ from \eqref{eq:solrho} and of $P^2 \zeta(z)$ and $\frac{1}{g_{d}(z)}$ from \eqref{eq:Fourierzeta} and \eqref{eq:Fourierg}, respectively. We get 
\begin{equation}
\begin{aligned}
	\left. S_{\text{eff}} \right\rvert_{\text{on-shell}}  = \frac{1}{2} P^{(d-3)} \cdot (d-2)^{d-2} \int_0 ^{1} \dd z & \int_{- \infty} ^{+ \infty} \dd x \sum_{k=1}^{\infty} \pi k R_k \sin (\pi k z) \ee^{-2 \pi k \lvert x \rvert} \\
		& +\sum_{\ell=1}^{\infty} \sin (\pi \ell z) \left[ \ell^2 \pi^2 R_{\ell} ~ s_d \lvert x \rvert^{d-2} \ + \  8 \tilde{w}_d \pi^2 \gamma_{\ell} ~  x^2  \right] 
\end{aligned}
\label{eq:Sonshelldint}
\end{equation}
where the Fourier coefficients $\tilde{\gamma}_{\ell}$ of the cosine part of the gauge coupling drop out due to the integration against sines. According to the assumption we have made along the way, the part in \eqref{eq:Sonshelldint} proportional to the gauge coupling has to be multiplied by $\theta_{4 \le d<6}$, where the characteristic function $\theta_{\mathscr{S}}$ introduced in \eqref{eq:defthetaS} is $1$ if the condition $\mathscr{S}$ is satisfied and vanishes otherwise. Besides, we introduce the shorthand notation 
\begin{equation*}
	\hat{s}_d := s_d \frac{\pi^{4-d}}{2^d} \Gamma (d-1) = - \frac{\pi^{5-d}}{2^d \sin \left( \frac{\pi d}{2}\right)} .
\end{equation*}
Equipped with these definitions, we solve the integrals in \eqref{eq:Sonshelldint} and find 
\begin{equation}
\label{eq:SonshellD}
	\left. S_{\text{eff}} \right\rvert_{\text{on-shell}}  = P^{(d-3)} (d-2)^{(d-2)}  \sum_{k=1} ^{\infty} \left[ k^{4-d} R_k ^2 \hat{s}_d  +  k^{-2} R_k \gamma_k \tilde{w}_d \theta_{4 \le d < 6} \right] .
\end{equation}

\end{proof}
Let us briefly pause to remark on expression \eqref{eq:SonshellD}.
\begin{itemize}
	\item The appearance of the pre-factor $1/ \sin \left( \frac{\pi d}{2}\right)$ in $\ln \mz_{\cs^d}$ is consistent with the expectations from \cite{Giombi:2014xxa}, as we explain momentarily. This piece blows up at even integer $d$, and is finite in the admissible regions.
	\item The overall coefficient $ (d-2)^{(d-2)} $ is unavoidable, and generalises the number $27=(5-2)^{5-2}$ in the free energy at $d=5$, and the number $1$ in the free energy at $d=3$.
	\item If the gauge coupling is the same at all gauge nodes, $\gamma_k=0$ and the dependence on the gauge coupling drops out.
	\item Formula \eqref{eq:SonshellD} is expressed in terms of Fourier coefficients of the rank function $\mathcal{R}(\eta)$ and of the inverse gauge coupling. However, it was derived in the planar long quiver limit. It should be understood as factoring out $N^2$, with both $\frac{R_k}{N}$ and $\frac{\gamma_k}{N}$ having a well-defined large-$N$ limit, in which they are $\mathcal{O}(1)$.
\end{itemize}\par
\bigskip
Giombi and Klebanov proposed a function $\tilde{F}_d$ that interpolates between the sphere free energy in odd dimensions and the Weyl anomaly coefficient in even dimensions \cite{Giombi:2014xxa}. We introduce our notation and normalisation and then define $\tilde{F}_d$.\par
\begin{lemma}
	Let $d\in\N$. Consider a Euclidean SCFT on $\cs^d$ of radius $r$ with UV cutoff $\Lambda$, and denote $\mz_{\cs^d}$ its partition function. It has the form \cite{Gerchkovitz:2014gta}:
	\begin{equation}
	\label{eq:GGKZd}
		\ln \mz_{\cs^d} = \begin{cases} c_{d-2} (\Lambda r)^{d-2} + \cdots + c_1 (\Lambda r) + (-1)^{(d-1)/2}  F + \mathcal{O}((\Lambda r)^{-1}) & d \text{ odd} \\ c_{d-2} (\Lambda r)^{d-2} + \cdots + c_2 (\Lambda r)^2 - 4 a \ln (\Lambda r) - F + \mathcal{O}((\Lambda r)^{-1}) & d \text{ even} \end{cases}
	\end{equation}
	where $c_n$ are numerical coefficients.
\end{lemma}
Coupling the theory supersymmetrically to the round sphere $\mathbb{S}^d$ of radius $r$, the energy scale at which the gauge coupling is defined is $1/r$, and the dimensionless combination with the UV cutoff is $r \Lambda$.
\begin{defin}
	Consider \eqref{eq:GGKZd}. The finite quantity $F$ is called \emph{free energy} --- note the coefficient $(-1)^{(d-1)/2}$ when $d$ is odd and $-1$ when $d$ is even.\par
	The quantity $a$, that appears as $-\frac{1}{4}$ of the coefficient of $\ln (\Lambda r)$ in $d\in 2\N$, is called \emph{Weyl anomaly} coefficient. Our definition of $a$ differs from \cite{Giombi:2014xxa} by the factor $\frac{1}{4}$. In our conventions, a free scalar has $a=\frac{1}{360}$, whilst in the conventions of \cite{Giombi:2014xxa} a free scalar has $a=\frac{1}{90}$.
\end{defin}
\begin{defin} The interpolating free energy is \cite{Giombi:2014xxa}
\begin{equation}
\label{eq:GKFE}
	\tilde{F}_d := \sin \left( \frac{\pi d}{2}\right) \ln \mz_{\mathbb{S}^d} .
\end{equation}
When $d\in\N$, it gives
\begin{equation}
\label{eq:FtildetoA}
	\tilde{F}_d = \begin{cases} F & d \text{ odd} \\ (-1)^{d/2} 2\pi a  & d \text{ even} \end{cases}
\end{equation}
and, in particular, it regulates the poles of $c_{\mathrm{univ.}}(d)$ in even dimensions.
\end{defin}
We now state the main result of this section. 
\begin{stm}For long balanced linear quivers, we find 
\begin{center}\noindent\fbox{\parbox{0.98\linewidth}{%
\begin{equation}
\label{eq:Ftilded}
\begin{aligned}
	\tilde{F}_d & \simeq P^{d-3} (d-2)^{(d-2)}  \frac{\pi^{5-d}}{2^d}  \sum_{k=1} ^{\infty}  k^{4-d} R_k ^2  \\
		&- \theta_{4 \le d < 6} P^{d-3} (d-2)^2 \pi^{(d-3)/2} \frac{ \sin \left( \frac{\pi d}{2}\right)}{\Gamma \left( \frac{d-3}{2} \right)}  \sum_{k=1} ^{\infty} \frac{1}{k^2} R_k \gamma_k .
\end{aligned}
\end{equation}}}\end{center}
At integer $d$ the coefficients attain the values listed in Table \ref{tab:cFtilde}.
\end{stm}
\begin{table}[th]
\centering
	\begin{align*}
	\begin{tabular}{|c | c c c c |}
	\hline
	$d$ & 3 & 4 & 5 & 6 \\
	\hline
	$(d-2)^{(d-2)} \pi^{5-d} 2^{-d} $ & $\frac{\pi ^2}{8} $ & $\frac{\pi }{4}$ & $\frac{27}{32}$ & $\frac{4}{\pi }$ \\
	\hline
	\end{tabular} && 
	\begin{tabular}{|c | c c c c  |}
	\hline
	$d$ & 3 & 4 & 5 & 6  \\
	\hline
	$(d-2) \pi^{(d-3)/2} \frac{ \sin \left( \frac{\pi d}{2}\right)}{\Gamma \left( \frac{d-3}{2} \right)}$ & 0 & 0 & $3\pi$ & 0  \\
	\hline
	\end{tabular}
\end{align*}
\caption{Coefficients in the first and second line of \eqref{eq:Ftilded}, respectively, specialised at $d =3,4,5,6$.}
\label{tab:cFtilde}
\end{table}
Formula \eqref{eq:Ftilded} correctly reproduces the known results in $d=3$ and $d=5$, extending the $d=5$ result to finite gauge coupling along the way (the previous results assumed $g_5(z) \to \infty$).\par
At $d=4$, \eqref{eq:Ftilded} yields 
\begin{center}\noindent\fbox{\parbox{0.98\linewidth}{%
\begin{equation}
\label{eq:tildeF4d}
	\lim_{d\to 4} \tilde{F}_d \simeq \frac{\pi}{4} P \sum_{k=1} ^{\infty}  R_k ^2 .
\end{equation}}}\end{center}
Using \eqref{eq:FtildetoA}, this expression matches the anomaly coefficient $a$ predicted from the supergravity dual in \cite{Nunez:2019gbg}.\par
One important corollary of our continuation in $d$ is to extract the value $\tilde{F}_{d=6}$ for long linear quivers, which cannot be derived directly from localisation. We find:
\begin{equation*}
	\lim_{d\to 6^{-}} \tilde{F}_d \simeq \frac{4}{\pi} P^3 \sum_{k=1} ^{\infty}  \frac{R_k ^2}{k^2} .
\end{equation*}
\begin{digr}
We comment on the consistency of the derivation presented herein with three-dimensional $\mN=4$ mirror symmetry. Mirror symmetry in $d=3$ is the statement that two different-looking gauge theories with eight supercharges flow in the IR to the same SCFT \cite{Intriligator:1996ex}. The mirror of a linear balanced quiver is not balanced in general. However, if a balanced linear quiver has hypermultiplets in the fundamental representations charged under only one gauge factor, say $\mathrm{U}(N_J)$ for $J \in \left\{ 1 , \dots, P-1 \right\}$, its mirror is linear, balanced and with fundamental hypermultiplets charged under a single gauge node.\par
In practice, this leads to consider pairs of balanced linear quivers whose rank functions $\mathcal{R} (\eta)$ and $\mathcal{R}^{\vee} (\eta^{\vee})$ are triangular. If one quiver has length $P-1$, the gauge ranks $N_j$ are all proportional to $N \in \N$ and the hypermultiplets are  placed at the $J^{\text{th}}$ node, its mirror has length $P^{\vee} -1 = \frac{NP}{P-J} -1$, where $(P-J)$ always divides $NP$ as a consequence of the balancing condition.\par
Quivers of this kind have been considered in \cite{Akhond:2021ffz,Fatemiabhari:2022kpv}, where it was observed that the Fourier coefficients $R_k, R^{\vee}_k$ of the two rank functions are equal, $R_k=R^{\vee}_k$. We see from \eqref{eq:Ftilded} that the free energies of the two quivers are equal exactly at $d=3$. However, the lengths $P, P^{\vee}$ of the two quivers differ, and hence their overall coefficients in \eqref{eq:Ftilded} mismatch if $d>3$. This underlines how mirror symmetry is characteristic of $d=3$.
\end{digr}

\subsection{One-dimensional defects}
\label{sec:WLd}
The next observable we compute in arbitrary $d$ is the expectation value of a half-BPS Wilson loop in the rank-$\kk$ antisymmetric representation $\mathsf{A}_{\kk}$ of the $j_{\ast}^{\text{th}}$ gauge group $\mathrm{U}(N_{j_{\ast}})$, with $0 < \kk < N_{j_{\ast}}$. Correlation functions of such Wilson loops are computed exactly inserting 
\begin{equation*}
	\frac{\kk ! (N_{j_{\ast}}-\kk)!}{N_{j_{\ast}} !}\mathrm{Tr}_{\mathsf{A}_{\kk}} \ee^{2 \pi \phi^{(j)}} 
\end{equation*}
into the matrix model, where the binomial coefficient in front comes from $1/\dim \mathsf{A}_{\kk}$. We now study the expectation value of these Wilson loop operators  procedure closely follows and extends \cite{Uhlemann:2020bek,Fatemiabhari:2022kpv}.\par
\begin{stm}
	Let $\langle W_{\mathsf{A}_{\kk}} \rangle_d$ denote the expectation value of a Wilson loop in $d$ dimensions, with the dependence on $j_{\ast}$ implicit in the notation. Then, in the long quiver limit, the dependence on $d$ appears only through a pre-factor. Namely, 
	\begin{equation*}
		\ln \langle W_{\mathsf{A}_{\kk}}  \rangle_d \simeq \frac{d-2}{3} \ln \langle W_{\mathsf{A}_{\kk}} \rangle_{d=5} ,
	\end{equation*}
	with the right-hand side extensively analysed in \cite{Uhlemann:2020bek}.
\end{stm}
\begin{proof}
Let $z_{\ast} := j_{\ast}/P$ and
\begin{equation*}
	\kappa (z) := \frac{\kk}{N_{j}} , \qquad 0 < \kappa (z) < 1  \quad \forall ~0<z<1, 
\end{equation*}
so that $\nu (z_{\ast})\kappa (z_{\ast}) = \frac{\kk}{N}$. We want to isolate the $\kk$ largest eigenvalues at the node $j_{\ast}$ \cite{Uhlemann:2020bek}. First, let $0 \le \tilde{\kappa} \le \kappa (z_{\ast})$ and define the number $x_{\ast} \equiv x_{\ast} (z_{\ast}, \tilde{\kappa})$ such that a fraction $\tilde{\kappa}$ of eigenvalues is larger than $x_{\ast}$, and the remaining fraction $1-\tilde{\kappa}$ of eigenvalues is smaller than $x_{\ast}$. Explicitly, $x_{\ast}$ is defined via the relation 
\begin{equation}
\label{eq:kappaVSxast}
	\int_{x_{\ast}} ^{\infty} \dd x ~\varrho (z_{\ast},x) = \nu (z_{\ast}) \tilde{\kappa} .
\end{equation}
Note that $\tilde{\kappa} \to 1$ gives $x_{\ast} \to - \infty$ and $\tilde{\kappa} \to 0$ gives $x_{\ast} \to +\infty$.\par
By a straightforward adaptation of \cite{Uhlemann:2020bek} we find that the expectation value of the Wilson loop is 
\begin{equation*}
	\ln \langle W_{\mathsf{A}_{\kk}} (z_{\ast}) \rangle \simeq 2 \pi (d-2)P N \nu (z_{\ast})  \int_{0} ^{\kappa (z_{\ast})} x_{\ast} (z_{\ast}, \tilde{\kappa} ) \dd \tilde{\kappa} .
\end{equation*}
The overall factor originates as follows: the coefficient $ 2 \pi $ comes from the definition of the Wilson loop; the factor $(d-2)P$ is due to the long quiver scaling \eqref{eq:LQansatz} to pass from the variable $\phi$ to the $\mathcal{O}(1)$ variable $x$; finally, passing from the sum over the $N_{j_{\ast}}$ eigenvalues to an integral produces the coefficient $N_{j_{\ast}}=N \nu (z_{\ast}) $, as appropriate in the planar limit.\par
Using the inverse function theorem we pass from an integral over $\tilde{\kappa}$ to an integral over $x_{\ast}$:
\begin{align}
	\ln \langle W_{\mathsf{A}_{\kk}} (z_{\ast}) \rangle & \simeq 2 \pi (d-2) \nu (z_{\ast})~ P N \int_{x_{\ast} (z_{\ast},0)} ^{x_{\ast} (z_{\ast},\kappa (z_{\ast})) } \frac{\dd \tilde{\kappa} }{\dd x_{\ast} } ~x_{\ast} \dd x_{\ast} \notag \\
		&= 2 \pi (d-2) ~ P N \int^{x_{\ast} (z_{\ast},0)} _{x_{\ast} (z_{\ast},\kappa (z_{\ast})) } \varrho (z_{\ast}, x_{\ast}) ~x_{\ast} \dd x_{\ast}  \label{eq:WLdvev}
\end{align}
where in the second line we have differentiated both sides of \eqref{eq:kappaVSxast} with respect to $x_{\ast}$ and utilised the ensuing expression for $\frac{\dd \tilde{\kappa} }{\dd x_{\ast} }$ (a minus sign from the derivative is used to invert the integration domain).\par
With the due identifications, \eqref{eq:WLdvev} only differs from the $d=5$ computation of \cite[Eq.(2.18)]{Uhlemann:2020bek} by the replacement of the coefficient $3=(5-2) \mapsto (d-2)$. In other words, the expectation value of a half-BPS Wilson loop in $d$ dimensions, in the allowed regions for $d$, always has the same form except for the overall coefficient $(d-2)$.
\end{proof}
That the result must be uniform in $d$ could be predicted looking at the functional dependence of the Wilson loop expectation value. It is linear in $\varrho$ and does not involve any of the $d$-dependent functions $f_{\mathrm{h}},f_{\mathrm{v}}, f_0$. Therefore, the only $d$-dependence should appear through overall scaling, entirely dictated by \eqref{eq:LQansatz}, whence the $(d-2)$.\par
We can then go on and solve \eqref{eq:WLdvev} uniformly in $d$. 
\begin{itemize}
	\item[---] Integrating the left-hand side of \eqref{eq:kappaVSxast} explicitly, we fix $x_{\ast} (z_{\ast}, \kappa )$.
	\item[---] We then perform the integration in \eqref{eq:WLdvev} with the thus-obtained boundary values $x_{\ast} (z_{\ast},0)$ and $x_{\ast} (z_{\ast},\kappa (z_{\ast}))$.
\end{itemize}
This algorithm gives $\ln \langle W_{\mathsf{A}_\kk} (z_{\ast}) \rangle$ for every choice of $j_{\ast}$ and of representation $\kk$, in arbitrary $d$.\par
\bigskip
At this point, we continue the calculation following \cite{Fatemiabhari:2022kpv} and \cite[Sec.3.7]{Akhond:2022oaf}. We plug the solution \eqref{eq:solrho} for $\varrho(z,x_{\ast})$ into \eqref{eq:kappaVSxast} and evaluate the integral. The number $x_{\ast}=x_{\ast} (z_{\ast},\kappa)$ is fixed by solving the equation 
\begin{equation}
\label{eq:xastfromk}
	\sum_{k=1}^{\infty} R_k \sin (\pi k z_{\ast}) \left[ \theta (- x_{\ast} ) + \frac{1}{2} \mathrm{sign} (x_{\ast}) \ee^{-2\pi k \lvert x_{\ast} \rvert } \right] = \kk
\end{equation}
where $\theta (\cdot )$ is the Heaviside function and we have multiplied both sides by $N$. We then integrate \eqref{eq:WLdvev} explicitly.\footnote{The left-hand side of \eqref{eq:xastfromk} is continuous at $x_{\ast}=0$, with the square bracket evaluating to $\frac{1}{2}$ from both sides $x_{\ast} \to 0^{-}$ and $x_{\ast} \to 0^+$.}
\begin{stm}
The defect free energy for a Wilson loop in the rank-$\kk$ antisymmetric representation $\mathsf{A}_{\kk}$ of $\mathrm{U}(N_{j_{\ast}})$ supported on $\cs^1 \subset \cs^d$ is 
\begin{equation}
\label{eq:logWLdstm}
	\ln \langle W_{\mathsf{A}_{\kk}} (z_{\ast}) \rangle  \simeq P  \frac{(d-2)}{2} \sum_{k=1}^{\infty} k^{-1} R_k \sin (\pi k z_{\ast}) \ee^{-2\pi k \lvert x_{\ast} \rvert }\left( 2\pi k \lvert x_{\ast} \rvert +1 \right) ,
\end{equation}
where $x_{\ast}$ is fixed by \eqref{eq:xastfromk}.
\end{stm}
We observe that \eqref{eq:logWLdstm} only depends on the rank $\kk$ through the right-hand side of \eqref{eq:xastfromk}. Moreover, using the representation $\kk = \frac{N_{j_{\ast}}}{2}$ in the right-hand side of \eqref{eq:xastfromk} we get $x_{\ast}=0$ and, plugging in \eqref{eq:WLdvev}, we find
\begin{align}
	\ln \langle W_{\mathsf{A}_{N_{j_{\ast}}/2}} (z_{\ast}) \rangle & \simeq 2 \pi (d-2) ~ P N \int_{0}^{\infty} \varrho (z_{\ast}, x) ~x \dd x \label{eq:WLhalfsym}
\end{align}

\subsection{Two-dimensional defects}
\label{sec:2ddef}

Taking inspiration from the defects constructed and studied in \cite{Santilli:2023fuh}, we consider two-dimensional half-BPS defects consisting of a chiral and an anti-chiral multiplets of equal R-charge $q$ localised on a $\cs^2 \subset \cs^d$. These chiral multiplets interact with the $d$-dimensional long quiver through the gauge coupling to the $j_{\ast}^{\text{th}}$ gauge group $\mathrm{U}(N_{j_{\ast}})$.\par
The one-loop determinant for a chiral multiplet of R-charge $q$ in the fundamental representation and coupled to a vector multiplet is \cite{Benini:2012ui,Doroud:2012xw}
\begin{equation}
\label{eq:Z2d1loop}
	\mz_{\text{\rm 2d defect}} = \prod_{a=1}^{N_{j_{\ast}}} \frac{ \Gamma \left( \frac{q}{2} - \ii \phi_a ^{(j_{\ast})} + \frac{b_a ^{(j_{\ast})}}{2}  \right)}{ \Gamma \left( 1 - \frac{q}{2} + \ii \phi_a ^{(j_{\ast})} + \frac{b_a ^{(j_{\ast})}}{2} \right)}
\end{equation}
where $\vec{b} = \left\{ b_a ^{(j)} \right\}$ is the flux of the gauge connection through the $\cs^2$ on which the defect lives. Using the property $H_2 (\cs^d) = \emptyset $ if $d\in \N_{\ge 3}$, we define the second homology of $\cs^d$ to be trivial also for real $d \ge 3$. As a consequence, there is no gauge flux, $b_a ^{(j)} =0 $ $\forall a=1, \dots, N_j$, $j=1, \dots , P-1$. This would not be the case localising on $d$-dimensional manifolds other than $\cs^d$.\par
We \emph{assume} that the defect contribution be accounted for by importing the one-loop determinant \eqref{eq:Z2d1loop} evaluated in the background of the $d$-dimensional vector multiplet. This is a reasonable working assumption, analogous to and compatible with all know defect localisation results in the literature, but we do not have a derivation for it. The R-charge $q$ is taken with respect to the defect R-symmetry.\par
\begin{stm}
	Let $F_{\text{\rm 2d defect},d}$ denote the two-dimensional defect free energy in $d$ dimensions, with the dependence on $j_{\ast}$ implicit in the notation. In the long quiver limit it is given by 
	\begin{equation*}
		F_{\text{\rm 2d defect},d} \simeq 2 (1-q) N_{j_{\ast}} \ln P 
	\end{equation*}
\end{stm}
\begin{proof}
The computation is similar in spirit to \cite{Santilli:2023fuh}. The insertion of the 2d defect does not modify the leading order contribution in the long quiver limit. Let use define 
\begin{equation*}
	f_{\text{\rm 2d defect}} (\phi_a) := - \ln \left[ \frac{ \Gamma \left( \frac{q}{2} - \ii \phi_a  \right)}{ \Gamma \left( 1 - \frac{q}{2} + \ii \phi_a \right)} \cdot  \frac{ \Gamma \left( \frac{q}{2} + \ii \phi_a  \right)}{ \Gamma \left( 1 - \frac{q}{2} - \ii \phi_a  \right)} \right] ,
\end{equation*}
which accounts for the contribution from both the chiral and the anti-chiral multiplets. Using the asymptotic expression of the $\Gamma$ function 
\begin{equation*}
	\ln \Gamma (z) \simeq \left( z- \frac{1}{2} \right) \ln (z) -z ,
\end{equation*}
the behaviour of the defect contribution in the long quiver scaling is:
\begin{equation*}
	f_{\text{\rm 2d defect}} ((d-2)Px) \simeq 2(1-q) \left[\ln \left( (d-2)P \right) + \ln \lvert x \rvert \right] .
\end{equation*}
We note that, had we kept the dependence on the fluxes $\vec{b}$ through $\cs^2$, it would drop out at this stage, due to simplifications at leading order in the planar limit. We can thus write 
\begin{align*}
	F_{\text{\rm 2d defect},d} & \simeq - \left. \ln \mz_{\text{\rm 2d defect}} \right\rvert_{\text{\rm on shell}} \\
	&= N \int \dd x \varrho (z_{\ast},x) f_{\text{\rm 2d defect}} ((d-2)Px) \\ 
	&= 2 (1-q) N  \left[ \nu (z_{\ast})  \ln \left( (d-2)P \right) + \sum_{k=1}^{\infty} \frac{2 \pi k}{N} R_k \sin (\pi k z_{\ast}) \int_{- \infty}^{+\infty} \dd x ~\ee^{- 2 \pi k \lvert x \rvert }\ln \lvert x \rvert  \right] \\
	&= 2 (1-q) N_{j_{\ast}} \left[ \ln \left( (d-2)P \right)  -  \gamma_{\text{\tiny EM}} \right] - 2 (1-q) \sum_{k=1}^{\infty} R_k \sin (\pi k z_{\ast}) \ln (2 \pi k) ,
\end{align*}
where $ \gamma_{\text{\tiny EM}}$ is the Euler--Mascheroni constant. The last term is ambiguous due to the $\ln P$ dependence of the leading term, and likewise for the coefficients of $P$ inside the logarithm. Only the coefficient of $\ln P$ is unambiguous and yields the leading behaviour.
\end{proof}
\begin{digr}
Other types of two-dimensional holographic defects at large-$N$ have been studied in \cite{Jokela:2021evo}. In $d=4$, a famous class of two-dimensional defects is provided by the Gukov--Witten operators \cite{Gukov:2006jk}, studied at large-$N$ in \cite{Drukker:2008wr}. Our construction is genuinely different. On the one hand, the defects we study here hold for all $\mN=2$ quivers, whereas the analysis of \cite{Drukker:2008wr} requires maximal supersymmetry. On the other hand, the data specifying our defects only capture a simple class of half-BPS defects in 4d $\mN=2$ linear quivers. It would be interesting to perform a continuation in $d$ of the analysis of \cite{Drukker:2008wr}.
\end{digr}

\subsection{Three-dimensional defects}
\label{sec:3ddef}

A class of three-dimensional defects in $d=5$ long quivers was studied in \cite{Santilli:2023fuh}. The defects consist in a pair of 3d chiral and antichiral multiplets localised on a $\cs^3 \subset \cs^d$ and charged under the $j_{\ast}^{\text{th}}$ gauge group $\mathrm{U}(N_{j_{\ast}})$. We now formally extend the computation of \cite{Santilli:2023fuh} by promoting the $d=5$ gauge theory to arbitrary $d\ge 3$.\par
\begin{stm}
	Let $F_{\text{\rm 3d defect},d}$ denote the 3d defect free energy in $d$ dimensions, with the dependence on $j_{\ast}$ implicit in the notation. Then, in the long quiver limit, the dependence on $d$ appears only through a pre-factor. Namely, 
	\begin{equation*}
		F_{\text{\rm 3d defect},d} \simeq \frac{d-2}{3} F_{\text{\rm 3d defect},d=5} ,
	\end{equation*}
	with the right-hand side extensively analysed in \cite{Santilli:2023fuh}.
\end{stm}
\begin{proof}
The sphere partition function in presence of the defect is modified by the inclusion of the one-loop determinant of the chiral multiplets in the background of the $d=5$ vector multiplet. This fact was derived from localisation in \cite{Santilli:2023fuh} when $d=5$, and here we \emph{assume} that this form extends to other values of $d$. We proceed directly to the evaluation of the defect free energy
\begin{equation*}
	F_{\text{\rm 3d defect},d}  = - \ln \frac{\mz_{\cs^d}^{\text{\rm 3d defect}} }{\mz_{\cs^d}},
\end{equation*}
where the argument of the logaritm is the ratio between the partition functions with and without defects. This quantity isolates the contribution of the 3d defect.\par
In the long quiver regime, we write $z_{\ast}=j_{\ast}/P$ and $N_{j_{\ast}} = N \nu (z_{\ast}) $. Let us also denote $q$ the defect R-charge of the chiral and antichiral pair, which equals twice its scaling dimension. The 3d defect free energy is obtained as \cite{Santilli:2023fuh}
\begin{equation}
\label{eq:F3dDint}
	F_{\text{\rm 3d defect},d} (z_{\ast}) \simeq 2 \pi \left(1- \frac{q}{2} \right) (d-2) P N \nu (z_{\ast})  \int_{- \infty}^{\infty} \dd x ~\varrho (z_{\ast},x) \lvert x \rvert .
\end{equation}
\end{proof}
Plugging \eqref{eq:solrho} into \eqref{eq:F3dDint}, we arrive at the following statement.
\begin{stm} 
The defect free energy for a chiral and an antichiral multiplets of R-charge $q$ supported on $\cs^3 \subset \cs^d$ is 
\begin{equation*}
	F_{\text{\rm 3d defect},d} (z_{\ast}) \simeq \left(1- \frac{q}{2} \right) (d-2) P \nu (z_{\ast}) \sum_{k=1}^{\infty} k^{-1} R_k \sin (\pi k z_{\ast}) .
\end{equation*}
\end{stm}
Comparing \eqref{eq:F3dDint} with \eqref{eq:WLhalfsym} we also observe the identity
\begin{center}\noindent\fbox{\parbox{0.98\linewidth}{%
\begin{equation*}
	\frac{F_{\text{\rm 3d defect},d} (z_{\ast})}{\nu (z_{\ast}) } \simeq \left(2-q \right) \ln \langle W_{\mathsf{A}_{N_{j_{\ast}}/2}} (z_{\ast}) \rangle .
\end{equation*}}}\end{center}
For a fixed $j_{\ast}$ it is always possible to choose the normalisation $N=N_{j_{\ast}}$, so that the left-hand side is divided by $\nu (z_{\ast})=1$. The 1d and 3d defect free energies are then exactly equal at $q=1$, corresponding to a conformal dimension $\frac{1}{2}$ for the defect chiral multiplets.

\subsection{Mirror-type identities}
\label{sec:mirror}

Recall the rank function $\mathcal{R} (\eta)$ with Fourier coefficients $\left\{ R_k \right\}_{k \ge 1}$, which takes the explicit form 
\begin{equation*}
	\mathcal{R} (\eta) = \begin{cases} N_1 \eta & \quad 0 \le \eta < 1 \\ N_j + (N_{j+1} - N_j) (\eta -j) & \quad j \le \eta < j+1 \quad \forall j=1, \dots, P-2 \\ N_{P-1} (P-\eta) & \quad P-1 \le \eta \le P . \end{cases}
\end{equation*}
Let $K:= \sum_{j=1}^{P-1} K_j$ be the total number of fundamental flavours in the quiver. Note that, using the balancing condition \eqref{eq:balance}, this number is easily obtained by 
\begin{align*}
	K &= \int_0 ^P \dd \eta \sum_{j=1}^{P-1} K_j \delta (\eta - j) \\
		& = - \int_0 ^P \dd \eta \partial_{\eta} ^2 \mathcal{R} (\eta) = \partial_{\eta} \mathcal{R} (0) - \partial_{\eta} \mathcal{R} (P) .
\end{align*}
Only throughout this subsection we impose the assumptions: 
\begin{equation}
\label{eq:assumptionMir}
(i) \ \frac{K}{P} \in \N \qquad \text{ and } \qquad (ii) \ \frac{P}{K} N_j \in \N \ \forall j=1, \dots, P-1. 
\end{equation}\par
Let us now define the function $\mathcal{R}^{\vee} (\eta^{\vee})$ as 
\begin{equation*}
	\mathcal{R}^{\vee} (\eta^{\vee}) = \begin{cases} N_1^{\vee} \eta^{\vee} & \quad 0 \le \eta < \frac{K}{P} \\ \frac{K}{P} N_j^{\vee} + (N_{j+1}^{\vee} - N_j^{\vee}) (\eta^{\vee} -j) & \quad j\frac{K}{P} \le \eta^{\vee} < (j+1) \frac{K}{P} \quad \forall j=1, \dots, P-2 \\ N_{K-1}^{\vee} (K-\eta^{\vee}) & \quad K \left( 1-\frac{1}{P}\right) \le \eta^{\vee} \le K \end{cases}
\end{equation*}
where $N_j ^{\vee} = N_j \frac{P}{K}$. The requirements \eqref{eq:assumptionMir} ensure that $\mathcal{R}^{\vee} (\eta^{\vee})$ is the rank function of a quiver of length $P^{\vee} :=K$. We will denote the Fourier coefficients of $\mathcal{R}^{\vee} (\eta^{\vee})$ by $\left\{ R_k^{\vee} \right\}_{k \ge 1}$.\par

\begin{figure}[htb]
\centering
\begin{tikzpicture}[auto,square/.style={regular polygon,regular polygon sides=4},scale=0.9]

 \path[->] (-1,0) edge node {$\vee$} (1,0);
 \path[->] (-1,-3) edge node {$\vee$} (1,-3);
 \path[->] (-1,-6) edge node {$\vee$} (1,-6);
 \path[->] (-1,-9) edge node {$\vee$} (1,-9);
 
 \node[circle, draw] (a) at (-2,0) {$\scriptscriptstyle 8$};
 \node[circle, draw] (b) at (-3,0) {$\scriptscriptstyle 8$};
 \node[circle, draw] (c) at (-4,0) {$\scriptscriptstyle 8$};
 \node[square, draw] (af) at (-2,1) {$\scriptscriptstyle 8$};
 \node[square, draw] (cf) at (-4,1) {$\scriptscriptstyle 8$};
 \draw (a) -- (af); \draw (c) -- (cf); \draw (a) -- (b); \draw (b) -- (c);
 
 \node[circle, draw] (av) at (2,0) {$\scriptscriptstyle 2$};
 \node[circle, draw] (bv) at (3,0) {$\scriptscriptstyle 4$};
 \node[circle, draw] (cv) at (4,0) {$\scriptscriptstyle 6$};
 \node[circle, draw] (dv) at (5,0) {$\scriptscriptstyle 8$};
 \node[circle, draw] (ev) at (6,0) {$\scriptscriptstyle 8$};
 \node[] (fv) at (7,0) {$\scriptstyle \cdots $};
 \node[circle, draw] (gv) at (8,0) {$\scriptscriptstyle 8$};
 \node[circle, draw] (kv) at (11,0) {$\scriptscriptstyle 2$};
 \node[circle, draw] (jv) at (10,0) {$\scriptscriptstyle 4$};
 \node[circle, draw] (iv) at (9,0) {$\scriptscriptstyle 6$};
 \node[square, draw] (xf) at (8,1) {$\scriptscriptstyle 2$};
 \node[square, draw] (yf) at (5,1) {$\scriptscriptstyle 2$};
 \draw (dv) -- (yf); \draw (gv) -- (xf); \draw (av) -- (bv); \draw (bv) -- (cv); \draw (cv) -- (dv); \draw (dv) -- (ev); \draw (ev) -- (fv); \draw (fv) -- (gv); \draw (gv) -- (iv); \draw (jv) -- (iv); \draw (jv) -- (kv);

 \node[circle, draw] (a2) at (-2,-3) {$\scriptscriptstyle 6$};
 \node[circle, draw] (b2) at (-3,-3) {$\scriptscriptstyle 4$};
 \node[circle, draw] (c2) at (-4,-3) {$\scriptscriptstyle 2$};
 \node[square, draw] (a2f) at (-2,-2) {$\scriptscriptstyle 8$};
 \draw (a2) -- (a2f); \draw (a2) -- (b2); \draw (b2) -- (c2);
 
 \node[circle, draw] (av2) at (2,-3) {$\scriptscriptstyle 1$};
 \node[circle, draw] (bv2) at (3,-3) {$\scriptscriptstyle 2$};
 \node[circle, draw] (cv2) at (4,-3) {$\scriptscriptstyle 3$};
 \node[circle, draw] (dv2) at (5,-3) {$\scriptscriptstyle 4$};
 \node[circle, draw] (ev2) at (6,-3) {$\scriptscriptstyle 5$};
 \node[circle, draw] (fv2) at (7,-3) {$\scriptscriptstyle 6$};
 \node[circle, draw] (gv2) at (8,-3) {$\scriptscriptstyle 3$};
 \node[square, draw] (yf2) at (7,-2) {$\scriptscriptstyle 4$};
 \draw (fv2) -- (yf2); \draw (av2) -- (bv2); \draw (bv2) -- (cv2); \draw (cv2) -- (dv2); \draw (dv2) -- (ev2); \draw (ev2) -- (fv2); \draw (fv2) -- (gv2);

 \node[circle, draw] (a3) at (-2,-6) {$\scriptscriptstyle 4$};
 \node[circle, draw] (b3) at (-3,-6) {$\scriptscriptstyle 8$};
 \node[circle, draw] (c3) at (-4,-6) {$\scriptscriptstyle 4$};
 \node[square, draw] (a3f) at (-3,-5) {$\scriptscriptstyle 8$};
 \draw (b3) -- (a3f); \draw (a3) -- (b3); \draw (b3) -- (c3);
 
 \node[circle, draw] (av3) at (2,-6) {$\scriptscriptstyle 2$};
 \node[circle, draw] (bv3) at (3,-6) {$\scriptscriptstyle 4$};
 \node[circle, draw] (cv3) at (4,-6) {$\scriptscriptstyle 6$};
 \node[circle, draw] (dv3) at (5,-6) {$\scriptscriptstyle 8$};
 \node[circle, draw] (ev3) at (6,-6) {$\scriptscriptstyle 6$};
 \node[circle, draw] (fv3) at (7,-6) {$\scriptscriptstyle 4$};
 \node[circle, draw] (gv3) at (8,-6) {$\scriptscriptstyle 2$};
 \node[square, draw] (yf3) at (5,-5) {$\scriptscriptstyle 4$};
 \draw (dv3) -- (yf3); \draw (av3) -- (bv3); \draw (bv3) -- (cv3); \draw (cv3) -- (dv3); \draw (dv3) -- (ev3); \draw (ev3) -- (fv3); \draw (fv3) -- (gv3); 
 
 \node[circle, draw] (a4) at (-2,-9) {$\scriptscriptstyle 6$};
 \node[circle, draw] (b4) at (-3,-9) {$\scriptscriptstyle 12$};
 \node[circle, draw] (c4) at (-4,-9) {$\scriptscriptstyle 12$};
 \node[circle, draw] (d4) at (-5,-9) {$\scriptscriptstyle 12$};
 \node[circle, draw] (e4) at (-6,-9) {$\scriptscriptstyle 6$};
 \node[square, draw] (af4) at (-3,-8) {$\scriptscriptstyle 6$};
 \node[square, draw] (cf4) at (-5,-8) {$\scriptscriptstyle 6$};
 \draw (b4) -- (af4); \draw (d4) -- (cf4); \draw (a4) -- (b4); \draw (b4) -- (c4);  \draw (c4) -- (d4); \draw (d4) -- (e4);
 
 \node[circle, draw] (av4) at (2,-9) {$\scriptscriptstyle 3$};
 \node[circle, draw] (bv4) at (3,-9) {$\scriptscriptstyle 6$};
 \node[circle, draw] (cv4) at (4,-9) {$\scriptscriptstyle 9$};
 \node[circle, draw] (dv4) at (5,-9) {$\scriptscriptstyle 12$};
 \node[circle, draw] (ev4) at (6,-9) {$\scriptscriptstyle 12$};
 \node[] (fv4) at (7,-9) {$\scriptstyle \cdots $};
 \node[circle, draw] (gv4) at (8,-9) {$\scriptscriptstyle 12$};
 \node[circle, draw] (kv4) at (11,-9) {$\scriptscriptstyle 3$};
 \node[circle, draw] (jv4) at (10,-9) {$\scriptscriptstyle 6$};
 \node[circle, draw] (iv4) at (9,-9) {$\scriptscriptstyle 9$};
 \node[square, draw] (xf4) at (8,-8) {$\scriptscriptstyle 3$};
 \node[square, draw] (yf4) at (5,-8) {$\scriptscriptstyle 3$};
 \draw (dv4) -- (yf4); \draw (gv4) -- (xf4); \draw (av4) -- (bv4); \draw (bv4) -- (cv4); \draw (cv4) -- (dv4); \draw (dv4) -- (ev4); \draw (ev4) -- (fv4); \draw (fv4) -- (gv4); \draw (gv4) -- (iv4); \draw (jv4) -- (iv4); \draw (jv4) -- (kv4);

\end{tikzpicture}
\caption{Left: balanced linear quivers that satisfy \eqref{eq:assumptionMir}. Right: the corresponding quivers characterised by $\mathcal{R}^{\vee} (\eta^{\vee})$. From top to bottom: $(P,K)=(4,16),(4,8),(4,8),(6,12)$. In the right quiver of the first line the dots omit 6 gauge nodes of rank 8; in the right quiver of the fourth line the dots omit 2 gauge nodes of rank 12. In $d=3$ the second and third pairs of quivers are mirror pairs.}
\label{fig:mirrortype}
\end{figure}
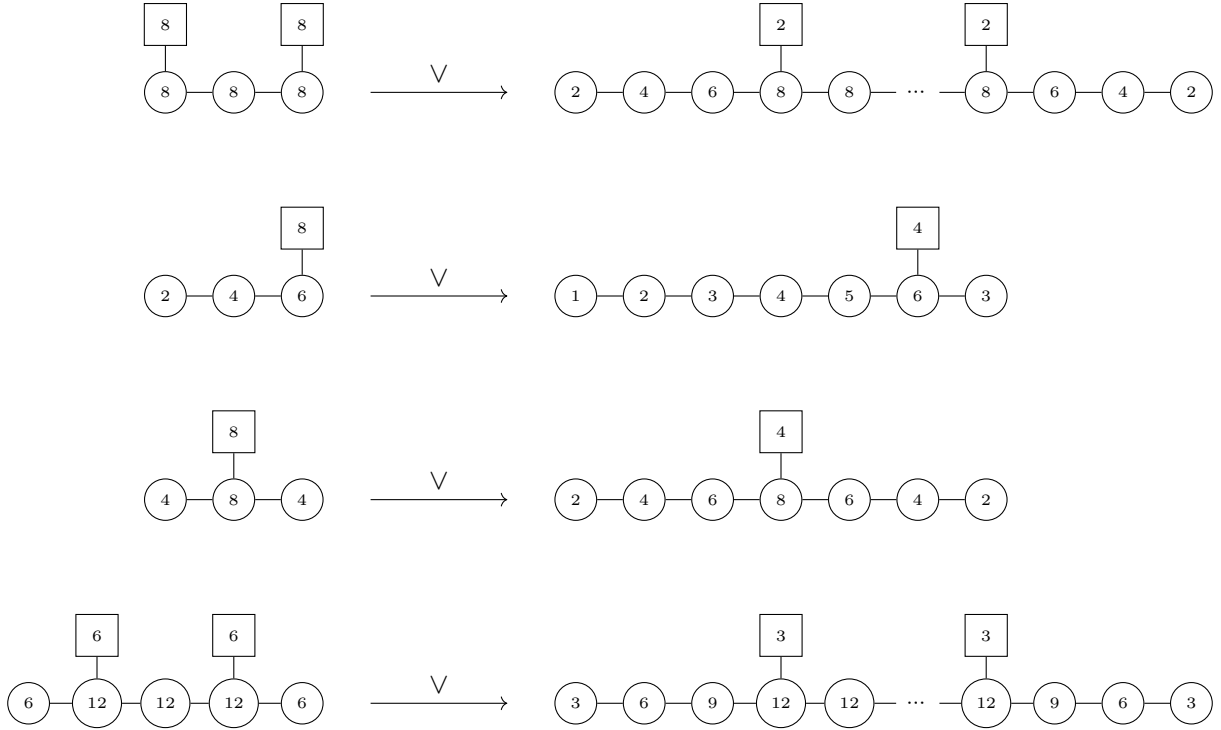\par

The functions $\mathcal{R} (\eta)$ and $\mathcal{R}^{\vee} (\eta^{\vee}) $ are in general different, unless $K=P$, and provide different linear quivers. Examples of quivers related by this transformation are given in Figure \ref{fig:mirrortype}. However, they have equal Fourier coefficients, 
\begin{align*}
	R_k &= \frac{2}{P} \int_0 ^{P} \dd \eta \mathcal{R} (\eta) \sin \left( \frac{\pi k \eta}{P} \right) \\
		&= \frac{2}{K} \int_0 ^{K} \dd \eta^{\vee} \mathcal{R}^{\vee} (\eta) \sin \left( \frac{\pi k \eta^{\vee}}{P} \right) = R_k ^{\vee} ,
\end{align*}
using the change of variables $\eta^{\vee} =\frac{K}{P} \eta$. From this simple identity and the fact that the quantities we compute are expressed mainly in terms of Fourier coefficients, we obtain the following mirror-type identities.
\begin{stm}
	Consider a balanced linear quiver that satisfies \eqref{eq:assumptionMir}, and denote with a symbol ${}^{\vee}$ the quantities in the quiver constructed from $\mathcal{R}^{\vee} (\eta^{\vee})$. It holds that
	\begin{align}
	\label{eq:FMir}
		\frac{\tilde{F}_d}{P^{d-3}} &= \frac{\tilde{F}^{\vee}}{K^{d-3}} , \\
		\frac{ \ln \langle W_{\mathsf{A}_{\kk}} (z_{\ast}) \rangle }{P} &= \frac{\ln \langle W_{\mathsf{A}_{\kk}}^{\vee} (z_{\ast}) \rangle }{K} , \qquad F_{\text{\rm 3d defect},d} (z_{\ast}) = F_{\text{\rm 3d defect},d} ^{\vee}(z_{\ast}) . \label{eq:WLandF3dMir}
	\end{align}
\end{stm}
At $d=3$ the relation \eqref{eq:FMir} becomes an exact equality of free energies from two distinct quiver gauge theories, first noted in \cite{Akhond:2021ffz}. When the initial quiver has $K$ hypermultiplets at a single node, then the two rank functions $\mathcal{R} (\eta)$ and $\mathcal{R}^{\vee} (\eta ^{\vee})$ describe genuine three-dimensional mirror theories and the equality \eqref{eq:FMir} becomes a check of mirror symmetry \cite{Akhond:2021ffz}. On the other hand, the identity \eqref{eq:WLandF3dMir} is not a statement on mirror symmetry, since we are equating two Wilson loops, instead of a Wilson and a vortex loop \cite{Assel:2015oxa}.\par
Let us now elaborate on aspects of \eqref{eq:FMir}-\eqref{eq:WLandF3dMir}. The equality of the free energies divided by a power of the length of the quiver holds directly at the infinite coupling point. Moreover, for every choice of gauge couplings $g_{d,j}$ in the initial quiver, there always exists a suitable choice of gauge couplings $g_{d,j}^{\vee}$ such that they have equal Fourier coefficients, thus extending \eqref{eq:FMir} to finite gauge couplings. We also stress that in \eqref{eq:WLandF3dMir} the defects are located at the same relative position $z_{\ast}$ along the quiver, which will correspond to the nodes $j_{\ast} = \lfloor P z_{\ast} \rfloor$ and $j_{\ast}^{\vee} = \lfloor K z_{\ast} \rfloor = \frac{K}{P} j_{\ast}$. The value of the parameter $x_{\ast}$ that appears in the evaluation of the Wilson loop expectation value is uniquely fixed by $z_{\ast}$, $\left\{ R_k \right\}$ and the representation $\kk$, thus it will be equal on the two sides of the map. Furthermore, observe that the non-universal part of $F_{\text{3d defect},d}$ is contained in the pre-factor $P \nu (z_{\ast})$. Using $P^{\vee} =K$ and $N_{K j_{\ast}/P}^{\vee} =\frac{P}{K} N_{j_{\ast}}$, this pre-factor is left invariant under the transformation, namely $P \nu (z_{\ast}) = P^{\vee} \nu^{\vee} (z_{\ast})$.\par
\begin{digr}
Partial Higgsing of a linear quiver preserves the balancing condition. Therefore, one can in principle always start with pairs of quivers with flavours at a single node, for which the graph of the rank function is triangular, and perform subsequent partial Higgsing to reach other balanced quivers, see Figure \ref{fig:Higgsing}. It would be interesting to see if this observation can be used to provide a physical derivation of \eqref{eq:FMir}.
\end{digr}
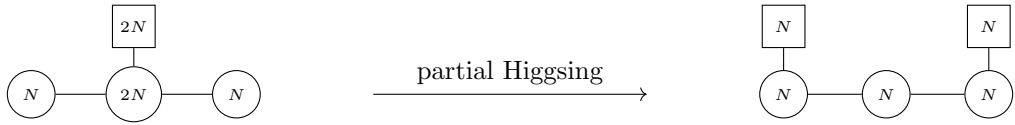
\begin{figure}[htb]
\centering
\begin{tikzpicture}[auto,square/.style={regular polygon,regular polygon sides=4},scale=0.9]

 \path[->] (-2,0) edge node {\small partial Higgsing} (2,0);
 
 \node[circle, draw] (a) at (4,0) {$\scriptscriptstyle N$};
 \node[circle, draw] (b) at (5.5,0) {$\scriptscriptstyle N$};
 \node[circle, draw] (c) at (7,0) {$\scriptscriptstyle N$};
 \node[square, draw] (af) at (4,1) { \hspace{4pt} } ;
 \node[square, draw] (cf) at (7,1) { \hspace{4pt} };
 \node[] (uf) at (4,1) {$\scriptscriptstyle N$};
 \node[] (vf) at (7,1) {$\scriptscriptstyle N$};
 \draw (a) -- (af); \draw (c) -- (cf); \draw (a) -- (b); \draw (b) -- (c);
 
 \node[circle, draw] (a3) at (-4,0) {$\scriptscriptstyle N$};
 \node[circle, draw] (b3) at (-5.5,0) {$\scriptscriptstyle 2N$};
 \node[circle, draw] (c3) at (-7,0) {$\scriptscriptstyle N$};
 \node[square, draw] (a3f) at (-5.5,1) { \hspace{4pt} };
 \node[] (u3f) at (-5.5,1) {$\scriptscriptstyle 2N$};
 \draw (b3) -- (a3f); \draw (a3) -- (b3); \draw (b3) -- (c3);
 
 \end{tikzpicture}
\caption{Two quivers that satisfy \eqref{eq:assumptionMir} are related by partial Higgsing. Left: The graph of the rank function $\mathcal{R}(\eta)$ is triangular, hence $\mathcal{R}^{\vee} (\eta^{\vee})$ describes the genuine mirror in $d=3$. Right: The rank functions $\mathcal{R} (\eta)$ and $\mathcal{R}^{\vee} (\eta^{\vee})$ are not related by duality.}
\label{fig:Higgsing}
\end{figure}\par

\subsection{Mass deformation}
Next, we study the long quiver limit when the fundamental hypermultiplets are given a real mass, closely following \cite[Sec.3]{Akhond:2022oaf}.\par
Let $\mathscr{F} \in \N$ be an $\mathcal{O}(1)$ number, in the sense that it does not scale with $N$ nor $P$ in the long quiver limit. At every $j$, we split the number $K_j$ of fundamental hypermultiplets into a collection 
\begin{equation*}
	\left\{ K_{j,\alpha} \right\}_{\alpha=1, \dots, \mathscr{F}} , \qquad \sum_{\alpha=1}^{\mathscr{F}}  K_{j,\alpha} = K_j.
\end{equation*}
The $\alpha^{\text{th}}$ collection at every $j$ is assigned a mass $m_{\alpha}$, uniform along the quiver (that is, $m_{\alpha}$ does not depend on the index $j=1, \dots, P$). The canonical way to introduce a real mass is to couple the hypermultiplet to a background vector multiplet for the flavour symmetry group. The real mass parameters $m_{\alpha}$ enter the matrix model on the same footing as the dynamical scalar zero-mode $\vec{\phi}$, although they are not integrated over. Therefore, by the long quiver scaling \eqref{eq:LQansatz}, the masses scale as 
\begin{equation*}
	m_{\alpha}  = (d-2)P \mu_{\alpha},
\end{equation*}
where $\mu_{\alpha}$ are $\mathcal{O}(1)$ mass parameters. Below we employ the notation $\mu_{\alpha\beta} := \mu_{\alpha}-\mu_{\beta}$ and $\vec{\mu}:=(\mu_1, \dots, \mu_{\mathscr{F}})$.\par
To the $\alpha^{\text{th}}$ collection of hypermultiplets we associate a Veneziano function $\zeta_{\alpha}(z)$ in the obvious way. Imposing that the balancing condition \eqref{eq:balance} is satisfied by every collection of hypermultiplets, the rank function splits into 
\begin{equation*}
	\mathcal{R} (\eta) \ \mapsto \ \sum_{\alpha=1}^{\mathscr{F}} \mathcal{R}_{\alpha} (\eta) .
\end{equation*}
We omit the details of the derivation, because they are identical to \cite[Sec.3]{Akhond:2022oaf}. The upshot is that the derivation in Subsections \ref{sec:LQlimit}-\ref{sec:LQsolved} goes through with only mild modifications, and eventually the eigenvalue density takes the form 
\begin{equation*}
	\varrho (z,x) = \sum_{\alpha=1}^{\mathscr{F}} \varrho_{\alpha} (z,x) \ , \qquad \varrho_{\alpha} (z,x)= \frac{\pi}{N}\sum_{k=1}^{\infty} k R_{\alpha,k} \sin (\pi k z) \ee^{- 2\pi k \lvert x \rvert}.
\end{equation*}
Evaluating the interpolating free energy with this solution we obtain 
\begin{equation*}
\begin{aligned}
	\tilde{F}_d (\vec{\mu}) \simeq - \frac{1}{2}  N^2 P^{(d-3)} \sin \left( \frac{\pi d}{2}\right) (d-2)^{d-2} & \int_0 ^{1} \dd z \int_{- \infty} ^{+ \infty} \dd x \sum_{\alpha=1}^{\mathscr{F}}\varrho_{\alpha} (z,x) \\
	& \times \left[  \sum_{\beta=1}^{\mathscr{F}} P^2 \zeta_{\beta} (z) s_d \lvert x - \mu_{\beta\alpha} \rvert^{d-2} + \frac{\lvert w_d \rvert }{2 \tilde{t}_d (z)} x^2 \right].
\end{aligned}
\end{equation*}
In the notation we are leaving implicit the fact that $\tilde{F}_d $ depends not only on the values of the masses, but also on the way of assigning them to the hypermultiplets, i.e. on the splitting $\sum_{\alpha=1}^{\mathscr{F}}  K_{j,\alpha} =K_j$. Using $\sum_{\alpha=1}^{\mathscr{F}} R_{\alpha,k}=R_k$, the part of $\tilde{F}_d (\vec{\mu})$ that depends on the gauge couplings is identical to the second line of \eqref{eq:Ftilded}, thus it is not affected by the mass deformation. Denoting 
\begin{equation*}
	\delta \tilde{F}_d (\vec{\mu}) := \tilde{F}_d (\vec{\mu}) -\tilde{F}_d (0) 
\end{equation*}
the mass-dependent part of $\tilde{F}_d$, we get 
\begin{equation*}
\begin{aligned}
	\delta \tilde{F}_d (\vec{\mu}) &\simeq - P^{(d-3)} \sin \left( \frac{\pi d}{2}\right) (d-2)^{(d-2)} \frac{\pi^3}{4} s_d \\
	& \times \sum_{\alpha, \beta =1}^{\mathscr{F}} \sum_{k=1} ^{\infty} k^{3} R_{\alpha,k}  R_{\beta,k}  \int_{- \infty} ^{+ \infty} \dd x ~\ee^{-2 \pi k \lvert x \rvert} \left( \lvert x-\mu_{\beta\alpha} \rvert^{d-2} -  \lvert x\rvert^{d-2}\right)  .
\end{aligned}
\end{equation*}
The integral can be performed explicitly for arbitrary $d$, but the resulting expression is rather cumbersome:
\begin{equation*}
\begin{aligned}
	\delta \tilde{F}_d (\vec{\mu}) & \simeq P^{(d-3)} \frac{(d-2)^{(d-2)} \sin \left( \pi d\right) \Gamma (2-d)}{4(d-1)}  \sum_{\alpha, \beta =1}^{\mathscr{F}} \sum_{k=1} ^{\infty} (\pi k)^{4} R_{\alpha,k}  R_{\beta,k}  \\
	& \times \left[ \frac{2\mu_{\beta\alpha}^d}{d}  {}_1 F_2 \left( 1; \frac{d+1}{2}, \frac{d+2}{2} ; (\pi k \mu_{\beta\alpha})^2 \right) -  \frac{\mu_{\beta \alpha}^{d-1}}{k \pi} {}_1 F_2 \left( 1; \frac{d+1}{2}, \frac{d}{2} ; (\pi k \mu_{\beta \alpha})^2 \right)  + \cdots \right]
\end{aligned}	
\end{equation*}
with the ellipsis containing additional terms involving Meijer's $G$-function with argument $(\pi k \mu_{\beta\alpha})^2$. Evaluating $\delta \tilde{F}_d (\vec{\mu}) $ at integer $d$, we find 
\begin{equation*}
	\delta \tilde{F}_d (\vec{\mu}) \simeq P^{(d-3)}(d-2)^{(d-2)} \frac{\pi^{5-d}}{2^d} \sum_{\alpha, \beta =1}^{\mathscr{F}} \sum_{k=1} ^{\infty} k^{4-d} R_{\alpha,k}  R_{\beta,k} ~ \wp_d (2\pi k\mu_{\beta\alpha}) ,
\end{equation*}
where $\wp_d (\cdot)$ is 
\begin{align*}
	\wp_d (y) = \begin{cases} \ee^{-y} + c_1 y + c_3 y^3 + \cdots + c_{d-2} y^{d-2} &  d \text{ odd } \\ 1 + c_2 y^2 + \cdots + c_{d-2} y^{d-2} & d \text{ even }  . \end{cases}
\end{align*}
That is, we find a polynomial in $2 \pi k \mu_{\alpha \beta}$, with the powers having precisely the structure discussed in \cite{Gerchkovitz:2014gta}. These polynomial pieces in $\mu_{\alpha \beta}$ are scheme-dependent and can be cancelled by local counterterms. Additionally, at odd $d$, we observe the exponential dependence on the masses already found in \cite{Akhond:2022oaf}.

\section{Four-dimensional long quiver SCFTs}
\label{sec:4dLQ}

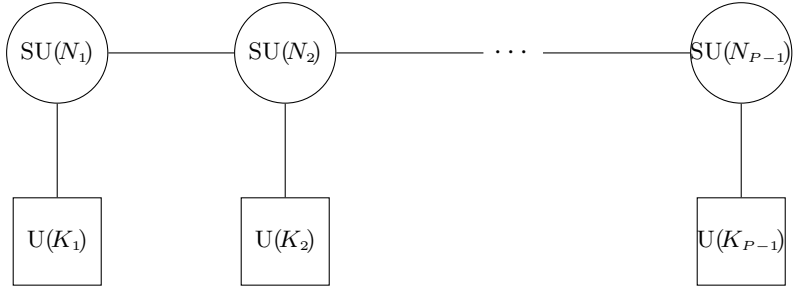
\begin{figure}[t]
\centering
\begin{tikzpicture}[auto,square/.style={regular polygon,regular polygon sides=4}]
	\node[circle,draw] (gauge1) at (5,0) { \hspace{30pt} };
	\node (a1) at (5,0) {\footnotesize $\mathrm{SU}(\!N_{\scriptscriptstyle P-1}\!)$};
	\node[draw=none] (gaugemid) at (2,0) {$\cdots$};
	\node[circle,draw] (gauge3) at (-1,0) { \hspace{30pt} };
	\node[circle,draw] (gauge4) at (-4,0) { \hspace{30pt} };
	\node (a2) at (-4,0) {\footnotesize $\mathrm{SU}(\!N_{\scriptscriptstyle 1}\!)$};
	\node (a3) at (-1,0) {\footnotesize $\mathrm{SU}(\!N_{\scriptscriptstyle 2}\!)$};
	\node[square,draw] (fl1) at (5,-2.5) { \hspace{16pt} };
	\node[square,draw] (fl2) at (-4,-2.5) { \hspace{16pt} };
	\node[square,draw] (fl3) at (-1,-2.5) { \hspace{16pt} };
	\node[draw=none] (aux1) at (5,-2.5) {\footnotesize $\mathrm{U}(\!K_{\scriptscriptstyle P-1}\!)$};
	\node[draw=none] (aux2) at (-4,-2.5) {\footnotesize $\mathrm{U}(\!K_{\scriptscriptstyle 1}\!)$};
	\node[draw=none] (aux3) at (-1,-2.5) {\footnotesize $\mathrm{U}(\!K_{\scriptscriptstyle 2}\!)$};
	\draw[-](gauge1)--(gaugemid);
	\draw[-](gaugemid)--(gauge3);
	\draw[-](gauge4)--(gauge3);
	\draw[-](gauge1)--(fl1);
	\draw[-](gauge4)--(fl2);
	\draw[-](gauge3)--(fl3);
\end{tikzpicture}
\caption{Linear quiver of length $P-1$. Circular nodes indicate gauge groups, square nodes indicate flavour symmetries. When the quiver is balanced, it describes a four-dimensional $\mN=2$ SCFT.}
\label{fig:quiver4d}
\end{figure}

In this section we study four-dimensional $\mN=2$ linear quivers shown in Figure \ref{fig:quiver4d}. The gauge group is a product of special unitary groups $\mathrm{SU}(N_j)$, $j=1,\dots,P-1$.\par
We assume that the gauge theory is balanced, which, in $d=4$, implies it is conformal. One consequence of this fact is that there are two ways of taking the limit:
\begin{itemize}
	\item the large-$P$ limit at fixed 't Hooft coupling;
	\item going to strong coupling first, and only then take the large-$P$ limit.
\end{itemize}
We will analyse both prescriptions and show that they explore different regimes.\par
\par

\subsection{Anomaly coefficients from counting supermultiplets}
Before delving into the study of the partition function of the gauge theory on $\cs^4$, we derive the anomaly coefficients $a$ and $c$ of the four-dimensional long quiver SCFTs from the relations \cite{Shapere:2008zf}
\begin{equation}
\label{eq:STac}
	a = \frac{5 n_{\rm v} + n_{\rm h}}{24} , \qquad c= \frac{2 n_{\rm v} + n_{\rm h}}{12}.
\end{equation}
In these expressions, $n_{\rm v}$ and $ n_{\rm h}$ are, respectively, the number of vector multiplet and hypermultiplet modes. 
\begin{stm}
	In the long quiver limit, \eqref{eq:STac} yields 
	\begin{center}\noindent\fbox{\parbox{0.98\linewidth}{%
	\begin{equation}
	\label{eq:acfromST}
		a\simeq c \simeq \frac{P}{8}  \sum_{k=1}^{\infty} R_k^2
	\end{equation}}}\end{center}
	where we recall that $R_k$ are the Fourier coefficients of the rank function.
\end{stm}
\begin{proof}
In the linear quivers we are considering, the number of vector multiplet and hypermultiplet modes are given by 
\begin{equation*}
	n_{\rm v} = \sum_{j=1} ^{P-1} \left( N_j ^2 -1 \right) , \qquad n_{\rm h} = \sum_{j=1} ^{P-2} N_j N_{j+1}  +  \sum_{j=1} ^{P-1} N_j K_{j}  ,
\end{equation*}
where the first and second contributions in $n_{\rm h}$ come from bifundamental and fundamental hypermultiplets. The contribution from the bifundamental hypermultiplets is rewritten in a more symmetric form:
\begin{equation*}
	\sum_{j=1} ^{P-2} N_j N_{j+1} = \frac{1}{2} \sum_{j=1} ^{P-1} N_j\left(  N_{j+1}+ N_{j-1} \right) , \qquad N_0 := 0 =: N_P .
\end{equation*}
Imposing the balancing condition \eqref{eq:balance}, we get 
\begin{equation*}
	n_{\rm v} = \sum_{j=1} ^{P-1} \left( N_j ^2 -1 \right) , \qquad n_{\rm h} =  \sum_{j=1} ^{P-1} N_j \left( 2 N_j -  \frac{1}{2} N_{j-1} -  \frac{1}{2} N_{j+1}  \right) .
\end{equation*}
Computing the $a$ and $c$ anomaly coefficients with these values we obtain 
\begin{align*}
	48 a &= \sum_{j=1} ^{P-1} \left[ 12 N_j^2 - N_j \left( N_{j+1} - 2 N_j + N_{j-1} \right) \right] - 10 (P-1) , \\
	48 c &= \sum_{j=1} ^{P-1} \left[ 12 N_j^2 -2  N_j \left( N_{j+1} - 2 N_j + N_{j-1} \right) \right] - 8 (P-1) .
\end{align*}
We now take the long quiver limit of these expressions, following the notation of Section \ref{sec:d}. We find 
\begin{align*}
	48 a &= N^2 P \int_0 ^{1} \dd z ~ \left[ 12 \nu (z)^2 - \frac{1}{P^2} \nu(z) \partial_z ^2 \nu (z) \right] , \\
	48 c &= N^2 P \int_0 ^{1} \dd z ~ \left[ 12 \nu (z)^2 - \frac{2}{P^2} \nu(z) \partial_z ^2 \nu (z) \right] .
\end{align*}
The scaled rank function $\nu (z)$ converges to a function of class $C^2 ((0,1))$ in the limit $P\to\infty$, hence the second piece drops out.\footnote{This ultimately stems from the definition of the long quiver limit. The balancing condition \eqref{eq:dnudziszeta} implies that $P^{-2} \partial_z ^2 \nu (z)$ vanishes almost everywhere in the limit, because it is not possible to obtain balanced linear quivers with $\mathcal{O}(P)$ flavour nodes. Furthermore, the flavour ranks must be independent of $P$ in order to fulfil the balancing condition when $N \to \infty$ independently of $P$. These two facts guarantee that $P^2 \zeta (z)$ is finite in the long quiver limit, thereby $\nu \in C^2 ((0,1))$ from \eqref{eq:dnudziszeta}.} We then expand $\nu (z)$ in Fourier series, recalling that $N\nu(z)= \mathcal{R}(\eta=zP)$, and integrate over $z$. We thus derive the identity 
\begin{equation*}
		a \simeq \frac{P}{8} \sum_{k=1}^{\infty} R_k^2 \simeq c ,
	\end{equation*}
for all four-dimensional linear quivers, at leading order in the planar and long-quiver limit.\par
This formula suggests a generic linear scaling of $a$ with $P$. However, the computation of $R_k$ at finite $P$ in concrete quivers, and the ensuing large-$P$ approximation, may yield corrections to this behaviour. This will be manifest in the examples in Section \ref{sec:ex}.
\end{proof}
The relation $a \simeq c$ is expected to hold in any SCFT with a supergravity dual at large-$N$ \cite{Henningson:1998gx}, and we have checked it explicitly for linear quivers. It matches the supergravity computation in \cite[Eq.(2.41)]{Nunez:2019gbg}. Recalling that $\tilde{F}_{d=4} = 2 \pi a$, \eqref{eq:acfromST} matches exactly with the derivation \eqref{eq:tildeF4d} using the iterpolating function $\tilde{F}_d$ of Giombi--Klebanov \cite{Giombi:2014xxa}.
\par
We should stress that both large-$N$ and large-$P$ are needed to ensure $a=c$. The simplest of all quivers, the $\mathcal{N}=2$ QCD, has $P=2$ and for it $a$ and $c$ differ by $\mathcal{O}(N^2)$. It has been known for a while that the string dual of $\mathcal{N}=2$ QCD must be a strongly coupled string theory, even in the limit of infinite 't~Hooft coupling \cite{Gadde:2009dj}. The central charge computation above demonstrates, in a simple way, that only long quivers can have a weakly coupled string dual well approximated by supergravity.  
\par
We can in fact compute the finite-$P$ corrections to $a-c$ while keeping the large-$N$ limit, and get:
\begin{equation*}
	a-c = \frac{\pi^2}{96 P}  \sum_{k=1}^{\infty} k^2 R_k^2 .
\end{equation*}
This value should arise from stringy corrections in the holographic dual background, and we have expressed it here using the rank function, which in turn characterises the backgrounds of Gaiotto--Maldacena \cite{Gaiotto:2009gz}.\par

\subsection{Four-dimensional long quiver partition function}
\label{sec:4dZlong}

The sphere partition function of the $d=4$, $\mN=2$ linear quiver is of the form introduced in Subsection \ref{sec:setup}. With the notation therein, we have \cite{Pestun:2007rz}
\begin{align*}
	f_{\mathrm{v}} (\phi ) & = - \ln (\phi) - \ln H (\phi ) \\
	f_{\mathrm{h}} (\phi) & = \ln H (\phi ) .
\end{align*}
The function $H(\cdot)$ is expressed using Barnes's $G$-function $G(\cdot)$ as\footnote{Our $H (\cdot )$ equals $H ( \sqrt{-1} \cdot )$ of \cite{Pestun:2007rz}.} 
\begin{equation*}
	H(\phi) := G(1+\ii\phi)G(1-\ii\phi),
\end{equation*}
and it admits an infinite product representation \cite[Eq.(4.49)]{Pestun:2007rz} which, taking the logarithm, reads
\begin{equation*}
	\ln H(\phi) = (1+\gamma_{\text{\tiny EM}})\phi^2 +  \sum_{n=1}^{\infty} \left[ - \frac{\phi^2}{n} + n \ln \left( 1+ \frac{\phi^2}{n^2}\right) \right] .
\end{equation*}
Here $\gamma_{\text{\tiny EM}}$ is the Euler--Mascheroni constant.\par
After using the substitutions explained in full generality in Subsection \ref{sec:setup}, the matrix model effective action becomes 
\begin{align*}
	S_{\mathrm{eff}} \simeq P N^2 \int_0 ^1 \dd z \int \dd \phi \rho (z, \phi) & \left\{  \frac{1}{2 t_4(z)} \phi^2 + \zeta (z) \ln H(\phi) \right. \\
	- & \dashint\dd \sigma  \rho (z, \sigma) \left[  \ln \lvert H (\phi- \sigma)\rvert  + \ln \lvert \phi- \sigma \rvert \right] \\
	+ & \left. \int \dd \sigma \left[ \frac{1}{2P^2} \partial_z ^2 \rho (z, \sigma) + \rho (z, \sigma) \right] \ln \lvert H (\phi- \sigma) \rvert  \right\}
\end{align*}
where $\dashint$ indicates the principal value integral. We have also used $u_{d=4}=1$. This expression is the direct four-dimensional analogue of \eqref{eq:Seffdlong1}.\par
We combine part of the terms coming from vector multiplets and hypermultiplets:
\begin{equation*}
\begin{aligned}
	- \dashint \dd \sigma  \rho (z, \sigma) \ln \lvert H (\phi- \sigma)\rvert + \int \dd \sigma  \rho (z, \sigma) \ln \lvert H (\phi- \sigma)\rvert & =  \int \dd \sigma  \rho (z, \sigma)  \ln \lvert H (\phi- \sigma)\rvert \delta(\sigma - \phi) \\
	&=  \ln \lvert H (0)\rvert \\
	&= 0 ,
\end{aligned}
\end{equation*}
using that $H(0)=G(1)^2=1$. With this simplification we have 
\begin{equation}
\label{eq:Seff4d}
\begin{aligned}
	S_{\mathrm{eff}} \simeq P N^2 \int_0 ^1 \dd z \int \dd \phi \rho (z, \phi) & \left\{  \frac{1}{2 t_4(z)} \phi^2 + \zeta (z) \ln H(\phi)  - \dashint \dd \sigma  \rho (z, \sigma) \ln \lvert \phi- \sigma \rvert \right. \\
		& \left. + \frac{1}{2P^2}  \int \dd \sigma ~ \ln \lvert H (\phi- \sigma) \rvert  ~\partial_z ^2 \rho (z,\sigma)   \right\} .
\end{aligned}
\end{equation}
Let us define the function 
\begin{equation*}
	\mathscr{K} (\phi) := - \partial_{\phi} \ln H(\phi) = \phi \left[ \psi (1+\ii\phi) + \psi (1-\ii\phi) -2 \right] ,
\end{equation*}
where $\psi (\cdot)$ is the digamma function.\par
The saddle point equation derived from the effective action \eqref{eq:Seff4d} is 
\begin{equation}
\label{eq:spefull4d}
	\dashint \dd \sigma \frac{\rho (z, \sigma)}{\phi - \sigma} + \frac{1}{2P^2} \int \dd \sigma \mathscr{K} (\phi - \sigma) \partial_z^2\rho (z, \sigma) = \frac{ \phi}{2t_4 (z)} - \frac{1}{2P^2} [P^2 \zeta (z)] \mathscr{K} (\phi) .
\end{equation}

\begin{digr}
Along the way, we observe that \eqref{eq:spefull4d} is invariant under redefinitions 
\begin{equation*}
	\mathscr{K} (\phi) \ \mapsto \ \mathscr{K} (\phi) + q(z) \phi 
\end{equation*}
for every continuous function $q:[0,1] \to \R$. Indeed, gathering the two terms in \eqref{eq:spefull4d} that bear the function $\mathscr{K} (\cdot )$ and performing the shift, we find 
\begin{equation*}
\begin{aligned}
	& \frac{1}{2P^2} \left[ \int \dd \sigma \mathscr{K} (\phi - \sigma) \partial_z^2\rho (z, \sigma) + [P^2 \zeta (z)] \mathscr{K} (\phi) \right] \\
	& \hspace{2.5cm} +  \frac{q(z) }{2P^2} \left[ \phi \partial_z^2 \int \dd \sigma \rho (z, \sigma) - \partial_z^2 \int \dd \sigma \rho (z, \sigma) \sigma  + [P^2 \zeta (z)] \phi \right] .
\end{aligned}
\end{equation*}
We claim that the second square bracket vanishes. The matrix model we started with is even in the eigenvalues, thus $\rho(z,\sigma)$ is even and has vanishing first moment:
\begin{equation*}
	 \int \dd \sigma \rho (z, \sigma) \sigma = 0 \qquad \forall~0<z<1.
\end{equation*}
On the other hand, using the normalisation condition $\int \dd \sigma \rho (z, \sigma) = \nu(z)$, which stems from the definition of $\rho$, we have that 
\begin{equation*}
	q(z) \phi \left[ \partial_z^2 \int \dd \sigma \rho (z, \sigma)  + P^2\zeta (z) \right] = q(z) \phi  \left[ \partial_z^2 \nu (z)  + P^2\zeta (z) \right] =0,
\end{equation*}
which vanishes due to the balancing condition \eqref{eq:dnudziszeta}.
\end{digr}

\subsection{Recovering \texorpdfstring{$a$}{a} from the matrix model}
\label{sec:a4d}
To solve \eqref{eq:spefull4d} we need to understand the scaling of $\phi$ in the long quiver limit. 
The behavior at large $P$ and strong coupling depends on the order of limits. If $P$ is taken large first, the interaction between different nodes cancels against the non-Abelian vector multiplet of each node and the solution behaves as in the $\mathcal{N}=4$ super-Yang--Mills with the Wigner distribution of eigenvalues, while taking the strong coupling limit first results in the same solution as in $d<4$, even if the action has a ln singularity as $d$ approaches $4$. Here we take $P\rightarrow \infty $ first and return to the second case later, in Subsection \ref{sec:4dStrong}.
\begin{ansatz}
	In $d=4$, $\phi$ remains $\mathcal{O}(1)$ as $P\to\infty$.
\end{ansatz}
We will see that the computation of $\ln \mz_{\cs^4}$ at finite coupling reproduces the value of the Weyl anomaly coefficient. With this no-scaling ansatz, \eqref{eq:spefull4d} drastically simplifies. The terms that carry the function $\mathscr{K} (\cdot)$ drop out, as both of them are suppressed by $1/P^2$. The saddle point equation reduces to 
\begin{equation}
\label{eq:spe4d}
	\dashint \dd \sigma \frac{ \rho (z, \sigma) }{\phi - \sigma} = \frac{\phi }{2 t_4 (z)}  ,
\end{equation}
The latter is precisely the saddle point equation of $\mN=4$ super-Yang--Mills theory on $\cs^4$, except for the $z$-dependence. It appears in two ways:
\begin{itemize}
	\item The $z$-dependent normalisation 
		\begin{equation*}
			\int_{-\infty}^{\infty}\dd\phi \rho (z,\phi) = \nu (z).
		\end{equation*}
		We denote $\rho(z,\phi)=\nu(z) \bar{\rho}(z,\phi)$, with $\bar{\rho}$ canonically normalised to 1.
	\item The dependence of the 't Hooft coupling $t_4 (z)$ on the gauge node. We introduce an effective 't Hooft coupling 
		\begin{equation*}
			\lambda (z) := \nu (z)t_4 (z).
		\end{equation*}
\end{itemize}
With these redefinitions, \eqref{eq:spe4d} becomes 
\begin{equation}
\label{eq:GUESPE}
	\dashint \dd \sigma \frac{ \bar{\rho} (z, \sigma) }{\phi - \sigma} = \frac{\phi }{2 \lambda (z)}  ,
\end{equation}
which is now the canonically normalised saddle point equation for $\mN=4$ super-Yang--Mills at every $0<z<1$. The solution is well-known to be the Wigner semicircle 
\begin{equation*}
	\bar{\rho} (z,\phi) = \frac{1}{2 \pi \lambda (z) } \sqrt{ 4\lambda(z) - \phi^2 } , \qquad \qquad \phi \in \left[ - 2\sqrt{\lambda (z)} , 2 \sqrt{\lambda (z)} \right] ,
\end{equation*}\par
\bigskip
Equipped with the solution for the eigenvalue density we can compute the leading order contribution to the partition function. We get 
\begin{align*}
	\left. S_{\mathrm{eff}} \right\rvert_{\text{on-shell}} &= \frac{1}{2} N^2 P \int_0 ^1 \dd z \int \dd \phi ~ \rho(z,\phi) \frac{\phi^2}{2 t_d(z)} \\
		&= \frac{1}{4} P N^2 \int_0 ^1 \dd z \nu(z)^2 \\
		&= \frac{P}{8} \sum_{k=1}^{\infty} R_k^2  .
\end{align*}
We thus reproduce the evaluation of the Weyl anomaly coefficient already obtained in \eqref{eq:acfromST}, 
\begin{equation*}
	a \simeq \left. S_{\mathrm{eff}} \right\rvert_{\text{on-shell}}  .
\end{equation*}
Notice that there is no dependence on $t_4(z)$, which are marginal parameters in the conformal gauge theory, despite $\rho(z,\phi)$ depends on it. This result is consistent with the fact that $a$ is invariant under marginal deformations.

\subsection{Extremal correlators}

Correlation functions of Coulomb branch chiral ring operators in any $\mN=2$ conformal gauge theory in four dimensions were considered in \cite{Gerchkovitz:2016gxx}. There it was shown that the correlation functions of any number of chiral primary and a single anti-chiral primary operators can be obtained from a deformed matrix model for the 4d $\mN=2$ theory on $\cs^4$. Other recent studies of chiral operators using the matrix model derived from localisation include \cite{Rodriguez-Gomez:2016ijh,Baggio:2016skg,Fiol:2020bhf,Fiol:2020ojn,Beccaria:2021hvt,Fiol:2021icm,Billo:2021rdb,Billo:2022gmq,Fiol:2022vvv,Billo:2022fnb}.
\begin{defin}
Consider a balanced four-dimensional SCFT. Correlation functions of an arbitrary number of chiral primary operators and one anti-chiral primary operator are called \emph{extremal correlators}.
\end{defin}
The master formula of \cite{Gerchkovitz:2016gxx} states that the generating function of extremal correlators is:
\begin{equation}
\label{eq:genfun4dExtCorr}
	\mz_{\cs^4} \left[ \left\{ \hat{\lambda}_{j,A} \right\} \right] = \mz_{\cs^4} \left\langle \exp \left( \sum_{j=1}^{P-1}\sum_{A=3}^{N_j} \pi^{A/2} \hat{\lambda}_{j,A} \sum_{a=1}^{N_j} \left( \phi_a ^{(j)} \right)^{A} \right)\right\rangle_{\cs^4} ,
\end{equation}
where, on the right-hand side, $\langle \cdot \rangle_{\cs^4}$ means the expectation value taken in the matrix model derived from localisation on the sphere. The couplings $\left\{ \hat{\lambda}_{j,A} = - 2 \Im (\tau_{j,A} ) \right\}$ are parameters which are assumed to live in a neighborhood of the origin, and $\tau_{j,A}$ are the deformation parameters that accompany the deformation by chiral operators in the Lagrangian. The correlation functions are taken by differentiating \eqref{eq:genfun4dExtCorr} with respect to various $\tau, \bar{\tau}$ and eventually setting $\hat{\lambda}_{j,A}=0$ for all $j,A$. The normalisation of the couplings is chosen to match with \cite[Sec.3.2]{Gerchkovitz:2016gxx}.\par
We can evaluate \eqref{eq:genfun4dExtCorr} in the long quiver limit. To do so, we do not impose any scaling on the additional coupling, so that they do not spoil the planar limit, and evaluate the expectation value in the background of the long quiver solution.
\begin{stm}
	Let
	\begin{equation*}
		f_{\text{\rm extremal}} \left[ \left\{ \hat{\lambda}_{A} (z) \right\} \right] := \lim \frac{1}{PN} \ln  \frac{\mz_{\cs^4} \left[ \left\{ \hat{\lambda}_{j,A} \right\} \right] }{\mz_{\cs^4}} ,
	\end{equation*}
	where the limit operation on right-hand side means the long quiver limit. Then
	\begin{equation}
	\label{eq:funcdiffFextremal}
		\frac{\delta f_{\text{\rm extremal}} }{\delta \hat{\lambda}_{A} (z_{\ast})} \simeq \begin{cases} 0 & A \text{ odd} \\ \pi^{(A-1)/2} 2^{A}   \frac{ \Gamma \left( \frac{A+1}{2} \right)}{ \Gamma \left( \frac{A}{2} +2\right)} t_4 (z_{\ast} )^{A/2} \nu (z_{\ast})^{1+A/2} & A \text{ even}. \end{cases}
	\end{equation}
	In particular, the generators associated to odd $A$ give a sub-leading contribution to the extremal correlators.
\end{stm}
\begin{proof}
	We use the fact that, at large-$N$, the generators of the chiral ring are not subject to the trace relations. In the long quiver limit we write 
	\begin{align*}
		\sum_{j=1}^{P-1}\sum_{A=3}^{N_j} \pi^{A/2} \hat{\lambda}_{j,A} \sum_{a=1}^{N_j} \left( \phi_a ^{(j)} \right)^{A} & \simeq P N \sum_{A \ge 3} \pi^{A/2}  \int_0 ^{1} \dd z \hat{\lambda}_{A} (z) \int \dd \phi \rho (z,\phi) \phi^{A} \\
			& = P N \sum_{A \ge 3} \pi^{A/2}  \int_0 ^{1} \dd z \frac{\hat{\lambda}_{A} (z) \nu (z)}{2 \pi \lambda (z)} \int_{-2 \sqrt{\lambda (z)}}^{2 \sqrt{\lambda (z)}} \dd \phi \sqrt{4 \lambda (z) - \phi^2} \phi^{A} \\
			&=P N \sum_{A \ge 3}\left( \frac{ 1+ (-1)^{A}}{2} \right) \mathfrak{c}_A \int_0 ^{1} \dd z \hat{\lambda}_{A} (z)  t_4 (z)^{A/2} \nu (z)^{1+A/2} ,
	\end{align*}
	where we recall that $\lambda (z)=\nu (z) t_4 (z)$. The $A$-dependent coefficient in the last line is
	\begin{equation}
	\label{eq:defcoefExtCor}
		 \mathfrak{c}_A = \frac{  \pi^{(A-1)/2} 2^{A}  \Gamma \left( \frac{A+1}{2} \right)}{ \Gamma \left( \frac{A}{2} +2\right)} .
	\end{equation}
	Due to the parity of the eigenvalue density, the integral vanishes if $A \in 1+ 2 \Z$, while if $A \in 2 \Z$ the coefficient is in $\pi^{A/2} \N$, that is, it evaluates to an integer times an integer power of $\pi$. The first few non-trivial values of the coefficient $\mathfrak{c}_A$ are listed in Table \ref{tab:extremalCA}.\par
	
	\begin{table}[thb]
	\centering
	\begin{tabular}{|c | c c c c c |}
	\hline
	$A$ & 4 & 6 & 8 & 10 & 12 \\
	\hline
	$ \mathfrak{c}_A  $ & $2 \pi^2$ & $5\pi^3$ & $14 \pi^4$ & $42 \pi^5 $ & $132 \pi^6$ \\
	\hline
	\end{tabular}
	\caption{Explicit values of the numerical coefficient $\mathfrak{c}_A$ in \eqref{eq:defcoefExtCor}.}
	\label{tab:extremalCA}
	\end{table}\par
	
	We then use the large-$N$ factorisation property of multi-trace correlators in matrix models, which implies the standard relation 
	\begin{equation*}
		\ln \left\langle \exp \left( \sum_{j=1}^{P-1}\sum_{A=3}^{N_j} \pi^{A/2} \hat{\lambda}_{j,A} \sum_{a=1}^{N_j} \left( \phi_a ^{(j)} \right)^{A} \right)\right\rangle_{\cs^4} =  \left. \left( \sum_{j=1}^{P-1}\sum_{A=3}^{N_j} \pi^{A/2} \hat{\lambda}_{j,A} \sum_{a=1}^{N_j} \left( \phi_a ^{(j)} \right)^{A} \right) \right\rvert_{\text{saddle}} 
	\end{equation*}
	to evaluate 
	\begin{equation*}
		f_{\text{\rm extremal}} \left[ \left\{ \hat{\lambda}_{A} (z) \right\} \right] \simeq \sum_{\substack{ A \ge 3 \\ A \in 2\N }} \frac{  \pi^{(A-1)/2} 2^{A}  \Gamma \left( \frac{A+1}{2} \right)}{ \Gamma \left( \frac{A}{2} +2\right)}\int_0 ^{1} \dd z \hat{\lambda}_{A} (z)  t_4 (z)^{A/2} \nu (z)^{1+A/2}  .
	\end{equation*}
	The conclusion follows from functional differentiation of this expression.
\end{proof}
Construct the matrix of derivatives:
\begin{equation*}
	\mec_{\vec{A},\vec{B}} := \frac{1}{\mz_{\cs^4}} \frac{\partial \ }{\partial \tau_{j_1,A_1}} \cdots  \frac{\partial \ }{\partial \tau_{j_m,A_m}} \frac{\partial \ }{\partial \bar{\tau}_{k_1,B_1}} \cdots  \frac{\partial \ }{\partial \bar{\tau}_{k_m,B_m}} \mz_{\cs^4} \left[ \left\{ \hat{\lambda}_{j,A} \right\} \right] , 
\end{equation*}
where $\vec{A},\vec{B}$ are collections of pairs $\vec{A}=\left\{ (j_1,A_1), \dots, (j_m, A_m) \right\} , \vec{B}=\left\{ (k_1,B_1), \dots, (k_m, B_m) \right\}$ for some integer $m \in \N$. We can introduce a lexicographic order on the set of arrays $\vec{A}$, and use the map from this ordered set to $\N$ to treat each $\vec{A}$ as an integer index. The extremal correlators are constructed algorithmically from the Gram--Schmidt diagonalisation of $\mec$ \cite{Gerchkovitz:2016gxx}.\par
Formula \eqref{eq:funcdiffFextremal} allows to easily evaluate $\mec_{\vec{A},\vec{B}}$. Using the leading order approximation 
\begin{equation*}
	 \mz_{\cs^4} \left[ \left\{ \hat{\lambda}_{j,A} \right\} \right] \simeq \mz_{\cs^4} \ee^{NP f_{\text{\rm extremal}} \left[ \left\{ \hat{\lambda}_{A} (z) \right\} \right] }, 
\end{equation*}
we obtain 
\begin{equation*}
	\mec_{\vec{A},\vec{B}} \simeq \prod_{s=1}^{m} \left[ N_{j_s} \mathfrak{c}_{A_s} t_{j_s}^{A_s/2}  \right] \times \left[ N_{k_s} \mathfrak{c}_{B_s} t_{k_s}^{B_s/2}  \right] 
\end{equation*}
if all $A_s,B_s$ are even, and is sub-leading otherwise. Note that there is a factor $\ii^{m}$ in passing from $\frac{\partial \ }{\partial \tau_{j,A} }$ to $\frac{\partial \ }{\partial \hat{\lambda}_{j,A} }$, and an analogous factor $(-\ii)^m$ from the derivatives $\frac{\partial \ }{\partial \bar{\tau}_{j,A} }$, and the two cancel out.\par
We remark that the simplicity of this expression is a consequence of the planar limit, and the lack of 't Hooft scaling of the auxiliary couplings $\left\{ \hat{\lambda}_A \right\}_{A \ge 3} $. The extremal correlators in the planar limit of $\mN=4$ super-Yang--Mills with the additional 't Hooft scaling of $\left\{ \hat{\lambda}_A \right\}_{A \ge 3} $ have been studied in \cite{Rodriguez-Gomez:2016ijh}, and the results therein may be straightforwardly exported to the present context.

\subsection{Wilson loops}
Consider a Wilson loop 
\begin{equation*}
	W_{\mathsf{R}} = \frac{1}{\dim \mathsf{R}} \mathrm{Tr}_{\mathsf{R}} \left( \ee^{2 \pi \vec{\phi}} \right)
\end{equation*}
in the irreducible representation $\mathsf{R}$ of the gauge group. Fixing $j_{\ast} \in \left\{1, \dots, P-1 \right\}$, $\mathsf{R}$ is the tensor products of an arbitrary $\mathrm{SU}(N_{j_{\ast}})$ representation and the trivial representation of $\mathrm{SU}(N_j)$ for $j \ne j_{\ast}$. We will focus on fundamental $\Box$, rank-$\kk$ antisymmetric $\mathsf{A}_{\kk}$ and rank-$\kk$ symmetric $\mathsf{S}_{\kk}$ representations of $\mathrm{SU}(N_{j_{\ast}})$. We denote $z_{\ast} :=\frac{j_{\ast}}{P}$.\par
The expectation value $\langle W_{\mathsf{R}} \rangle$ in the planar limit is obtained as an immediate corollary of the previous result. Indeed, we have found that, at every $z$, the normalised eigenvalue density $\bar{\rho}(z,\phi)$ equals the one of $\mN=4$ super-Yang--Mills with 't Hooft coupling $\lambda (z)$. Therefore, in the planar limit, the result for $\langle W_{\mathsf{R}} \rangle$ is exported from $\mN=4$ super-Yang--Mills.\par
For the fundamental representation we have the classical result 
\begin{align*}
	\langle W_{\Box_{j_{\ast}}} \rangle &\simeq 
	\int \dd \phi \bar{\rho} (z_{\ast},\phi) ~\ee^{2\pi \phi} 
	\\
	&= 
	\frac{1}{2\pi \sqrt{ \lambda (z_{\ast})}}\,I_1\left(4\pi \sqrt{\lambda (z_{\ast})}\right)
	= \sum_{n=0} ^{\infty}  
	\frac{\left( 4 \pi^2 \lambda (z_{\ast}) \right)^{n}}{(n+1)! n!} ,
\end{align*}
where in the second line $I_{1} (\cdot )$ is the modified Bessel function. This expression only differs from $\mN=4$ super-Yang--Mills in the dependence $\lambda (z_{\ast}) = t_{4} (z_{\ast}) \nu (z_{\ast})$, which varies as we vary the gauge node under which the line operator is charged.\par
The computation for the $\mathsf{A}_{\kk}$ and $\mathsf{S}_{\kk}$ Wilson loops also follows from $\mN=4$, and can be exported step by step from \cite{Hartnoll:2006is}.

\subsection{Solution at strong coupling}
\label{sec:4dStrong}
We wish to go beyond the computation of the coefficient $a$ and extract more interesting behaviour from the sphere free energy. To do so, we consider the strong 't Hooft coupling limit $t_4 (z) \to \infty$. This is the regime in which we expect a more precise match with supergravity, beyond the purely kinematical anomaly coefficient $a$.\par
In the strong coupling limit, \eqref{eq:spe4d} trivialises. This indicates that the ansatz that $\phi$ does not scale with $P$ misses all the information at strong 't Hooft coupling. We therefore abandon the no-scaling ansatz, and plug the long quiver scaling ansatz \eqref{eq:LQansatz} into \eqref{eq:spefull4d}.\par
The saddle point equation at strong coupling reads 
\begin{equation}
\label{eq:spestrong4d}
	\dashint \dd \sigma \frac{\rho (z, \sigma)}{\phi - \sigma} + \frac{1}{2P^2} \int \dd \sigma \mathscr{K} (\phi - \sigma) \partial_z^2\rho (z, \sigma) + \frac{1}{2P^2} [P^2 \zeta (z)] \mathscr{K} (\phi) =0.
\end{equation}
For large argument, 
\begin{equation*}
	\mathscr{K}(\phi) \approx 2 \phi \ln \lvert \phi \rvert - 2 \phi + \frac{1}{6 \phi} ,
\end{equation*}
and therefore, replacing $\phi$ with $2P^{\chi}x$ as prescribed by \eqref{eq:LQansatz}, we have 
\begin{equation}
\label{eq:Klargephi}
	\frac{1}{2P^2} \mathscr{K}(\phi) \approx  2\chi P^{\chi-2} \ln (P) x  +  2P^{\chi-2} \left(  x \ln \lvert 2x \rvert - x \right)+ \frac{1}{24 x}P^{-\chi-2} .
\end{equation}
With the substitutions \eqref{eq:LQansatz}-\eqref{eq:Klargephi}, eq.~\eqref{eq:spestrong4d} becomes 
\begin{equation*}
\begin{aligned}
	0=2\chi P^{\chi-2} \ln (P) & \left[ \int \dd y \partial_z^2 \varrho(z,y) (x-y) + [P^2 \zeta (z)] x\right] \\
	+  2P^{\chi-2} & \left[  \int \dd y \partial_z^2 \varrho(z,y)  (x-y) \left( \ln \lvert 2x -2y\rvert -1 \right) + [P^2 \zeta (z)] x  \left( \ln \lvert 2x\rvert -1 \right)\right] \\
	+ \frac{1}{2} P^{-\chi} & \dashint \dd y \frac{\varrho (z, y)}{x-y} + \mathcal{O}(P^{-\chi-2}) .
\end{aligned}
\end{equation*}
The first line vanishes as a consequence of the balancing condition \eqref{eq:dnudziszeta}. The shifts by $\ln(2)-1$ in the second line cancel for the same reason. To have competition between the terms in the second and third line, so to reach a non-trivial saddle point configuration, the matrix model requires $\chi-2=-\chi$. We thus recover in $d=4$ the value $\chi=1$, that we have found in Section \ref{sec:d} $\forall ~d \in \R_{\ge 3} \setminus 2\Z $.\par
We arrive at
\begin{equation*}
	\frac{1}{4} \dashint \dd y \frac{\varrho (z, y)}{x-y} + \int \dd y (x-y) \ln \lvert x -y\rvert \partial_z^2 \varrho(z,y)   + [P^2 \zeta (z)] x   \ln \lvert x\rvert =0.
\end{equation*}
We now use the trick of writing 
\begin{equation*}
	\frac{1}{x-y} = \partial_y^2 \left( (x-y) \ln \lvert x -y\rvert \right)
\end{equation*}
and integrate twice by parts,
\begin{equation*}
	\dashint \dd y \frac{\varrho (z, y)}{x-y} = \dashint \dd y  (x-y) \ln \lvert x -y\rvert \partial_y^2 \varrho (z, y) .
\end{equation*}
Besides, we use the trivial equalities 
\begin{equation*}
\begin{aligned}
	 x   \ln \lvert x\rvert & = \dashint \dd y (x-y) \ln \lvert x -y\rvert  \delta (y), \\
	 \int \dd y (x-y) \ln \lvert x -y\rvert f(y) &= \dashint \dd y (x-y) \ln \lvert x -y\rvert f(y),
\end{aligned}
\end{equation*}
with the second identity valid for every smooth function $f: \R \to \R$. The saddle point equation is thus recast in the form 
\begin{equation}
\label{eq:integralofPoisson}
	\dashint \dd y (x-y) \ln \lvert x -y\rvert ~ \left[ \frac{1}{4}  \partial_y^2 \varrho (z, y) + \partial_z^2  \varrho (z, y)  + [P^2 \zeta (z)]  \delta (y)\right] =0,
\end{equation}
that must hold $\forall~x\in \R$ and $0<z<1$. Therefore, the solution to \eqref{eq:integralofPoisson} satisfies the Poisson equation 
\begin{center}\noindent\fbox{\parbox{0.98\linewidth}{%
\begin{equation*}
	\frac{1}{4}  \partial_x^2 \varrho (z, x) + \partial_z^2  \varrho (z, x)  + P^2 \zeta (z) \delta (x) =0.
\end{equation*}}}\end{center}
This exact same equation was already found in \eqref{eq:Poisson} for $d \ne 4$, and we have derived it directly in $d=4$ at strong coupling.\par
\begin{digr}
	The effect for which, at large-$N$, the eigenvalues scale differently at finite versus infinite 't Hooft coupling already appears in $\mN=4$ super-Yang--Mills.\par
	The partition function of $\mN=4$ super-Yang--Mills on $\cs^4$ is the Gaussian Unitary Ensemble. In the planar limit with 't Hooft coupling $\lambda=Ng_4$, one obtains the saddle point equation \eqref{eq:GUESPE}, solved by the Wigner semicircle. The finite part of the partition function is 
	\begin{equation}
	\label{eq:FN4SYM}
		F_{\mN=4} = N^2 \ln \frac{1}{\lambda}.
	\end{equation}
	However, one may also take the limit with fixed gauge coupling $g_4$. This corresponds to $\lambda\to \infty$ linearly with $N\to \infty$. The only way for the action to admit an equilibrium configuration is if the eigenvalues grow as $\phi=\sqrt{N}x$ with $x$ being $\mathcal{O} (1)$. One then proceeds to the computation with fixed $g_4$ and find the same answer as directly replacing $\lambda = Ng_4$ in \eqref{eq:FN4SYM}.
\end{digr}

\subsection{Free energy in four-dimensional long quivers}
\label{sec:4dlogZstrong}

\begin{stm}
	In the long quiver limit and at infinite 't Hooft coupling, 
\begin{equation}\label{free-quiver}
-\ln \mz_{\cs^4} \simeq \frac{1}{2} P 
\sum_{k=1}^{\infty} R_k^2 \,\ln\frac{\pi k \ee^{\gamma _{\text{\tiny EM}}} }{P} +  \mathcal{O} (\ln P) .
\end{equation}
\end{stm}
\begin{proof}
We compute 
\begin{equation*}
	-\ln \mz_{\cs^4} \simeq \left. S_{\mathrm{eff}} \right\rvert_{\text{on-shell}} .
\end{equation*}
The on-shell effective action is given by 
\begin{equation*}
\begin{aligned}
	\left. S_{\mathrm{eff}} \right\rvert_{\text{on-shell}} = \frac{1}{2}P N^2 \int_0 ^1 \dd z \int \dd x \varrho (z, x) & \left[  \zeta (z) \ln H(2Px)  - \dashint\dd y \varrho (z, y) \ln \lvert 2P(x-y) \rvert \right. \\
		& \left. -\frac{1}{8P^2}  \dashint \dd y \varrho (z,y) \partial_y ^2 \left( \ln \lvert H (2P(x-y)) \rvert  \right)  \right] ,
\end{aligned}
\end{equation*}
where we have used the fact that $\varrho (z,y)$ solves the saddle point equation to manipulate the expression. 
At large values of the argument,
\begin{equation*}
 \ln H(\phi )\simeq -\frac{\phi ^2}{2}\left(\ln \phi ^2-3\right),
\end{equation*}
and $\partial ^2_\phi \ln H(\phi )\simeq -\ln\phi ^2$. We thus find cancellations between the double integrals in $S_{\mathrm{eff}}$. 
The remaining single integral can be computed with the help of the Fourier series \eqref{eq:Fourierzeta} and \eqref{eq:solrho}:
\begin{equation*}
 \left. S_{\mathrm{eff}} \right\rvert_{\text{on-shell}} = -\frac{\pi ^3P}{2}\sum_{k=1}^{\infty }k^3R_k^2\int_{\R} \dd x\,x^2\left[ \ln\left(2Px\right)^2-3\right]\ee^{-2\pi k|x|}.
\end{equation*}
An integration over $x$ gives \eqref{free-quiver}.\par
Before concluding, let us stress that, using $a \propto P \sum_{k=1}^{\infty}R_k^2$, the $\mathcal{O}(P)$ constant in $\ln \mz_{\cs^4}$ is scheme-dependent.
Recall from \eqref{eq:GGKZd} that the partition function on $\cs^4$ contains a UV divergence controlled by the central charge: $\ln \mz_{\cs^4}=-4a\ln ( \Lambda r)  +{\rm finite}$, for $r$ the radius of $\cs^4$ and $\Lambda$ the UV cutoff. Changing renormalisation scheme shifts the free energy by $\,{\rm const}\, P \sum_{k=1}^{\infty}R_k^2$, effectively multiplying the the argument of the logarithm in \eqref{free-quiver} by a constant. The constant we computed ($\pi \ee^{\gamma _{\text{\tiny EM}}}$) corresponds to a particular scheme used in \cite{Pestun:2007rz} and in practice has no particular meaning. On the other hand, the dependence on $P$ and on the mode number $k$, does not change with the scheme and can be faithfully compared to supergravity, which we will do later on in Subsection \ref{sec:toysugra}.

\end{proof}

\subsection{Four-dimensional linear quivers of arbitrary length}
\label{sec:4dshortQ}

It is possible to solve the planar limit of any linear $d=4$ $\mN=2$ quiver for arbitrary $P$. We omit the details of the derivation of the saddle point equation, since they are conceptually analogous to what we have done so far, and only highlight the differences.
\begin{itemize}
	\item The discrete index $j$ labelling the nodes of the quiver is not replaced by a parameter $z=j/P$.
	\item The gauge ranks $N_j$ scaled into rational numbers $\left\{ \nu_j\right\}_j,$ $\nu _j=N_j/N$ are not collected into a function $\nu (z)$.
	\item Likewise, the Veneziano parameters $\left\{\zeta_j \right\}$ are not collected into a function $\zeta (z)$. In the planar limit, $\zeta_j $ are $\mathcal{O}(1)$.
	\item The eigenvalue densities are normalised as 
		\begin{equation*}
			\int \dd \phi \rho_j (\phi) = \nu_j , \qquad \forall j=1, \dots, P-1 .
		\end{equation*}
	\item Related to the above, being $P$ finite, the contribution from the bifundamental hypermultiplets is a difference 
		\begin{equation*}
			\frac{1}{2} \left[ \rho_{j+1} -2 \rho_j + \rho_{j-1} \right] ,
		\end{equation*}
		and does not produce a differential $\partial_z^2$.
	\item To encode the above fact, we define the $(P-1)\times (P-1)$ matrix 
		\begin{equation*}
			\mathsf{M} := \left( \begin{matrix} 2  & -1 &  &  &  & & \\ 
			 -1 & 2 & -1 & & & &  \\
			  & -1 & 2 & -1 &  & & & \\ 
 			  & & & & \ddots &  & & \\ 
			  & & & & -1 & 2 & -1 \\ 
			&  & & & & -1 & 2 \\  \end{matrix} \right)
		\end{equation*}
		That is, $\mathsf{M}$ is the Cartan matrix of the $A_{P-1}$ Lie algebra, which equals twice the unit matrix minus the adjacency matrix of the linear quiver.
	\item The balancing condition is written in matrix form as $K_j = \mathsf{M}_{ji} N_i $, or more compactly 
		\begin{equation*}
			\vec{K} = \mathsf{M} \vec{N} .
		\end{equation*}
	\item As a notational reminder, the indices $i,j \in \left\{ 1, \dots, P-1 \right\}$ will indicate the nodes of the quiver. Moreover, sum over repeated indices will be understood to lighten the formulae. The imaginary unit is denoted $\ii =\sqrt{-1}$ to avoid confusion with the index $i$.
\end{itemize}
To diagonalise the Cartan matrix we introduce its eigenvectors $\vec{v}^{(p)}= (v_1 ^{(p)}, \cdots, v_{P-1} ^{(p)})$, $p=1, \dots, P-1$. To be more explicit, the components are 
\begin{equation}\label{eigenvectors}
 v_j^{(p)}=\sin\frac{\pi p j}{P} , \qquad j=1 ,\dots, P-1,
\end{equation}
and, after elementary trigonometric manipulations, the corresponding eigenvalues are
\begin{equation*}
	\mathsf{M}_{ji}v_i^{(p)}=\left(2\sin\frac{\pi p }{2P}\right)^2v_j^{(p)} 
\end{equation*}
or, in matrix notation, 
\begin{equation*}
	\mathsf{M} \vec{v}^{(p)} = \left(2\sin\frac{\pi p }{2P}\right)^2 \vec{v}^{(p)} , \qquad \forall p=1, \dots, P-1.
\end{equation*}
These vectors form a basis that satisfies orthogonality and completeness conditions
\begin{equation}\label{completeness}
	\sum_{j=1}^{P-1}v_j^{(p)}v_j^{(p^{\prime})}=\frac{P}{2}\,\delta ^{p p^{\prime}}, \qquad  \sum_{p=1}^{P-1}v_j^{(p)}v_i^{(p)}=\frac{P}{2}\,\delta _{ji}.
\end{equation}
We can express the vector of gauge ranks $\vec{N}$ in this basis:
\begin{equation*}
	\vec{N} = \sum_{p=1}^{P-1} \tilde{N}_p \vec{v}^{(p)} ,
\end{equation*}
which in components reads 
\begin{equation*}
 N_j=\sum_{p=1}^{P-1} \tilde{N}_p \sin\frac{\pi p j}{P}\,.
\end{equation*}
That is to say, $\vec{N}$ has components $N_i$ in the canonical basis $\left\{ \vec{e}^{(i)} \right\}_{i=1,\dots,P-1}$ with $e^{(i)}_j = \delta_{ji}$, and it has components $\tilde{N}_p$ in the eigenbasis of $\mathsf{M}$ $\left\{ \vec{v}^{(p)} \right\}_{p=1,\dots,P-1}$. The balancing condition \eqref{eq:balance} results in
\begin{equation}\label{K-to-R}
 K_j=\sum_{p=1}^{P-1} \tilde{N}_p \left(2\sin\frac{\pi p }{2P}\right)^2\sin\frac{\pi p j}{P} .
\end{equation}

\begin{stm}
	Consider a balanced linear quiver in $d=4$, and take the planar limit with arbitrary $P \in \N_{\ge 2}$. At infinite 't Hooft coupling, the densities of eigenvalues are 
	\begin{equation}
	\label{eq:solrho4dsmallP}
		\rho_j (\phi) = \frac{1}{\sinh \pi \phi }\sum_{p=1}^{P-1} \frac{ \sinh \left( \frac{\pi \phi (P-p)}{P} \right) \sin \left(\frac{\pi p j}{P} \right)  }{\sin\left(\frac{\pi p}{P}\right)} ~ \frac{1}{P} \sum_{\ell=1}^{P-1} \sin \left(\frac{\pi p \ell }{P} \right) \zeta_{\ell}    ,
	\end{equation}
	$\forall ~j=1, \dots, P-1$.
\end{stm}
A corollary and cross-check of this statement is that for $P=2$, namely $\mathrm{SU}(N)$ gauge theory and $2N$ fundamental hypermultiplets, this expression reproduces the strong coupling solution of conformal $\mN=2$ QCD \cite[Eq.(4.4)]{Passerini:2011fe}:
	\begin{equation*}
 		\rho _{{\rm SQCD}}(\phi)=\frac{1}{2\cosh\frac{\pi \phi}{2}} .
	\end{equation*}
\begin{proof}
The finite-$P$ version of the saddle point equation \eqref{eq:spefull4d} reads 
\begin{equation}
\label{eq:spe4dfiniteP}
	\dashint \dd \sigma \frac{\rho_j (\sigma)}{\phi - \sigma} - \frac{1}{2} \mathsf{M}_{ji}\int \dd \sigma \mathscr{K} (\phi - \sigma)\rho_i(\sigma)  + \frac{\zeta_j}{2} \mathscr{K} (\phi) =0,
\end{equation}
where again we restrict ourselves to the infinite 't Hooft coupling limit.\par
We take the Fourier transform of \eqref{eq:spe4dfiniteP} and denote $\tilde{\rho}_j (\omega)$ the Fourier transform of $\rho_j (\phi)$. We then simplify the expression using that $(2+\gamma_{\text{\tiny EM}}) (\mathsf{M}_{ji}\nu_i - \zeta_j ) =0 $ by virtue of the balancing condition. After these manipulations we obtain 
\begin{equation*}
	\int \dd \omega  \ee^{-\ii \omega (\phi - \sigma)} \mathrm{sign}(\omega)  \left[ \tilde{\rho}_j (\omega) + \frac{1}{4}  \mathsf{M}_{ji}  \tilde{\rho}_i (\omega) \frac{1}{\left(\sinh \frac{\omega}{2}\right)^2}- \frac{\zeta_j}{4} \frac{1}{\left(\sinh \frac{\omega}{2}\right)^2}\right] =0 ,
\end{equation*}
which is valid if and only if 
\begin{equation}
\label{eq:spe4dFourier}
	\left[ \left(2 \sinh \frac{\omega}{2}\right)^2 \delta_{ji} + \mathsf{M}_{ji} \right] \tilde{\rho}_i (\omega) = \zeta_j ,
\end{equation}
$\forall ~j=1, \dots, P-1 $, where now $P\in \N_{\ge 2}$ is arbitrary and finite.\par
We now expand the vector of densities $(\tilde{\rho}_1 (\omega), \dots, \tilde{\rho}_{P-1} (\omega))$ in the basis \eqref{eigenvectors},
\begin{equation*}
	\tilde{\rho}_j (\omega) = \sum_{p=1}^{P-1} \tilde{r}_p (\omega) v_j ^{(p)} = \sum_{p=1}^{P-1} \tilde{r}_p (\omega) \sin \frac{\pi p j}{P} .
\end{equation*}
Plugging this formula in \eqref{eq:spe4dFourier}, we obtain: 
\begin{equation*}
	\tilde{r}_p(\omega )=\frac{1}{2P}\,\,\frac{1}{\left(\sinh\frac{\omega }{2}\right)^2+\left(\sin\frac{\pi p}{2P}\right)^2}\,\sum_{\ell=1}^{P-1} \zeta_{\ell}\sin\frac{\pi p \ell}{P}\,.
\end{equation*}
Therefore, the Fourier transform of the eigenvalue densities at large-$N$ but finite length of the quiver is 
\begin{equation*}
	\tilde{\rho}_j (\omega) = \sum_{p=1}^{P-1}\left( \frac{1}{2P}\sum_{\ell=1}^{P-1} \frac{\zeta_{\ell}  \sin \frac{\pi p \ell}{P} }{ \left(\sinh \frac{\omega}{2}\right)^2 + \left(\sin\frac{\pi p }{2P}\right)^2} \right) \sin \frac{\pi p j}{P}\, .
\end{equation*}
After the inverse Fourier transform, we find the eigenvalue densities \eqref{eq:solrho4dsmallP}. Using \eqref{K-to-R} we can also write the solution as
\begin{equation}\label{rho-R}
 	\tilde{\rho }_j(\omega )=\sum_{p=1}^{P-1}\frac{\left(\sin\frac{\pi k }{2P}\right)^2}{\left(\sinh \frac{\omega}{2}\right)^2 + \left(\sin\frac{\pi p }{2P}\right)^2} \frac{\tilde{N}_p}{N}\sin \frac{\pi p j}{P} ,
\end{equation}
with $\left\{ \frac{\tilde{N}_p}{N} \right\}_{p=1,\dots, P-1}$ being just a rotation of $\left\{ \nu_j \right\}_{j=1, \dots, P-1}$.
\end{proof}\par
Equipped with the strong coupling solution \eqref{eq:solrho4dsmallP}, we can now evaluate the free energy.
\begin{stm}
	Consider a balanced linear quiver in $d=4$, and take the planar limit with arbitrary $P \in \N_{\ge 2}$. At infinite 't~Hooft coupling we have: 
\begin{center}\noindent\fbox{\parbox{0.98\linewidth}{%
\begin{equation}\label{logZ-fourier}
  -\ln \mz_{\cs^4} =\frac{P}{2}\sum_{p=1}^{P-1}\tilde{N}_p ^2\left[ 2\tan\frac{\pi p}{2P}\, \sum_{a=1}^{2P-1}\sin\frac{\pi p a}{P}\, \ln\Gamma \left(\frac{a}{2P}\right) -\ln(2P) \right] .
\end{equation}
}}
\end{center}
\end{stm}
\begin{proof}
We compute 
	\begin{equation*}
		-\ln \mz_{\cs^4} \simeq \left. S_{\mathrm{eff}} \right\rvert_{\text{on-shell}} ,
	\end{equation*}
	which in the strict infinite coupling limit becomes (sum over repeated indices understood)
	\begin{align*}
		 -\frac{1}{N^2} \ln \mz_{\cs^4} = -\frac{1}{2} & \int \dd \phi \int \dd \sigma \,\rho _j(\phi )\rho _i(\sigma )\left[ \mathsf{M}_{ji}\ln H(\phi -\sigma )+\delta _{ji}\ln(\phi -\sigma )^2  \right] \\
			+ & \int \dd \phi \rho _j(\phi ) \zeta_j \ln H(\phi ).
\end{align*}
Mimicking our derivation of the eigenvalue density, we take the Fourier transform of the integrals:
	\begin{align}
 		-\frac{1}{N^2} \ln \mz_{\cs^4} &=\int \frac{\dd\omega }{|\omega |\left(2\sinh\frac{\omega }{2}\right)^2} \left\{ \frac{1}{2}\,\tilde{\rho }_j(\omega )\tilde{\rho }_i(\omega ) \left[\left(2\sinh\frac{\omega }{2}\right)^2\delta _{ji}+\mathsf{M}_{ji}\right]-\zeta_j \tilde{\rho }_j(\omega ) \right\} \notag \\
			&\simeq -\int\frac{\dd\omega }{2|\omega |\left(2\sinh\frac{\omega }{2}\right)^2} \zeta_j\tilde{\rho }_j(\omega ),  \label{F-Fourier}
	\end{align}
where in the last step we have used the saddle point equations in the form \eqref{eq:spe4dFourier}.
This expression is ill-defined because of a bad divergence at $\omega =0$. The divergence arises because $\ln H(\phi )$ and $\ln \phi ^2$ are growing too fast and their Fourier images, strictly speaking, do not exist. Subtractions are needed to define them properly.\par
More accurate definition of the Fourier transform involves regularisation:\footnote{We mention in passing that regularisation was not necessary in the saddle point equations. The latter contain once-differentiated kernels, that brings in an extra  $\omega $ in the Fourier space softening the $1/\omega ^3$ behavior of the integrand. The remaining $1/\omega ^2$ divergence is cancelled by the balancing condition. Because of these cancellations we could use the Fourier transforms without subtractions in deriving the saddle point eigenvalue densities.}
\begin{align*}
 \ln\phi ^2&= \lim_{\alpha \to 0}\left( -\int \dd\omega \,|\omega |^{\alpha -1} \ee^{-\ii\omega \phi }+\frac{2}{\alpha }-2\gamma_{\text{\tiny EM}}\right) \\
	\ln H(\phi )&=-\int\frac{\dd \omega }{|\omega |} \frac{ \ee^{-\ii\omega \phi }-1+\frac{\omega ^2\phi ^2}{2}}{\left(2\sinh\frac{\omega }{2}\right)^2}\,.
\end{align*}
Using these formulae we get a regularised version of \eqref{F-Fourier}:
\begin{align*}
 -\frac{1}{N^2} \ln \mz_{\cs^4} = \lim_{\alpha \to 0} & \left\{  -\int\frac{\dd \omega }{2|\omega |^{1-\alpha }\left(2\sinh\frac{\omega }{2}\right)^2} \left[ \zeta_j\left(\tilde{\rho }_j(\omega )- \nu_j+\frac{\omega ^2}{2}\,\left\langle \phi ^2\right\rangle_j\right) \right. \right. \\
	& \ \left. \left. -\frac{\omega ^2}{2} \left(2\mathsf{M}_{ji}\nu_i-\zeta_j\right)\left\langle \phi ^2\right\rangle_j \right] +\left(\gamma_{\text{\tiny EM}}-\frac{1}{\alpha }\right)\nu_j \nu_j  \right\} ,
\end{align*}
with the obvious notation for the second moment of the density:
\begin{equation*}
 \left\langle \phi ^2\right\rangle_j :=\int_{}^{}\dd\phi \,\rho _j(\phi )\phi ^2=-\tilde{\rho }''_j(0).
\end{equation*}
The latter can be actually found from the saddle point equation \eqref{eq:spe4dFourier}. When expanded to the second order in $\omega $, it gives: 
\begin{equation*}
	\mathsf{M}_{ji}\left\langle \phi ^2\right\rangle_i=2 \nu_j.
\end{equation*}
This relation and the balancing condition simplify the second line: 
\begin{align*}
 -\frac{1}{N^2}\ln \mz_{\cs^4} &= -\int \frac{\dd\omega }{2|\omega |\left(2\sinh\frac{\omega }{2}\right)^2} \zeta_j\left(\tilde{\rho }_j(\omega )-\tilde{\rho }_j(0)-\frac{\omega ^2}{2}\,\tilde{\rho }''_j(0)\right) \\
	& +\lim_{\alpha \to 0} \left( \int_{}^{}\frac{\dd\omega |\omega |^{1+\alpha }}{2\left(2\sinh\frac{\omega }{2}\right)^2}+\gamma_{\text{\tiny EM}}-\frac{1}{\alpha }\right)\nu_j \nu_j.
\end{align*}
This expression has a finite $\alpha \rightarrow 0$ limit:
\begin{equation}\label{F-seminfinal}
 -\frac{1}{N^2}\ln \mz_{\cs^4} = -\int \frac{\dd\omega }{2|\omega |\left(2\sinh\frac{\omega }{2}\right)^2} \zeta_j\left(\tilde{\rho }_j(\omega )-\tilde{\rho }_j(0)-\frac{\omega ^2}{2}\,\tilde{\rho }''_j(0)\right) +(1+\gamma _{\text{\tiny EM}})\nu_j \nu_j.
\end{equation}
This is a well-defined, properly regularised version of \eqref{F-Fourier} with the integral in the first term manifestly convergent.\par
It remains to substitute the solution for the density \eqref{rho-R}. That gives rise to an integral of the form
\begin{equation*}
 \mathcal{I}(m):=\int_{0}^{\infty }\frac{\dd\omega }{\omega \left(\sinh\frac{\omega }{2}\right)^2}\, \left[\frac{1}{\left(\sinh\frac{\omega }{2}\right)^2+m^2}-\frac{1}{m^2}+\frac{\omega ^2}{4m^4} \right].
\end{equation*}
This integral can be calculated in closed form. After lengthy but straightforward manipulations we find:
\begin{equation}\label{I(sin)}
	\mathcal{I}\left(\sin\frac{\pi p}{2P}\right)=\frac{1+\gamma_{\text{\tiny EM}}+\ln(2P)}{\left(\sin\frac{\pi p}{2P}\right)^4}-\frac{4}{\left(\sin\frac{\pi p}{2P}\right)^2} \sum_{a=1}^{2P-1}\frac{\sin\frac{\pi p a}{P}}{\sin\frac{\pi p}{P}}\, \ln\Gamma \left(\frac{a}{2P}\right).
\end{equation}
Substitution of \eqref{rho-R} into \eqref{F-seminfinal} and the use of the basis \eqref{eigenvectors} result in
\begin{equation*} 
	-\ln \mz_{\cs^4} = \frac{P}{2}\sum_{p=1}^{P-1} \tilde{N}_p^2\left[1+\gamma_{\text{\tiny EM}}-\left(\sin\frac{\pi p}{2P}\right)^4\mathcal{I}\left(\sin\frac{\pi p}{2P}\right)\right],
\end{equation*}
which finally gives \eqref{logZ-fourier} upon using the explicit expression \eqref{I(sin)} for $\mathcal{I}$.
\end{proof}
The thermodynamic limit of the free energy \eqref{free-quiver} can be recovered from the exact expression \eqref{logZ-fourier} by taking the $P\rightarrow \infty $ limit. Indeed,
\begin{align*}
 	2 \tilde{N}_p^2 \tan\frac{\pi p}{2P}\sum_{a=1}^{2P-1}\sin\frac{\pi p a}{P}\,\ln\Gamma \left(\frac{a}{2P}\right) &\simeq 2\pi p \tilde{N}^2_p \int_{0}^{1}d\alpha \,\sin(2\pi p \alpha )\ln\Gamma (\alpha ) \\
		& =\tilde{N}^2_p \ln\left(2\pi p \ee^{\gamma _{\text{\tiny EM}}} \right),
\end{align*}
and the exact free energy reduces to our previous result. For the first approximation we have used that the rank function converges to a smooth function in the limit, thus either $p \ll P$ and $ 2\sin \frac{\pi p}{2P} \simeq \frac{\pi p}{P}$, or otherwise $\tilde{N}_p$ is vanishingly small for large $p$.\par
So far we have expressed the free energy in terms of $\tilde{N}_p$, but it can be also written in the `coordinate' representation. A suggestive form valid in the long quiver limit is
\begin{equation*}
 	-\frac{2}{P} \ln \mz_{\cs^4}\stackrel{P\rightarrow \infty }{\simeq }\left\langle \vec{N} \right| \gamma _{\text{\tiny EM}}+\ln \mathsf{M} \left| \vec{N}\right\rangle .
\end{equation*}
Indeed, at large $P$,
\begin{align*}
	\left\langle \vec{N} \right| \frac{1}{2}\ln \mathsf{M} \left| \vec{N}\right\rangle &= \frac{1}{2} \sum_{p=1}^{P-1} \tilde{N}_p^2 \ln \left( 2\sin \frac{\pi p}{2P}\right)^2 \\
		& \stackrel{P\rightarrow \infty }{\simeq } \frac{1}{2} \sum_{p=1}^{\infty} \tilde{N}_p^2 \ln \frac{\pi p}{P} ,
\end{align*}
where we have used that the eigenvalues of the matrix $\mathsf{M}$ are $\left( 2\sin \frac{\pi p}{2P}\right)^2$ and $\tilde{N}_p$ are the components of the vector $\vec{N}$ in the basis where $\mathsf{M}$ is diagonal. To pass to the second line, we have used once again that only values of $p \ll P$ contribute at leading order in the limit. Therefore, using $ \left\{ \tilde{N}_p \right\}_{1 \le p \le P-1} \stackrel{P\rightarrow \infty }{\simeq } \left\{ R_k \right\}_{k \ge 1}$ in the long quiver limit, the free energy \eqref{free-quiver} coincides with the matrix element above. In the same approximation,
\begin{equation}
	\mathsf{M}\stackrel{P\rightarrow \infty }{\simeq } \frac{\Delta }{P^2}\,,
\end{equation}
where $\Delta $ is the Laplace operator $\Delta =-\frac{\dd^2}{\dd z^2}$ on the interval $[0,1]$ with Dirichlet boundary conditions. The free energy can thus be expressed through the Green's function of the operator $\ln\Delta $, properly defined.\par
\bigskip
It is possible to compute the exact Green's function without taking the long quiver limit. By examining \eqref{logZ-fourier} we arrive at the following result:
\begin{stm}
For every $\ell \in \Z$ let 
	\begin{equation*}
		\Xi_{\ell}  := \sum_{r=-\infty }^{+\infty }\left(N_{\ell+2Pr} - N_{-\ell-2Pr} \right),
	\end{equation*}
	with the understanding that $N_j :=0$ if $j \notin \left\{1, \dots, P-1 \right\}$. The sum is formally infinite, but in practice only one term contributes for any $\ell$. In other words, $\Xi_\ell $ is a continuation of $N_\ell$ outside of its natural domain of defintion $1\leq\ell\leq P-1$ that satisfies 
	\begin{itemize}
		\item[(i)] $\Xi_j=N_j$ and $\Xi_{-j}=-N_{j}$ for $j=1,\dots ,P-1$; and
		\item[(ii)] $\Xi_{j+2P}=\Xi_j$.
	\end{itemize}
The free energy can then be written as
	\begin{center}\noindent\fbox{\parbox{0.98\linewidth}{%
	\begin{equation}
	\label{eq:lnZfiniteP}
	\begin{aligned}
		-\ln \mz_{\cs^4} =\sum_{j=1}^{P-1} & \left[K_j \sum_{a=1}^{2P-1} \ln \Gamma\left(\frac{a}{2P}\right) \sum_{s=1}^{a} \Xi_{j+2s-a-1}  -\ln (2P) N_j ^2  \right] .
	\end{aligned}
	\end{equation}}}\end{center}
\end{stm}
\begin{proof}
Using completeness relations \eqref{completeness} for the basis \eqref{eigenvectors} we can re-write \eqref{logZ-fourier} as
\begin{equation}\label{almost-final}
  -\ln \mz_{\cs^4} = \frac{2}{P}\,K_jN_{\ell}\sum_{p=1}^{P-1}\sin\frac{\pi p j}{P}\,\,\sin\frac{\pi p \ell }{P}\, \sum_{a=1}^{2P-1}\frac{\sin\frac{\pi p a}{P}}{\sin\frac{\pi p}{P}}\, \ln\Gamma \left(\frac{a}{2P}\right) -N_jN_j\ln(2P),
\end{equation}
with the summation over $j$ and $\ell$ implied. Now,
\begin{equation*}
	\frac{\sin\frac{\pi p a}{P}}{\sin\frac{\pi p }{P}}=\sum_{s=1}^{a}\cos\frac{(2s-a-1)\pi p}{P}\,,
\end{equation*}
and 
\begin{align*}
	\sum_{p=1}^{P-1}\sin\frac{\pi p j}{P} \sin\frac{\pi p \ell }{P} \cos\frac{\pi p m}{P} & = -\frac{1}{8} \sum_{p=1}^{P-1} \sum_{\epsilon_1, \epsilon_2, \epsilon_3 \in \left\{ \pm 1 \right\} }  \epsilon_1 \epsilon_2 \ee^{\frac{\ii\pi p }{P} ( \epsilon_1 j +  \epsilon_2 \ell + \epsilon_3 m )} \\
		&= - \frac{P}{4}  \sum_{\epsilon_1, \epsilon_3 \in \left\{ \pm 1 \right\} } \epsilon_1 \delta_{\epsilon_1 j +  \ell + \epsilon_3 m,0 } \\
		&= \frac{P}{4}\left(\delta _{j-\ell,m }+ \delta _{-j+\ell,m } - \delta _{j+\ell,m } -\delta _{-j-\ell,m} \right) .
\end{align*}
with all Kronecker deltas understood $\mod 2P$. To get the second line we have rearranged the terms in the sum and noted a symmetry $p \mapsto -p$, then added the $p=0$ term for free since the summand vanishes, and finally observed that the non-trivial phases cancel out in the sum. Applying these formulae to \eqref{almost-final} gives \eqref{eq:lnZfiniteP}.
\end{proof}
The free energy is a quadratic form in $K_j$ and $N_j$. When it is written as \eqref{eq:lnZfiniteP}, however, a given monimial $K_jN_i$ appears more than once in the sum. One may want to collect all the terms with the same  $K_jN_i$. This leads to the following result.
\begin{stm}
Define
\begin{equation*}
\mathbf{I}_{\ell}:=\ln\left( \prod_{m=0}^{\left\lfloor P-\frac{\ell}{2}-1\right\rfloor}\Gamma \left(\frac{2m+1+\ell}{2P}\right)\prod_{m=0}^{\left\lfloor\frac{\ell}{2}\right\rfloor-1}\Gamma \left(\frac{2m+2P+1-\ell}{2P}\right)
\right)
\end{equation*}
and
\begin{equation*}
 \mathsf{G}_{jk}:=\mathbf{I}_{|j-k|}-\mathbf{I}_{j+k}.
\end{equation*}
Then
\begin{equation*}
  -\ln \mz_{\cs^4} = \vec{K}^{\top} \mathsf{G} \vec{N} - \vec{N}^{\top} \vec{N} \ln(2P).
\end{equation*}
\end{stm}\par

\subsection{Towards comparison with supergravity: Simplified holography}
\label{sec:toysugra}
The supergravity dual of a 4d quiver CFT is the Gaiotto--Maldacena background \cite{Gaiotto:2009gz}, with the supergravity approximation valid at infinite 't Hooft coupling and in the long quiver limit.\footnote{See however \cite{Aharony:2012tz} for comments on finite gauge couplings.} The sphere free energy is holographic dual to the on-shell Einstein--Hilbert action, evaluated on the classical solution and properly regularised. The regularisation unavoidably introduces scheme dependence, and both in CFT and supergravity the free energy is defined up to a finite counterterm proportional to the central charge: 
\begin{equation*}
  S_{\text{\rm counterterm}}= \text{\rm const} \cdot a.
\end{equation*}
The supergravity action and the free energy of the matrix model thus agree possibly up to a counterterm of this form. Since $a$ is a fairly simple function of the parameters, this ambiguity  is easy to factor out and faithfully compare localisation to supergravity.\par
\bigskip
We now quickly review the generalities of the supergravity solution, and then propose a heuristic approach to evaluate the on-shell action, finding agreement with the computation in the long quiver SCFT.\par
The starting point is eleven-dimensional supergravity on AdS$_5 \times M_6$, where the internal manifold $M_6$ is a fibration over a Riemann surface $\Sigma$ that preserves the isometry $\mathrm{SU}(2)_R \times \mathrm{U}(1)_r \times \mathrm{U}(1)_{\mathrm{M}}$. The first two factors realise the holographic dual to the R-symmetry of the four-dimensional $\mN=2$ SCFT. The last factor $\mathrm{U}(1)_{\mathrm{M}}$ is identified with the isometry of the M-theory circle and will be used to reduce from M-theory to Type IIA string theory.\par
Here we follow the presentation in \cite{Nunez:2018qcj,Nunez:2019gbg} and work in Type IIA supergravity. The reduction of the eleven-dimensional Gaiotto--Maldacena background along $\mathrm{U}(1)_{\mathrm{M}}$ has been carried out in \cite{Aharony:2012tz}. The background is
\begin{equation*}
	\text{AdS}_5 \times \cs^2 \times \cs^1 \times \Sigma 
\end{equation*}
with metric
\begin{equation*}
	\dd s_{\text{IIA}}^2 = f_0 \left[ \dd s_{\text{AdS}_5}^2 + \sigma^2 f_{\cs^2} \dd s_{\cs^2} ^2 + f_{\cs^1} \dd \chi^2 + f_{\Sigma} \left( \dd \sigma^2 + \dd \eta^2 \right)\right] ,
\end{equation*}
descending from \cite[Eq.(3.17)]{Gaiotto:2009gz}. In this expression,
\begin{itemize}
	\item $\chi$ is the coordinate along the circle $\cs^1$;
	\item $\dd s_{\cs^2}$ is the round metric on $\cs^2$;
	\item The local coordinates on $\Sigma$ are achieved using a diffeomorphism to the upper-half plane, with $\Sigma$ identified with the half-strip $0 \le \eta \le P$ and $\sigma >0$. The coordinate $\eta \in [0,P]$ is a holographic image of the quiver direction and is identified with the argument of the rank function from Subsection~\ref{sec:rkfn} straight away.
	\item The functions $f_{\bullet}$ depend \emph{only} on first- and second-order partial derivatives of a scalar function $V= V(\sigma, \eta)$ (and not explicitly on $\sigma$ nor $\eta$). They can be found in \cite[Eq.(3.17)]{Aharony:2012tz} or \cite[Eq.(21)]{Nunez:2018qcj}, together with the expressions for $B_2$ and the dilaton, but we will not use their explicit form. 
\end{itemize}
The metric is thus entirely characterised by $V(\sigma, \eta)$. Writing $\dot{V}:= \sigma \partial_{\sigma} V $ and $V^{\prime}:= \partial_{\eta} V$, the system is governed by a linear equation \cite{Gaiotto:2009gz}
\begin{equation}
\label{eq:EOMVholo}
	\ddot{V} + \sigma^2 V^{\prime \prime} =0 .
\end{equation}
The quiver data enter through the boundary conditions
\begin{equation}
\label{eq:BCVholo}
	\dot{V}\Big\rvert_{\sigma =0}= \mathcal{R}(\eta ), \qquad  V\Big\rvert_{ \sigma \to \infty }=0 , \qquad  \dot{V}\Big\rvert_{ \eta=0 }=0=\dot{V}\Big\rvert_{ \eta=P }.
\end{equation}\par
The explicit relation between $V (\sigma, \eta)$ and the eigenvalue density, together with their respective equations of motion \eqref{eq:EOMVholo} and its analogue \eqref{eq:Poisson}, was explained in \cite{Legramandi:2021uds,Akhond:2021ffz,Akhond:2022awd,Akhond:2022oaf} for $d=3,5$. Building on that analogy, one may perhaps think of $\sigma$ as a direction giving vacuum expectation values to certain fields in the dual SCFT. However, it is not known whether such holographic relation explored in \cite{Akhond:2022awd,Akhond:2022oaf} holds in $d=4$.\par 
Now that we have reviewed the supergravity setup, we go on and solve for $V$ directly.\par
\begin{stm}
	The solution to \eqref{eq:EOMVholo}-\eqref{eq:BCVholo} is 
	\begin{equation}
	\label{eq:SolVholo}
		V(\sigma ,\eta )=-\sum_{k=1}^{\infty }R_k\, K_0\left(\frac{\pi k\sigma }{P}\right) \sin\frac{\pi k\eta }{P}\,.
\end{equation}
\end{stm}
\begin{proof}
The Fourier expansion of $V(\sigma, \eta)$ in the compact $\eta$-direction is
\begin{equation*}
	V(\sigma ,\eta )=\sum_{k=1}^{\infty }V_k(\sigma )\sin\frac{\pi k\eta }{P} ,
\end{equation*}
where we have already imposed the third condition in \eqref{eq:BCVholo}. Plugging this expansion into \eqref{eq:EOMVholo} we get:
\begin{equation*}
 \left[ \left(\sigma \partial _{\sigma} \right)^2 -\frac{\pi ^2k^2}{P^2} \sigma^2\right] V_k (\sigma)=0 .
\end{equation*}
The solution to this equation is:
\begin{equation*}
	V_k(\sigma )= V_k ^{(0)} \cdot K_0\left(\frac{\pi k\sigma }{P}\right),
\end{equation*}
where $K_0 $ is the modified Bessel function of the second kind, and $V_k ^{(0)}$ is an integration constant. The second boundary condition in \eqref{eq:BCVholo} is automatically satisfied, while the first one combined with \eqref{Fourier-R} fixes $V_k ^{(0)}= -R_k$. We thus get \eqref{eq:SolVholo}.
\end{proof}
It remains to substitute this into the action of Type IIA supergravity.\par
The 10d supergravity fields depend on $V$ in a complicated way, or more precisely, they depend on $\dot{V}, V^{\prime}$ and higher derivatives. Instead of an honest supergravity calculation we take a simpler phenomenological approach. We consider the quadratic action
\begin{equation}
\label{eq:SIIAshortcut}
	S=\int_{\Sigma}\dd\eta \dd \sigma ~\sigma \left[ \left(\partial_{\sigma} V\right)^2+\left(\partial _\eta V\right)^2 \right] 
\end{equation}
and observe that it leads to the equation of motion \eqref{eq:EOMVholo}. We then evaluate \eqref{eq:SIIAshortcut} on the classical solution applying the holographic renomalisation, here extremely simple. That is to say: instead of plugging the on-shell solution into the honest Type IIA action, we consider a much simpler action \eqref{eq:SIIAshortcut}, which however leads to the same equation \eqref{eq:EOMVholo}. In this section we limit ourselves to this observation, which should be supplemented with a proof of the equality of \eqref{eq:SIIAshortcut} with the Ype IIA action when both are on-shell, to yield a complete derivation. To put our proposal on firm ground, it remains as an open problem to provide a first principles derivation of \eqref{eq:SIIAshortcut}.\par
\begin{stm}
	Let $S_{\rm R}$ denote the finite part of \eqref{eq:SIIAshortcut}, evaluated on shell with cutoff at $\sigma=\varepsilon$ and after holographic renormalisation. With a suitable choice of renormalisation scheme, we find 
	\begin{equation}
	\label{eq:SsugraRen}
  		S_{\rm R}=-\frac{P}{2}\sum_{k=1}^{\infty }R_k^2\ln\frac{\pi k \ee^{\gamma_{\text{\tiny\rm EM}}}}{P} ,
	\end{equation}
	in agreement with the free energy \eqref{free-quiver} of the holographic dual long quiver.
\end{stm}
\begin{proof}
We regulate the radial integral with a cutoff at $\sigma =\varepsilon $. Plugging the solution \eqref{eq:SolVholo} into \eqref{eq:SIIAshortcut}, we find 
\begin{align*}
	S_{\text{on-shell}} (\varepsilon) = \int_0^{\infty} \dd \eta \int_{\varepsilon}^{\infty} \sigma  \dd \sigma & \sum_{k=1}^{\infty} \sum_{\ell=1}^{\infty} R_k R_{\ell}  \left( \frac{\pi k}{P} \right)\left( \frac{\pi  \ell}{P} \right) \\
	\times & \left[ \cos \left( \frac{\pi k \eta}{P} \right) \cos \left( \frac{\pi \ell \eta}{P} \right)  K_0\left(\frac{\pi k\sigma }{P}\right)  K_0\left(\frac{\pi \ell\sigma }{P}\right) \right. \\
	+ & \left. ~  \sin\left( \frac{\pi k \eta}{P} \right) \sin \left( \frac{\pi \ell \eta}{P} \right)  K_1\left(\frac{\pi k\sigma }{P}\right)  K_1\left(\frac{\pi \ell\sigma }{P}\right)\right] .
\end{align*}
We integrate over $\eta$ and use the orthogonality relations. We then pull the remaining sum outside of the $\sigma$-integral and change variable to $\hat{\sigma}= \frac{\pi k}{P} \sigma$. We arrive at:
\begin{align*}
	S_{\text{on-shell}}(\varepsilon) = \frac{P}{2} \sum_{k=1}^{\infty} R_k^2 \int_{\frac{\pi k}{P}\varepsilon}^{\infty} \dd \hat{\sigma} ~  \hat{\sigma} \left[ K_0 ( \hat{\sigma})^2 + K_1 ( \hat{\sigma})^2  \right] .
\end{align*}
The integral is performed explicitly and features a logarithmic divergence in the small-$\varepsilon$ limit:
\begin{equation*}
	S_{\text{on-shell}}(\varepsilon) = \frac{P}{2} \sum_{k=1}^{\infty} R_k^2 \left(  - \ln \left(\frac{\pi k\varepsilon }{2P}\right) - \gamma_{\text{\tiny EM}} + \mathcal{O} (\varepsilon)\right) .
\end{equation*}
The divergence, which originates from the term $\hat{\sigma} K_0(\hat{\sigma})^2$ in the integrand, is removed by adding a counterterm:
\begin{equation*}
 S_{\text{\rm counterterm}} (\varepsilon)=\frac{P}{2}\,\ln \frac{\varepsilon}{2}\sum_{k=1}^{\infty }R_k^2 .
\end{equation*}
This expression is achieved from a local functional of $\dot{V}$ only, with a suitably chosen coefficient. As expected, the counterterm is proportional to the Weyl anomaly coefficient, by virtue of \eqref{eq:acfromST}. The renormalised action 
\begin{equation*}
	S_{\rm R} := \lim_{\varepsilon \to 0^{+}}  \left[  S_{\text{on-shell}}(\varepsilon) + S_{\rm counterterm} (\varepsilon) \right]
\end{equation*}
yields \eqref{eq:SsugraRen}.
\end{proof}
The metric falls off exponentially as $\sigma \to \infty$, thus the non-compactness of the integration domain, which is a semi-infinite strip, poses no problem. The regularisation scheme introduced a cutoff at at $\sigma = \varepsilon>0$, to prevent the singularities due to the source branes located at $\sigma=0$ in the Type IIA setup.\par
We conclude the supergravity analysis with three remarks.
\begin{itemize}
	\item[---] The numerical value $\pi \ee^{\gamma_{\text{\tiny\rm EM}}}$ inside the logarithm in \eqref{eq:SsugraRen} is scheme-dependent and thus unphysical. We could have chosen a different counterterm to remove the divergence (say, using $\ln \varepsilon$ instead of $\ln \frac{\varepsilon}{2}$) and obtain a different number. The dependence on $P$ and on the Fourier modes $(k, R_k)_{k \ge 1}$, on the contrary, is universal and unambiguous.
	\item[---] To fully establish agreement with Type IIA supergravity it remains to show that the full supergravity action is equivalent on-shell to \eqref{eq:SIIAshortcut}. Because of this, in this subsection we have only provided a heuristic argument rather than a derivation.
	\item[---] Five-dimensional $\mathrm{SU}(2) \times \mathrm{U}(1)$ gauged supergravity \cite{Romans:1985ps} admits an uplift to ten-dimensional Type IIA supergravity which includes the Gaiotto--Maldacena solution \cite{Gauntlett:2007sm}. In the consistent truncation of the solution reviewed above to 5d gauged supergravity, the Fourier coefficients $R_k$ of the rank function enter through the definition of the Newton constant in AdS$_5$. We have computed $a$ holographically in this formalism (similarly to \cite[Sec.4.2]{Macpherson:2014eza}), finding perfect agreement with our previous result, as expected. The consistent truncation to 5d gauged supergravity may as well provide a route to complete the match of the free energy with an holographic computation, as elaborated in the previous point.
\end{itemize}

\section{Examples}
\label{sec:ex}

The aim of this section is to illustrate the general setup in two simple and concrete examples. They are associated to the two rank functions plotted in Figure \ref{fig:rankplusTN}.

\subsection{Example 1}
\label{sec:ex1}

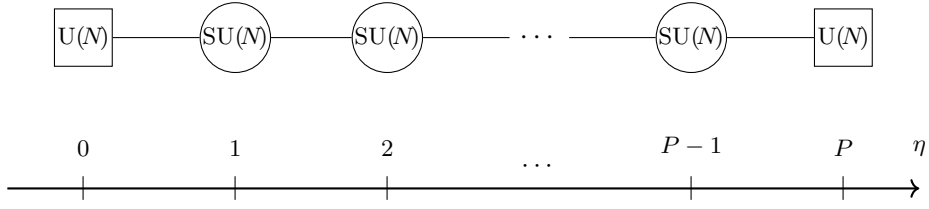
\begin{figure}[t]
\centering
\begin{tikzpicture}[auto,square/.style={regular polygon,regular polygon sides=4}]
	\node[circle,draw] (gauge1) at (4,0) { \hspace{18pt} };
	\node (a1) at (4,0) {\footnotesize $\mathrm{SU}(\!N\!)$};
	\node[draw=none] (gaugemid) at (2,0) {$\cdots$};
	\node[circle,draw] (gauge3) at (0,0) { \hspace{18pt} };
	\node[circle,draw] (gauge4) at (-2,0) { \hspace{18pt} };
	\node (a2) at (-2,0) {\footnotesize $\mathrm{SU}(\!N\!)$};
	\node (a3) at (0,0) {\footnotesize $\mathrm{SU}(\!N\!)$};
	\node[square,draw] (fl1) at (6,0) { \hspace{8pt} };
	\node[square,draw] (fl2) at (-4,0) { \hspace{8pt} };
	\node[draw=none] (aux1) at (6,0) {\footnotesize $\mathrm{U}(\!N\!)$};
	\node[draw=none] (aux2) at (-4,0) {\footnotesize $\mathrm{U}(\!N\!)$};
	\draw[-](gauge1)--(gaugemid);
	\draw[-](gaugemid)--(gauge3);
	\draw[-](gauge4)--(gauge3);
	\draw[-](gauge1)--(fl1);
	\draw[-](gauge4)--(fl2);
	
	\draw[->,thick] (-5,-2) -- (7,-2);
	\node[anchor=south] at (7,-1.7) {\footnotesize $\eta$};
	\node[anchor=south] at (6,-1.7) {\footnotesize  $P$};
	\node[anchor=south] at (4,-1.7) {\footnotesize  $P-1$};
	\node[anchor=south] at (-4,-1.7) {\footnotesize  $0$};
	\node[anchor=south] at (-2,-1.7) {\footnotesize  $1$};
	\node[anchor=south] at (0,-1.7) {\footnotesize  $2$};
	\node[draw=none] (gaugemid) at (2,-1.7) {\small $\cdots$};
	\node[] at (6,-2) {\footnotesize  $\vert$};
	\node[] at (-4,-2) {\footnotesize  $\vert$};
	\node[] at (4,-2) {\footnotesize  $\vert$};
	\node[] at (0,-2) {\footnotesize  $\vert$};
	\node[] at (-2,-2) {\footnotesize  $\vert$};
\end{tikzpicture}
\caption{Example 1: balanced linear quiver with gauge group $\mathrm{SU}(N)^{P-1}$.}
\label{fig:plunNP4d}
\end{figure}\par

The first linear quiver SCFT we consider consists of the gauge group $\mathrm{SU}(N)^{P-1}$ with $N$ hypermultiplets in the fundamental representation of the first and last $\mathrm{SU}(N)$ gauge factors. The quiver is drawn in Figure \ref{fig:plunNP4d}, and its $d=5$ incarnation was studied in \cite{Bergman:2018hin} and subsequent works.\par
The rank function is 
\begin{equation*}
	\mathcal{R}_{\text{Ex 1}} (\eta) = \begin{cases} N \eta & \hspace{25pt} 0 \le \eta < 1 \\ N & \hspace{25pt} 1 \le \eta < P-1 \\ N (P-\eta) & \hspace{4pt} P-1\le \eta \le P , \end{cases}
\end{equation*}
whose Fourier coefficients are 
\begin{equation*}
	R_k =\begin{cases} N P \frac{4}{ k^2 \pi^2} \sin\left( \frac{\pi k}{P}\right) & k \text{ odd} \\ 0  & k \text{ even}. \end{cases}
\end{equation*}
We begin with the analysis in arbitrary dimension $d \ge 3$ and specialise to $d=4$ below.\par
\bigskip
\begin{stm}
	The interpolating free energy $\tilde{F}_d$ at infinite 't Hooft coupling is 
	\begin{center}\noindent\fbox{\parbox{0.98\linewidth}{%
	\begin{equation}
	\label{eq:FtildeEx1}
		\tilde{F}_{d,\text{\rm Ex 1}} \simeq N^2 P^{d-3} \zeta (d-2) \frac{(d-2)^{d-2} (2^{d-2}-1)}{(4\pi)^{d-3}} 
	\end{equation}}}\end{center}
	for $d >3$, and with the replacement $P^{d-3} \zeta (d-2)  \to \ln P$ in the limit $d \to 3$.
\end{stm}
\begin{proof}
We plug the Fourier coefficients $R_k$ in formula \eqref{eq:Ftilded} and compute 
\begin{equation*}
\begin{aligned}
	\frac{P^{d-3}}{2^d \pi^{d-5}}\sum_{k=1}^{\infty} k^{4-d}R_k^2 = N^2P^{d-1} \frac{2}{(4\pi)^{d-1} } \left[ (2^d-1)\zeta (d) - \Re \left\{ \ee^{\frac{\ii 2\pi}{P}}\Phi \left( \ee^{\frac{\ii 4\pi}{P}} , d, \frac{1}{2} \right)\right\} \right] ,
\end{aligned}
\end{equation*}
where 
\begin{equation*}
	\Phi (z,d,a):= \sum_{k=0}^{\infty} \frac{z^k}{(k+a)^d}
\end{equation*}
is Lerch's transcendent \cite{Lerch}. See Appendix \ref{app:lerch} for a brief overview of this function. For integer $d \in \N_{> 3}$, it enjoys the property
\begin{equation*}
	\ee^{\frac{\ii 2\pi}{P}}\Phi \left( \ee^{\frac{\ii 4\pi}{P}} , d, \frac{1}{2} \right) = \zeta \left( d,\frac{1}{2} \right) + \zeta \left( d-1,\frac{1}{2} \right)  \left( \frac{\ii 4\pi}{P} \right) + \frac{\zeta \left( d-2,\frac{1}{2} \right)}{2!} \left( \frac{\ii 4\pi}{P} \right)^2 + \mathcal{O} (P^{-3}),
\end{equation*}
with the coefficients given in terms of Hurwitz's $\zeta$-function. The case $d=3$ will be discussed separately below. Using 
\begin{equation*}
	\zeta \left( d,\frac{1}{2} \right)  = (2^d -1) \zeta (d)
\end{equation*}
we obtain 
\begin{equation*}
	(2^d-1)\zeta (d) - \Re \left\{ \ee^{\frac{\ii 2\pi}{P}}\Phi \left( \ee^{\frac{\ii 4\pi}{P}} , d, \frac{1}{2} \right)\right\} = - \zeta (d-2) \frac{(2^{d-2}-1)}{2} \left( \frac{\ii 4\pi}{P}\right)^2 .
\end{equation*}
Therefore 
\begin{equation*}
	\frac{\tilde{F}_{d,\text{\rm Ex 1}} }{(d-2)^{d-2}}= N^2 P^{d-3} \zeta (d-2) (4\pi)^{3-d} (2^{d-2}-1) ,
\end{equation*}
as claimed in \eqref{eq:FtildeEx1}. However, we have proven this identity assuming $d \in \N_{d > 3}$. More precisely, the pre-factor $N^2P^{d-3} (d-2)^{d-2}$ has already been shown to be valid in $d > 3$, thus we have to prove the validity of  
\begin{equation*}
	f_{\text{\rm Ex 1}}(d) := \frac{\tilde{F}_{d,\text{\rm Ex 1}}}{N^2 P^{d-3} (d-2)^{d-2}}  ,
\end{equation*}
computed by the large-$P$ approximation of Lerch's transcendent. Combining the analyticity property of $f_{\text{\rm Ex 1}}(d)$ for $d \in \C \setminus \left\{3 \right\}$ with Carlson's theorem, we obtain that \eqref{eq:FtildeEx1} holds by analytic continuation in $d > 3$.\par
More rigorously, we fix an arbitrarily small $\varepsilon >0$ and define 
\begin{equation*}
	f_{\text{\rm Ex 1}} ^{(\varepsilon)} (d^{\prime}) := \begin{cases} f_{\text{\rm Ex 1}}(d^{\prime}+3) & \lvert d^{\prime} \rvert \ge  \varepsilon \\ f_{\text{\rm Ex 1}}(\varepsilon+3) & \lvert d^{\prime} \rvert < \varepsilon  \end{cases}
\end{equation*}	
as a function of $d^{\prime} \in \C $. Then, $f_{\text{\rm Ex 1}} ^{(\varepsilon)} (d^{\prime})$ is continuous everywhere, analytic in $\Re (d^{\prime}) > \varepsilon$, and bounded by exponentials, $\forall \varepsilon >0$. Besides, it interpolates between the values of $f_{\text{\rm Ex 1}}$ for $d^{\prime} \in \N$. Sending $\varepsilon \to 0^{+}$, we have that Carlson's theorem holds, whence the conclusion for $d \ne 3$.\par
The case $d=3$ must be dealt with separately. We have 
\begin{equation*}
	\Re \left\{ \ee^{\frac{\ii 2\pi}{P}}\Phi \left( \ee^{\frac{\ii 4\pi}{P}} , d=3, \frac{1}{2} \right) \right\} = \zeta \left( 3,\frac{1}{2} \right) + \frac{1}{2} \left( \frac{3}{2} + 2 \ln (2) + \ln \left( \frac{P}{2\pi}\right)\right) \left( \frac{\ii 4\pi}{P} \right)^2 ,
\end{equation*}
and from this expression we get 
\begin{equation*}
	\tilde{F}_{d=3,\text{\rm Ex 1}} = N^2 \left[ \frac{3}{2} + \ln \left( \frac{2P}{\pi}\right) \right] .
\end{equation*}
The leading term is $N^2 \ln P$. This matches with the leading order in the large-$P$ expansion of $F_{d=3}$ obtained from \cite[Eq.(5.17)]{Akhond:2021ffz}.\footnote{Note that there is a misprint in \cite[Eq.(5.17)]{Akhond:2021ffz}, where the right-hand side should be divided by $\pi$. Moreover, that expression is for $c_{\mathrm{hol},d=3}= \frac{F_{d=3}}{4\pi}$, with the coefficient fixed in \cite[Sec.3.3.4]{Akhond:2022oaf}.} Therefore, the apparent divergence of $\tilde{F}_d$ in the limit $d \to 3^{+}$ due to the pole of $\zeta (d-2)$ is actually signalling an enhancement of the scaling from $N^2$ to $N^2 \ln P$.
\end{proof}\par
Taking the limit $d \to 6$ from below, the interpolating free energy \eqref{eq:FtildeEx1} gives a prediction for the anomaly coefficient in $d=6$ for the quiver of Figure \ref{fig:plunNP4d}:
\begin{equation*}
	\tilde{F}_{d=6,\text{\rm Ex 1}} \simeq \frac{2\pi}{3} N^2 P^{3} .
\end{equation*}\par
\medskip
We now consider odd-dimensional defects charged under a $\mathrm{SU}(N)$ group labelled by $j_{\ast}$. Let us introduce some notation. As above, $\Phi$ denotes Lerch's transcendent, and $D_{\mathrm{BW}} (\cdot)$ is the Bloch--Wigner dilogarithm \cite{Zagier2007}
\begin{equation*}
		D_{\mathrm{BW}} (u) := \Im \left\{ \mathrm{Li}_2 (u) \right\} + \mathrm{Arg} (1-u) \ln \lvert u \rvert .
\end{equation*}
The parameter $x_{\ast}$ is \cite{Uhlemann:2020bek}: 
\begin{equation*}
		 x_{\ast} (z_{\ast},\kappa)= \frac{1}{2\pi} \sinh^{-1} \left[ \cot (\pi \kappa) \sin (\pi z) \right] .
\end{equation*}\par
\begin{stm}
	Consider a Wilson loop in the rank-$\kk$ antisymmetric representation of the $j_{\ast}^{\text{th}}$ gauge node, supported on $\cs^1 \subset \cs^d$. Its defect free energy in the long quiver limit is 
	\begin{equation*}
		\ln \langle W_{\mathsf{A}_{\kk}} (j_{\ast}) \rangle \simeq NP \frac{(d-2)}{\pi} \left[ D_{\mathrm{BW}} \left(\ee^{2 \pi x_{\ast}+ \ii \pi z_{\ast}} \right) + D_{\mathrm{BW}} \left(\ee^{2 \pi x_{\ast} + \ii \pi (1-z_{\ast})} \right) \right] ,
	\end{equation*}
	with $z_{\ast}=\frac{j_{\ast}}{P}$ and $\kappa=\frac{\kk}{N}$.\par
	Consider a chiral and an antichiral multiplets of scaling dimension $\Delta$ charged under the $j_{\ast}^{\text{th}}$ gauge node, supported on $\cs^3 \subset \cs^d$. The defect free energy in the long quiver limit is 
	\begin{equation}
	\label{eq:F3dDefectEx1}
		F_{\text{\rm 3d defect},d} \simeq NP (1-\Delta)\frac{(d-2)}{\pi} \Im \left\{  \ee^{\ii \pi z_{\ast}} \Phi \left( \ee^{\ii 2\pi z_{\ast}} ,2 , \frac{1}{2}  \right) \right\} .
	\end{equation}
\end{stm}
\begin{proof}
The Wilson loop result follows directly from \cite[Eq.(2.23)]{Uhlemann:2020bek} and our general discussion on the $d$-dependence of the defect in Subsection \ref{sec:WLd}. We refer to \cite{Uhlemann:2020bek} for the details of the derivation. We have also computed the expectation value using our prescription with the Fourier coefficients $R_k$, which gives a complicated expression in terms $\Phi \left( \ee^{-\frac{\ii 2 \pi }{P}-4 \pi  x_{\ast}-2 \ii \pi  z_{\ast}},2, \frac{1}{2}\right)$, to be approximated at leading order in $P$.\par
Our analysis in Subsection \ref{sec:3ddef} predicts the analogous relation between three-dimensional defects in $d$ and $5$ dimensions. We now proceed to confirm the result explicitly.\par
To evaluate the three-dimensional defect free energy, we approximate $\cos\left( \frac{\pi k(P-2)}{2P}\right)$ at leading order for large $P$ and write the Fourier coefficients $R_k$ in the form 
\begin{equation}\label{Ex1-Rk}
	R_{k} \simeq \begin{cases} \frac{4N}{\pi k } & k \text{ odd} \\ 0 &  k \text{ even}.\end{cases}
\end{equation}
We utilise them to compute 
\begin{equation*}
\begin{aligned}
	\frac{F_{\text{\rm 3d defect},d}}{NP (d-2)\pi (1-\Delta)} &= \sum_{k=1}^{\infty} \frac{R_k}{\pi k} \sin (\pi k z_{\ast}) \\
	& = \frac{1}{\pi^2} \Im \left\{  \ee^{\ii \pi  z_{\ast}} \Phi \left( \ee^{\ii 2\pi z_{\ast}} ,2 , \frac{1}{2}  \right) \right\} .
\end{aligned}
\end{equation*}
The plot of the defect free energy is in Figure \ref{fig:F3dDefEx1}, and it matches \cite[Fig.8a]{Santilli:2023fuh}, as predicted.\par
\begin{figure}[tb]
\centering
\includegraphics[width=0.5\textwidth]{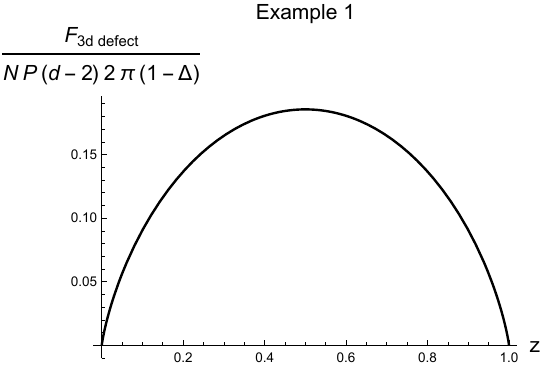}
\caption{Three-dimensional defect free energy in the balanced linear quiver with gauge group $\mathrm{SU}(N)^{P-1}$.}
\label{fig:F3dDefEx1}
\end{figure}\par
On the other hand, the argument in Subsection \ref{sec:3ddef} combined with \cite[Eq.(4.33)]{Santilli:2023fuh} yields 
\begin{equation*}
	\frac{F_{\text{\rm 3d defect},d}}{NP (d-2)\pi (1-\Delta)} = \frac{2}{\pi^2} \Im \left\{  \mathrm{Li}_2 (\ee^{\ii \pi z}) -  \mathrm{Li}_2 (-\ee^{\ii \pi z}) \right\} ,
\end{equation*}
Comparing the direct derivation, that yields \eqref{eq:F3dDefectEx1}, we observe that the two procedures are consistent if and only if the identity 
\begin{equation*}
	\Im \left\{  \ee^{\ii \pi z} \Phi \left( \ee^{\ii 2\pi z} ,2 , \frac{1}{2}  \right) \right\}  = 2 \Im \left\{  \mathrm{Li}_2 (\ee^{\ii \pi z}) -  \mathrm{Li}_2 (-\ee^{\ii \pi z}) \right\} 
\end{equation*}
holds. Using the series representation of $\Phi \left( \cdot, 2, \frac{1}{2}\right)$ and $\mathrm{Li}_2 (\cdot)$, it is straightforward to check that this is indeed the case.
\end{proof}
The linear quiver in Figure \ref{fig:plunNP4d} possesses a $\Z_2$ outer automorphism that acts as $z \mapsto 1-z$. The defect free energies are maximised at the fixed point $z_{\ast}=\frac{1}{2}$ of this automorphism, as first pointed out in \cite{Santilli:2023fuh}.\par
\bigskip
Let us now focus on the $\mN=2$ conformal gauge theory obtained from the quiver in Figure \ref{fig:plunNP4d} at $d=4$. An immediate corollary of the result \eqref{eq:FtildeEx1}, and the universal relation $\tilde{F}_{d=4}=2 \pi a $ from \eqref{eq:FtildetoA} is 
\begin{equation*}
	a_{\text{Ex 1}} = \frac{N^2P}{4} .
\end{equation*}
Let us count the central charge directly from the field content. We have $N^2-1$ vector multiplet modes at every $j=1,\dots,P-1$, $N^2$ hypermultiplet modes for each of the $(P-2)$ internal edges of the quiver, and $N^2$ hypermultiplet modes each of the two flavour nodes. 
This gives, to the leading large-$N$ order
\begin{equation*}
	n_{\mathrm{v},\text{Ex 1}} = N^2 (P-1), \qquad n_{\mathrm{h},\text{Ex 1}} = N^2 P ,
\end{equation*}
and therefore, from \eqref{eq:STac}, 
\begin{equation*}
	a_{\text{Ex 1}} = \frac{6P-5}{24}\,N^2 \simeq \frac{P}{4} N^2
\end{equation*}
in agreement with our derivation in the long quiver limit.
\par
 To compute the free energy, we use the representation \eqref{eq:lnZfiniteP}. We have:
\begin{equation*}
\frac{1}{N}\,\Xi _\ell=\mathop{\mathrm{sign}}\left((\ell-P)_{\mathop{\mathrm{mod}}2P} -P+\delta _{\ell_{\mathop{\mathrm{mod}}2P},P}\right),
\end{equation*}
and
\begin{equation*}
\sum_{s=1}^{a}\Xi _{2s-a-1}=\sum_{s=1}^{a}\Xi _{P+2s-a-2}=N\mathop{\mathrm{sign}}(P-a),\qquad {\rm for~}a=1,\ldots, 2P-1.
\end{equation*}
This gives:
\begin{equation*}
 \frac{1}{N^2}\ln \mz_{\cs^4} =2\sum_{\ell=1}^{P-1} \ln\frac{\Gamma \left(\frac{1}{2}+\frac{\ell}{2P}\right)}{\Gamma \left(\frac{1}{2}-\frac{\ell}{2P}\right)}+(P-1)\ln(2P).
\end{equation*}
Asympotically, at large $P$,
\begin{equation*}
  \frac{1}{N^2}\ln \mz_{\cs^4} \simeq 2P\int_{0}^{1}dx\,\ln\frac{\Gamma \left(\frac{1+x}{2}\right)}{\Gamma \left(\frac{1-x}{2}\right)}+P\ln(2P)  =P\ln\frac{2^{\frac{4}{3}}P}{A^{12}}\,,
\end{equation*}
where $A$ is the Glaisher constant. The same result follows upon substituting \eqref{Ex1-Rk} into \eqref{free-quiver}.

\subsection{Example 2}
\label{sec:ex3}
\begin{figure}[t]
\centering
\begin{tikzpicture}[auto,square/.style={regular polygon,regular polygon sides=4}]
	\node[circle,draw] (gauge1) at (4,0) { \hspace{34pt} };
	\node (a1) at (4,0) {$\scriptstyle {\mathrm{SU}(\!(\!P-1\!)N\!)}$};
	\node[draw=none] (gaugemid) at (2,0) {$\cdots$};
	\node[circle,draw] (gauge3) at (0,0) { \hspace{34pt} };
	\node[circle,draw] (gauge4) at (-2,0) { \hspace{34pt} };
	\node (a2) at (-2,0) {\footnotesize $\mathrm{SU}(\!N\!)$};
	\node (a3) at (0,0) {\footnotesize $\mathrm{SU}(\!2N\!)$};
	\node[square,draw] (fl1) at (6,0) { \hspace{16pt} };
	\node[draw=none] (aux1) at (6,0) {\footnotesize $\mathrm{U}(\!PN\!)$};
	\draw[-](gauge1)--(gaugemid);
	\draw[-](gaugemid)--(gauge3);
	\draw[-](gauge4)--(gauge3);
	\draw[-](gauge1)--(fl1);
	
	\draw[->,thick] (-5,-2) -- (7,-2);
	\node[anchor=south] at (7,-1.7) {\footnotesize $\eta$};
	\node[anchor=south] at (6,-1.7) {\footnotesize $P$};
	\node[anchor=south] at (4,-1.7) {\footnotesize $P-1$};
	\node[anchor=south] at (-4,-1.7) {\footnotesize $0$};
	\node[anchor=south] at (-2,-1.7) {\footnotesize $1$};
	\node[anchor=south] at (0,-1.7) {\footnotesize $2$};
	\node[draw=none] (gaugemid) at (2,-1.7) {\small $\cdots$};
	\node[] at (6,-2) {\footnotesize $\vert$};
	\node[] at (-4,-2) {\footnotesize $\vert$};
	\node[] at (4,-2) {\footnotesize $\vert$};
	\node[] at (0,-2) {\footnotesize $\vert$};
	\node[] at (-2,-2) {\footnotesize $\vert$};
\end{tikzpicture}
\caption{Example 2: balanced linear quiver with gauge group $\prod_{j=1}^{P-1}\mathrm{SU}(Nj)$.}
\label{fig:TNP4d}
\end{figure}
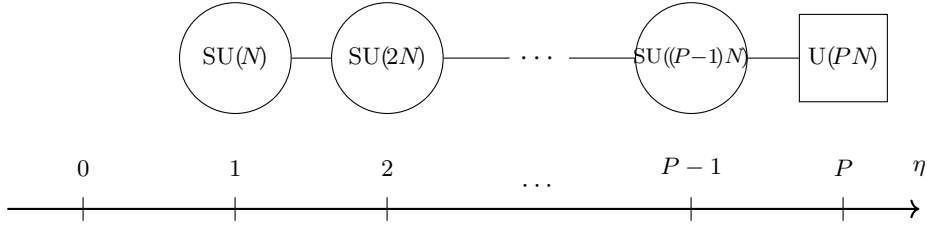\par

The next example we consider is the quiver in Figure \ref{fig:TNP4d}. For $N=1$, this is the $T_P$ theory \cite{Benini:2009gi}. The rank function is 
\begin{equation*}
	\mathcal{R} (\eta) = \begin{cases} N \eta & 0 \le \eta \le P-1 \\ N(P-1)(P-\eta) & P-1 < \eta \le P , \end{cases}
\end{equation*}
and its Fourier coefficients are 
\begin{equation}\label{Rk-Ex2}
	R_k = 2 N P^2 \frac{(-1)^{k-1}}{\pi^2 k^2} \sin \left( \frac{\pi k}{P}\right).
\end{equation}
\begin{stm}
	The interpolating free energy $\tilde{F}_d$ at infinite 't Hooft coupling is 
	\begin{center}\noindent\fbox{\parbox{0.98\linewidth}{%
	\begin{equation}
	\label{eq:FtildeEx2}
		\tilde{F}_{d,\text{\rm Ex 2}} \simeq N^2 P^{d-1} \zeta (d-2) \frac{(d-2)^{d-2}}{2 (2\pi)^{d-3}}
	\end{equation}}}\end{center}
	for $d >3$, and with the replacement $P^{d-3} \zeta (d-2)  \mapsto \ln P$ in the limit $d \to 3$.
\end{stm}
\begin{proof}
Plugging the Fourier coefficients $R_k$ in \eqref{eq:Ftilded} we find 
\begin{equation*}
	P^{d-3} \frac{(d-2)^{d-2}}{2^d \pi^{d-5}}\sum_{k=1}^{\infty} k^{4-d}R_k^2 = N^2 P^{d+1} \frac{(d-2)^{d-2}}{(2\pi)^{d-1}} \left[ \zeta (d) - \Re \left\{ \mathrm{Li}_d \left( \ee^{\frac{\ii 2\pi}{P}}\right)\right\} \right] .
\end{equation*}
Let us assume momentarily that $d>3$ and come back to $d=3$ below. Expanding the polylogarithm for large $P$, 
\begin{equation*}
	\mathrm{Li}_d \left( \ee^{\frac{\ii 2\pi}{P}}\right) = \zeta (d) + \zeta(d-1) \left( \frac{\ii 2\pi}{P} \right) + \frac{\zeta(d-2) }{2} \left( \frac{\ii 2\pi}{P} \right)^2 + \mathcal{O}(P^{-3}) , 
\end{equation*}
we arrive at \eqref{eq:FtildeEx2}. Considering the case $d=3$ separately and working it out directly, following identical steps as in Subsection \ref{sec:ex1} we find that $P^{d-3} \zeta (d-2) $ is replaced by $\ln P$. This correctly reproduces the enhanced scaling in $d=3$ \cite{Assel:2012cp}.
\end{proof}\par
The limit $d \to 6$ from below of \eqref{eq:FtildeEx2} gives a prediction for the anomaly coefficient in $d=6$ for the quiver of Figure \ref{fig:TNP4d}:
\begin{equation*}
	\tilde{F}_{d=6,\text{\rm Ex 2}} \simeq \frac{8\pi}{45} N^2 P^{5} .
\end{equation*}\par
\medskip
\begin{digr}
From our explicit results we observe 
\begin{equation*}
	\frac{ \left. \tilde{F}_{d,\text{\rm Ex 1}}\right\rvert_{N \mapsto NP} }{2 \tilde{F}_{d,\text{\rm Ex 2}}} = 2 - 2^{3-d}.
\end{equation*}
The inequality 
\begin{equation*}
	\left. \tilde{F}_{d,\text{\rm Ex 1}}\right\rvert_{N \mapsto NP} > 2 \tilde{F}_{d,\text{\rm Ex 2}} 
\end{equation*}
holds in strict form for $d>3$. This implies that there may exist an RG flow from the theory of Example 1, with the replacement $N \mapsto PN$, to two copies of the theory in Example 2. This is consistent with the proposal in \cite{Akhond:2022awd,Akhond:2022oaf}, and formally extends it to real values $d \in \R_{>3}$.
\end{digr}\par
\bigskip
Taking $d \to 4$ and using $\tilde{F}_{d=4}=2\pi a$, we find 
\begin{equation*}
	a_{\text{\rm Ex2}} = \frac{N^2 P^3 }{12} .
\end{equation*}\par
Counting the central charge directly gives the same result, as can be easily seen.
Indeed, the quiver in Figure \ref{fig:TNP4d} contains (at large-$N$ but any $P$)
\begin{equation*}
	\left. n_{\mathrm{v},\text{Ex 2}} \right\rvert_{\text{\rm node }j}= N^2 j^2 , \qquad \left. n_{\mathrm{h},\text{Ex 2}} \right\rvert_{\text{\rm edge }j \to j+1} = N^2j(j+1)  , \qquad \left. n_{\mathrm{h},\text{Ex 2}} \right\rvert_{\text{\rm tail }} = N^2 P(P-1) .
\end{equation*}
Therefore
\begin{equation*}
	n_{\mathrm{v},\text{Ex 2}} =\frac{N^2}{6} (P-1) P (2 P-1), \qquad n_{\mathrm{h},\text{Ex 2}} =\frac{N^2}{3} (P-1) P (P+1), 
\end{equation*}
from which we obtain 
\begin{equation*}
	a_{\text{Ex 2}} = \frac{P(P-1)(4P-1)}{48}N^2 \simeq \frac{P^3}{12}N^2\,, 
\end{equation*}
in perfect agreement with the long quiver formula.\par
The free energy is readly found from \eqref{eq:lnZfiniteP}. For the quiver in Figure~\ref{fig:TNP4d}, $N_j=jN$ and thus
\begin{equation*}
\frac{1}{N}\,\Xi _\ell= (\ell-P)_{\mathop{\mathrm{mod}}2P} -P+\delta _{\ell_{\mathop{\mathrm{mod}}2P},P}.
\end{equation*}
It is easy to see that
\begin{equation*}
\sum_{s=1}^{a}\Xi _{P+2s-a-2}=N(P-a),\qquad {\rm for~}a=1,\ldots, 2P-1,
\end{equation*}
which gives for the free energy:
\begin{equation*}
 \frac{1}{N^2}\ln \mz_{\cs^4} =P\sum_{\ell=1}^{P-1}\ell\,
 \ln\frac{\Gamma \left(\frac{1}{2}+\frac{\ell}{2P}\right)}{\Gamma \left(\frac{1}{2}-\frac{\ell}{2P}\right)}+
 \frac{P(P-1)(2P-1)}{6}\ln(2P).
\end{equation*}
In the long-quiver limit this expression becomes:
\begin{equation*}
  \frac{1}{N^2}\ln \mz_{\cs^4} \simeq P^3\int_{0}^{1}dx\,x\,\ln\frac{\Gamma \left(\frac{1+x}{2}\right)}{\Gamma \left(\frac{1-x}{2}\right)}+\frac{P^3}{3}\,\ln(2P)
  =\frac{P^3}{3}\ln\frac{2P}{A^{12}}\,.
\end{equation*}
Evaluation of the thermodynamic free energy \eqref{free-quiver} on \eqref{Rk-Ex2} gives of course the same result.

\section{Summary and outlook}
\label{sec:outlook}

In the present work we have exhaustively investigated linear quiver theories with eight supercharges.\par
The long quiver limit has been solved in arbitrary dimension $d \ge 3$ (Section \ref{sec:d}). The central ingredient in our method is the piece-wise linear rank function $\mathcal{R} (\eta)$, which encodes the specifics of the conformal linear quiver under consideration. This rank function is defined independently of the dimension $d$ and serves as input for the holographic dual problem in supergravity when $d$ is integer.\par
The function $\tilde{F}_d$ given in \eqref{eq:Ftilded} interpolates between the free energies in $d=3$ and $d=5$ and matches the Weyl anomaly coefficient $a$ in $d=4$. Besides, not only \eqref{eq:Ftilded} reproduces the previously known answers when evaluated at $d\in \left\{ 3,5 \right\}$, but it extends them to finite 't Hooft couplings.\par
The interpolation across dimensions successfully extends to defect one-point functions, including expectation values of half-BPS Wilson loops. We have computed the expectation value of one-, two-, and three-dimensional defects in $d \ge 3$, finding a universal behaviour.\par
\medskip
\begin{table}[th]
\centering
	\begin{tabular}{l|c|c|c|}
		& length & 't Hooft coupling & supergravity dual\\
		\hline 
		Subsec.~\ref{sec:4dZlong}-\ref{sec:a4d} & large & arbitrary & no \\
		Subsec.~\ref{sec:4dshortQ} & arbitrary & infinite & no \\
		Subsec.~\ref{sec:4dStrong}-\ref{sec:4dlogZstrong} & large & infinite & yes \\
		\hline
	\end{tabular}
\caption{Four-dimensional $\mN=2$ conformal linear quivers at large-$N$.}
\label{tab:4dlimits}
\end{table}
We have then tackled four-dimensional $\mN=2$ conformal linear quivers directly, without resorting to continuation of the spacetime dimension (Section \ref{sec:4dLQ}). Differently from their $d \ne 4$ analogues, balanced linear quivers in $d=4$ are conformal at any coupling. We have therefore addressed their large-$N$ limit both at arbitrary 't Hooft coupling and at infinite 't Hooft coupling, finding perfect agreement with the results from analytic continuation in $d$. Furthermore, we have solved the large-$N$ limit of conformal linear quivers of arbitrary length. The various regimes are summarised in Table \ref{tab:4dlimits}.\par
Considering the different large-$N$ regimes allowed us to go beyond the computation of the Weyl anomaly coefficient and compute the free energy. While the latter is in general scheme-dependent, we have identified and extracted an unambiguous part. To test our results we have proposed a simplified computation inspired by the on-shell action in the supergravity dual of Gaiotto--Maldacena, finding agreement in the universal part of the dependence on the gauge theory data (Subsection \ref{sec:toysugra}). Our general results have been exemplified in two explicit instances in Section \ref{sec:ex}.\par
A take-home summary of this work is:
\begin{center}\noindent\setlength{\fboxrule}{3pt}%
\fbox{\parbox{0.8\linewidth}{%
\vspace{6pt}
Long linear quivers at large-$N$ show a universal behaviour, are dealt with uniformly as functions of $d \ge 3$, and are amenable to possessing a supergravity dual.
\vspace{6pt}
}}\end{center}\par
\medskip
Superconformal gauge theories in four dimensions can be also modelled by circular quivers. A linear quiver is a circular quiver with one node erased, by setting one coupling to zero, but that necessarily takes the theory out of the strong-coupling regime. The circular quivers indeed have a very different holographic description, they are dual to orbifolds of AdS$_5\times \mathbb{S}^5$ at some point in their parameter space, and thus naturally belong to type IIB strings, while linear quivers have type IIA holographic duals. The difference is well refelcted in the matrix model. The large-$N$ solution of a linear quiver, as we saw, reaches a limiting shape at infinite coupling. No such limiting shape exists for a circular quiver, where the coupling dependence persists up to arbitrary large coupling \cite{Rey:2010ry}.\footnote{Precision and perturbative \cite{Galvagno:2020cgq,Galvagno:2021bbj} studies of circular quivers at the orbifold point \cite{Billo:2021rdb,Billo:2022gmq,Billo:2022lrv} and beyond \cite{Zarembo:2020tpf,Ouyang:2020hwd} detail this picture to a much degree. Circular quivers are not our main focus and we refer the reader to these references for more detail on their large-$N$ behavior.}\par
\medskip
A problem that remains open is to substantiate the comparison of our results with a full-fledged string theory computation.\par
It might also be possible to find supergravity solutions that hold for real $3 \le d \le 6$ and interpolate among the known solutions at integer $d$, thus completely paralleling the field theory calculations in this work.\par
Along different lines, one consequence of the long quiver limit working uniformly in $d$ is the existence of identities relating partition functions of different quivers for each $d$, as shown in Subsection \ref{sec:mirror}. These identities are reminiscent of three-dimensional mirror symmetry but, contrary to mirror symmetry, only relate balanced quivers with balanced quivers. It would be desirable to elucidate the origin of these relations and possibly derive them from an underlying automorphism of quantum field theories.

\vspace{0.4cm}
\subsubsection*{Acknowledgements}
We thank Stefano Cremonesi and Amihay Hanany for discussions.\\
C.N. is supported by STFC grants ST/X000648/1 and ST/T000813/1.
The work of L.S. was supported by the Shuimu Scholars program of Tsinghua University and by the Beijing Natural Science Foundation project IS23008 ``Exact results in algebraic geometry from supersymmetric field theory''.
The work of K.Z. was supported by VR grant 2021-04578. Nordita was partially supported by Nordforsk.
The authors have applied to a Creative Commons Attribution (CC-BY) licence.

\vspace{0.5cm}

\begin{appendix}

\section{Special functions}

\subsection{Analytically continued one-loop determinants}
\label{app:Minahan}
In this appendix we review the proposal of \cite{Minahan:2015any} (further substantiated in \cite{Minahan:2017wkz,Gorantis:2017vzz}) for the one-loop determinants on the round $\cs^d$ for arbitrary $3 \le d \le 7$. From there, we extract the functions $f_{\mathrm{v}},f_{\mathrm{h}}$. See also \cite{Pestun:2016jze} for an overview of localisation on $\cs^d$ in dimension $d \in [3,7] \cap \N$.\par
Let $G$ be a compact, semi-simple Lie group. Let $\Lambda_{\mathrm{r}}$ be the root lattice of $G$, $\triangle_{+}$ denote the set of positive roots, and $\langle \cdot , \cdot \rangle $ be the pairing between $\Lambda_{\mathrm{r}}$ and $\mathfrak{g}$. Let also $\mathfrak{R}$ be the (reducible) representation of $G$ in which the hypermultiplets transform. For linear quivers, $G$ is as in \eqref{eq:GaugelinearQ}, and $\mathfrak{R}$ is the direct sum of the bifundamental of $\mathrm{U}(N_j)\times \mathrm{U}(N_{j+1})$ with $K_j$ copies of the fundamental of $\mathrm{U}(N_j)$, for every $j=1,\dots, P-1$. We denote $\Lambda_{\mathfrak{R}}$ the weight lattice of $\mathfrak{R}$, and use the same symbol $\langle \cdot , \cdot \rangle $ for the pairing between $\Lambda_{\mathfrak{R}}$ and $\mathfrak{g}$. As in the main text, the zero-mode of the real scalar in the vector multiplet for $G$ is denoted $\vec{\phi}\in \mathfrak{t}$, and $\ii := \sqrt{-1}$.\par
The infinite products 
\begin{align}
	Z_{\mathrm{vec}} (\vec{\phi}) & = \prod_{\alpha \in \triangle_+} \prod_{n=0}^{\infty} \left[ \left(n^2 + \langle \alpha , \vec{\phi} \rangle^2 \right) \left((n+d-2)^2 + \langle \alpha , \vec{\phi} \rangle^2 \right)\right]^{\frac{\Gamma (n+d-2)}{\Gamma (n+1)\Gamma (d-2)}} \label{eq:ZvecdM}\\
	Z_{\mathrm{hyp}} (\vec{\phi}) & = \prod_{w \in \Lambda_{\mathfrak{R}}} \prod_{n=0}^{\infty} \left[ \left(n + \ii \langle w , \vec{\phi} \rangle +\frac{d-2}{2} \right) \left(n - \ii \langle w , \vec{\phi} \rangle +\frac{d-2}{2} \right)\right]^{- \frac{\Gamma (n+d-2)}{\Gamma (n+1)\Gamma (d-2)}} \label{eq:ZhypdM}
\end{align}
are defined for $d \ge 3$. The infinite products over $n$ are dealt with using $\zeta$-function regularisation exactly as for $d \in \N$.\par
\begin{lemma}[\cite{Minahan:2015any}]
	When $d \in [3,7] \cap \N$, expressions \eqref{eq:ZvecdM} and \eqref{eq:ZhypdM} reproduce the one-loop determinants of the vector multiplet and hypermultiplet, respectively, in the background of the trivial connection on $\cs^d$.
\end{lemma}
Note that $\frac{d-2}{2}$ in \eqref{eq:ZhypdM} comes from the scaling dimension of the hypermultiplet, which is protected by the non-Abelian R-symmetry when the gauge theory preserves eight supercharges.\par
Passing everything to the exponential, we define the functions $f_{\mathrm{v}}$ and $f_{\mathrm{h}}$ through, respectively, 
\begin{align*}
	Z_{\mathrm{vec}} (\vec{\phi}) & = \exp \left\{ - \sum_{\alpha \in \triangle_+} f_{\mathrm{v}} ( \langle \alpha , \vec{\phi} \rangle ) \right\}  \\
	Z_{\mathrm{hyp}} (\vec{\phi}) & = \exp \left\{ - \sum_{w \in \Lambda_{\mathfrak{R}}} f_{\mathrm{h}} ( \langle w , \vec{\phi} \rangle ) \right\} .
\end{align*}
Therefore, for an indeterminate $y\in \R$, 
\begin{align*}
	f_{\mathrm{v}} (y) &= - \sum_{n=0}^{\infty} \frac{\Gamma (n+d-2)}{\Gamma (n+1)\Gamma (d-2)} \ln \left[ \left(n^2 + y^2 \right) \left((n+d-2)^2 + y^2 \right) \right] \\
	f_{\mathrm{h}} (y) &= \sum_{n=0}^{\infty} \frac{\Gamma (n+d-2)}{\Gamma (n+1)\Gamma (d-2)} \ln \left[\left(n + \ii y +\frac{d-2}{2} \right) \left(n - \ii y +\frac{d-2}{2} \right)\right] .
\end{align*}
In particular they are even functions. Moreover, we will not need $f_{\mathrm{v}}(y),f_{\mathrm{h}}(y)$ explicitly, but rather $\partial_y f_{\mathrm{v}}$ and $\partial_y f_{\mathrm{h}}$:
\begin{align*}
	\partial_y f_{\mathrm{v}}(y) &= -2 y\sum_{n=0}^{\infty} \frac{\Gamma (n+d-2)}{\Gamma (n+1)\Gamma (d-2)} \left( \frac{1}{n^2+y^2} + \frac{1}{(n+d-2)^2+y^2} \right) \\
	\partial_y f_{\mathrm{h}}(y) &= 2 y \sum_{n=0}^{\infty} \frac{\Gamma (n+d-2)}{\Gamma (n+1)\Gamma (d-2)} \left( \frac{1}{\left(n+\frac{d-2}{2}\right)^2+y^2} \right) .
\end{align*}
The infinite sums over $n$ can be performed explicitly and, after simplification, one is left with \cite[Eq.(4.1) \& Eq.(4.7)]{Minahan:2015any}
\begin{align*}
	\partial_y f_{\mathrm{v}}(y) &= - \ii \Gamma (3-d) \left[ \frac{\Gamma (\ii y)}{\Gamma (\ii y + 3-d)} + \frac{\Gamma (\ii y +d-2)}{\Gamma (\ii y + 1)}  \ - \ (y \mapsto -y) \right] \\
	\partial_y f_{\mathrm{h}}(y) &= 2 \ii \Gamma (3-d) \left[ \frac{\Gamma \left( \ii y + \frac{d-2}{2}\right)}{\Gamma \left(\ii y + 2-\frac{d-2}{2}\right)}  \ - \ (y \mapsto -y) \right]
\end{align*}
where the notation $- (y \mapsto -y)$ means that the two summands are repeated with negative sign and with argument replacing $y$ with $-y$. Indeed, being $f_{\mathrm{v}},f_{\mathrm{h}}$ even functions of $y$, their derivatives must be odd.\par
For the linear quivers of the main text and an arbitrary function $f^{\prime}: \R \to \R$, the root and weight lattices are such that
\begin{align*}
	\sum_{\alpha \in \triangle_+} f^{\prime} (\langle \alpha , \vec{\phi} \rangle  ) & = \sum_{j=1}^{P-1} \sum_{a=1}^{N_j -1}\sum_{b=a+1  }^{N_j} f^{\prime} (\phi_a ^{(j)} - \phi_b ^{(j)} ) \\
	\sum_{w \in \Lambda_{\mathfrak{R}}} f^{\prime} ( \langle w , \vec{\phi} \rangle ) & = \sum_{j=1}^{P-1} \sum_{a=1}^{N_j }\left[ K_j f^{\prime}(\phi_a^{(j)}  ) + \sum_{b=1 }^{N_{j+1}} f^{\prime} ( \phi_a ^{(j)} - \phi_b ^{(j+1)}) \right] .
\end{align*}
Using these expressions we arrive at the formulae used in Section \ref{sec:d}.

\subsection{Lerch's transcendent}
\label{app:lerch}
In this appendix we collect the main properties of Lerch's transcendent function \cite{Lerch}
\begin{equation*}
	\Phi (z,d,a):= \sum_{k=0}^{\infty} \frac{z^k}{(k+a)^d} ,
\end{equation*}
defined for $a>0$ and $\lvert z \rvert <1$, or $\lvert z \rvert=1$ and $\Re (d)>1$. The latter case is the one realised by our gauge theories. It follows from the infinite series definition that Lerch's transcendent generalises both the Hurwitz $\zeta$-function and the polylogarithm: 
\begin{equation*}
	\Phi (1,d,a) = \zeta (d,a) , \qquad z \Phi (z,d,1)= \mathrm{Li}_d (z) .
\end{equation*}
It admits various integral representations, including 
\begin{equation*}
	\Phi (z,d,a) = \frac{1}{\Gamma (d)} \int_0^{\infty} \dd u ~\frac{u^{d-1} \ee^{-au}}{1-z\ee^{-u}} ,
\end{equation*}
which provides a definition valid outside the disk $\lvert z \rvert \le 1$. Another useful integral representation is 
\begin{equation*}
	\Phi (z,d,a) =-\Gamma (1-d) \int_{\mathcal{C}} \frac{\dd u}{2\pi \ii}~\frac{(-u)^{d-1} \ee^{-au}}{1-z\ee^{-u}} ,
\end{equation*}
with the integration contour $\mathcal{C}$ being a Hankel contour (encircles the positive real semi-axis counterclockwise) that leaves outside all the poles of the integrand at $u=\ln(z) + 2\pi \ii \Z$.\par
We need the following result.
\begin{lemma}
For $d \in \N$ and $\lvert z \rvert \le 1$,
\begin{equation*}
	z^{a} \Phi (z,d,a) = \sum_{\substack{k=0 \\ k \ne d-1} }^{\infty} \frac{\zeta (d-k,a) }{k!} (\ln z)^k + \left[ \psi (d)- \psi(a) - \ln (-\ln(z))\right] \frac{ (\ln z)^{d-1}}{(d-1)!},
\end{equation*}
where $\zeta (\cdot, \cdot)$ is Hurwitz's $\zeta$-function and $\psi (\cdot)$ is the digamma function.
\end{lemma}

\section{Three perspectives on the saddle point equation}
\label{app:3SPE}
We discuss three different approaches to derive the saddle point equation \eqref{eq:spe1}. The content of this appendix will certainly be known to most readers, but we wish to emphasise the physical significance of certain aspects of the computation.\par
The canonical way to derive the saddle point equation, reviewed for example in \cite{Marino:2004eq,Russo:2013sba}, goes as follows. One keeps $S_{\mathrm{eff}}$ written in terms of the eigenvalues $\phi^{(j)}_a$ of the matrix model, and keeps the sums instead of the integrals. The saddle point configuration is a point in $\R^{\mathrm{rk}(G)}$, where 
\begin{equation*}
	\mathrm{rk}(G) = \sum_{j=1}^{P-1} N_j .
\end{equation*}
It is determined by solving the system 
\begin{equation*}
	\frac{\partial S_{\mathrm{eff}} }{\partial \phi_a^{(j)}} =0 , \qquad \forall~a=1, \dots, N_j , \ \forall~j=1, \dots, P-1.
\end{equation*}
It is at this stage that one introduces the eigenvalue densities $\rho_j (\phi)$, which reduce the system of $\mathrm{rk}(G)$ equations to a system of $(P-1)$ integral equations for the $(P-1)$ measures $\rho_j (\phi)\dd \phi$ on $\R$, for $j=1, \dots, P-1$. In the large-$P$ limit, we can collect all of them into a single measure $\rho(z,\phi)\dd \phi \dd z$ on the strip $[0,1]\times\R$. We thus reduce the problem to a unique saddle point equation. Performing the steps explicitly, one finds \eqref{eq:spe1}.\par
\bigskip
There are at least two alternative methods. In both cases, one starts by approximating the sums $\sum_{a=1}^{N_j}$ with integrals over the variable 
\begin{equation*}
	\mathsf{a}=\frac{a}{N_j}, \qquad 0<\mathsf{a}\le \nu(z). 
\end{equation*}
The analogous substitution is made to replace $\sum_{j=1}^{P-1}$ with an integral over the variable $z=j/P$, $0<z<1$. The eigenvalues $\phi_a^{(j)}$ become functions 
\begin{equation}
\label{eq:phicontfunc}
\phi : [0,1]\times [0, \nu(z)] \to \R
\end{equation}
and, for every function $f(\phi)$, 
\begin{equation*}
	\sum_{j=1}^{P-1} \sum_{a=1}^{N_j} f\left( \phi_a^{(j)}\right) \simeq P \int_0^1 \dd z ~N \int_0^{\nu(z)} \dd \mathsf{a} ~f \left( \phi(z,\mathsf{a})\right) .
\end{equation*}
The approximation becomes exact as $N,P\to \infty$. The $\mathrm{rk}(G)$-fold integral over the eigenvalues in the matrix model is replaced by a functional integral over continuous functions \eqref{eq:phicontfunc}. One then implicitly inverts the relation $\phi=\phi(z,\mathsf{a})$ to write a function $\mathsf{a}(z,\phi)$. The density of eigenvalues is defined as the inverse of the derivative $\frac{\partial \phi}{\partial \mathsf{a}}$, namely 
\begin{equation*}
	\rho (z,\phi) := \frac{\partial \mathsf{a}(z,\phi)}{\partial \phi}.
\end{equation*}
This definition agrees with the definition using $\delta$-functions. Writing the latter relation as an equality of measures, one replaces the integrals 
\begin{equation*}
	\int_0^{\nu(z)} \dd \mathsf{a} ~f \left( \phi(z,\mathsf{a})\right) = \int_{- \infty}^{\infty} \rho (z,\phi) \dd\phi ~f \left( z,\phi\right).
\end{equation*}
This is the presentation given in the main text.\par
One way to obtain the saddle point equation from there is to take the functional derivative 
\begin{equation}
\label{eq:funcspe}
	\frac{\delta S_{\mathrm{eff}} }{\delta \rho } =0.
\end{equation}
However, the effective action $S_{\mathrm{eff}}$ can always be redefined by a (scheme-dependent) term of the form 
\begin{equation*}
	S_{\mathrm{c.t.}} = c \int_{0} ^{1} \dd z  \int \dd \phi \rho (z, \phi) ,
\end{equation*}
for a constant $c$. This action is a local counterterm that multiplies the partition function by a complex number, 
\begin{equation*}
	\mz_{\cs^d} \ \mapsto \ \ee^{-S_{\mathrm{c.t.}}} \mz_{\cs^d} ,
\end{equation*}
without altering the field content and interactions in the gauge theory. As usual, we are interested in studying the theory up to these counterterms.\par
By the above discussion, shifting $S_{\mathrm{eff}} \mapsto S_{\mathrm{eff}} + S_{\mathrm{c.t.}}$ produces a shift of the functional derivative by a constant $c$. We are therefore interested in \eqref{eq:funcspe} up to constant shifts, which leads us to solve 
\begin{equation}
\label{eq:funcspe2}
	\partial_{\phi}\left(\frac{\delta S_{\mathrm{eff}} }{\delta \rho } \right)=0.
\end{equation}
Computing the latter explicitly indeed yields \eqref{eq:spe1}.\par
\bigskip
The third approach is a variation of the above. One establishes a variational problem on the functional space of continuous functions $\mathsf{a}(z,\phi)$. One first (implicitly) inverts the relation $\phi=\phi(z,\mathsf{a})$ from \eqref{eq:phicontfunc}, introduces the density of eigenvalues, and writes $S_{\mathrm{eff}}$ as we have done in the main text. This procedure sets up the variational problem with respect to the function $\mathsf{a}(z,\phi)$. $S_{\mathrm{eff}}$ does not depend explicitly on $\mathsf{a}$, but depends on $\rho = \partial_{\phi}\mathsf{a}$. The functional Euler--Lagrange equation yields \eqref{eq:funcspe2}.

\end{appendix}	
\section*{Declarations}
\subsubsection*{Funding and/or Conflicts of interests/Competing interests}
The authors have no relevant financial or non-financial interests to disclose.
\bibliography{long4d}

\end{document}